\documentclass[prl,twocolumn,superscriptaddress]{revtex4-2}
\usepackage{bm}
\usepackage{qcircuit}
\usepackage{graphicx}
\usepackage{subcaption}
\usepackage[dvipsnames]{xcolor}
\usepackage{dcolumn}
\usepackage{braket}
\usepackage{amsmath}
\usepackage{color}
\usepackage{amssymb}
\usepackage{amsthm}
\usepackage{amsmath}
\usepackage{romannum}
\usepackage{dsfont}
\usepackage{adjustbox}
\usepackage{tikz}
\usetikzlibrary{decorations.markings}
\usepackage{graphicx}
\usepackage{subcaption}
\usepackage{pgfplots}
\pgfplotsset{compat=1.3}
\usepgfplotslibrary{fillbetween}
\usetikzlibrary{patterns}
\usepackage{mathtools}
\usepackage{xcolor}
\usepackage{enumerate}
\usepackage{soul}
\usepackage[many]{tcolorbox}    	% for COLORED BOXE
\usepackage{physics}
\usepackage[labelfont=bf,
   justification=Justified,
   format=plain]{caption}
\usepackage[hidelinks]{hyperref}% add hypertext capabilities
\hypersetup{
    colorlinks,
    linkcolor={red!50!black},
    citecolor={blue!50!black},
    urlcolor={blue!80!black}
}

\newtheorem{theorem}{Theorem}

\definecolor{amber}{rgb}{1.0, 0.75, 0.0}
\definecolor{color1}{RGB}{94, 201, 98}
\definecolor{color2}{RGB}{33, 145, 140}
\definecolor{color3}{RGB}{59, 82, 139}

\definecolor{plasma_color3}{RGB}{248, 149, 64}
\definecolor{plasma_color2}{RGB}{204, 71, 120}
\definecolor{plasma_color1}{RGB}{126, 3, 168}

\definecolor{colore}{rgb}{0.8, 0.8, 0.8}
\newtcolorbox{boxA}{
    colback = colore, % background color
    boxrule = 0pt  % no borders
}

\begin{document}

\pagenumbering{arabic}
\title{Symmetry-guided quantum state preparation: \\Branched-Subspaces Adiabatic Preparation (B-SAP)}

\author{Davide Cugini}
\email{davide.cugini01@universitadipavia.it}
\affiliation{Dipartimento di Fisica ``Alessandro Volta,'' Universit\`a di Pavia, via Bassi 6, 27100, Pavia, Italy}
\affiliation{Theoretical Division, Los Alamos National Laboratory, Los Alamos, NM 87545, USA}

\author{Giacomo Guarnieri}
\affiliation{Dipartimento di Fisica ``Alessandro Volta,'' Universit\`a di Pavia, via Bassi 6, 27100,  Pavia, Italy}
\author{Mario Motta}
\address{IBM Quantum, T. J. Watson Research Center, Yorktown Heights, NY
10598, USA}
\author{Dario Gerace}
\affiliation{Dipartimento di Fisica ``Alessandro Volta,'' Universit\`a di Pavia, via Bassi 6, 27100,  Pavia, Italy}

\begin{abstract}
Quantum state preparation lies at the heart of quantum computation and quantum simulations, enabling the investigation of complex manybody systems across physics, chemistry, and data science. 
While existing methods such as Variational Quantum Algorithms (VQAs) 
and Adiabatic Preparation (AP) offer viable pathways, both face substantial limitations.
Here we introduce a hybrid algorithm that integrates the conceptual strengths of both VQAs and AP, enhanced via the use of group-theoretic structures and classical post-processing to approximate ground and excited states of many-body Hamiltonian models.
We validate our approach by applying it to the one-dimensional XYZ Heisenberg model
with periodic boundary conditions,
evaluating its performance across a broad range of parameters and system sizes. Our results show accurate preparation of low-energy eigenstates, achieved with circuit depths with polynomial scaling versus system size.
\end{abstract}

\maketitle
\section{Introduction}\label{Introduction}

Preparing ground, excited, and thermal states of quantum many-body Hamiltonians is generally considered a central task in quantum computation, with profound implications for quantum simulation~\cite{aolita2015reliable,eisert2020quantum,Tacchino2020AQT} and data science~\cite{farhi2000quantum, hogg2003adiabatic}. 
Perspective applications range from the simulation of quantum field theories~\cite{preskill2018simulating, funcke2023review, martinez2016, klco2018quantum}  to equilibration and thermalization~\cite{eisert2015quantum,bertoni2025typical}, as well as
engineering of quantum materials~\cite{daley2022practical, julia2025catalysis} and quantum chemistry in general~\cite{bauer2020quantum, gonzalez2020quantum,roldao2022quantum,levine2017photochemistry}.
Having access to a fault-tolerant quantum computer would in principle allow for such quantum state preparation via, e.g., phase-estimation algorithms; 
however, currently available quantum hardware hinders their effectiveness 
and calls for alternative approaches that take into account the severly limited resources.
Variational Quantum Algorithms (VQAs) provide an alternative~\cite{cerezo2021variational}, 
but their reliance 
on an ansatz and
on high-dimensional classical optimization often limits scalability and convergence. 
Adiabatic Preparation (AP) ~\cite{born1928beweis,albash2018adiabatic} represents a conceptually distinct route, 
slowly evolving a given initial state 
under a time-dependent Hamiltonian toward the desired target state. 
However, this method is extremely sensitive to energy spectral-gaps closing, which often happens 
unless the adiabatic process can be tailored to the specific problem.
The aforementioned limitations often hinder the simultaneous preparation of the ground and the lowest excited states, 
which may be essential, e.g., for understanding key properties of physical and chemical systems~\cite{babbush2014adiabatic}. 
Moreover, they typically prevent any investigation of higher-order excited states, which are often physically relevant
to determine, e.g., optical spectra~\cite{griffith1957ligand, bredas2014mind}, charge-transfer processes~\cite{tiwary2016spectral}, and reaction intermediates~\cite{lupo2023two, deshpande2022importance}, 
providing insights into chemical reactivity and material properties that cannot be obtained from the ground state alone~\cite{somma2013spectral, arad2017rigorous}.
Therefore, balancing accuracy, resource efficiency, and robustness across these methods remains a critical open problem for unlocking practical quantum advantage in simulation and beyond.

In this work, 
we propose an algorithm,
which we call Branched-Subspaces Adiabatic Preparation (B-SAP),
designed to efficiently approximate 
target eigenstates of a many-body Hamiltonian $H_T$. 
By employing tools from group theory, 
we combine the strengths of AP 
and VQAs, 
while overcoming their respective limitations.
In particular, B-SAP avoids the need for circuit ansatzes and sidesteps common issues associated with classical optimization. At the same time, its validity depends on less stringent conditions than those required in conventional adiabatic procedures.
The first two sections provide brief overviews of VQAs and AP, respectively, 
while B-SAP is introduced in the following one.
Focusing on the Heisenberg Hamiltonian 
of an $L$-qubit system, 
we use our protocol to approximate its lowest eigenstates,
with a circuit depth scaling polynomially with $L$, 
across all ranges of values 
of the coupling constants in the Hamiltonian.
Our results validate the effectiveness of the proposed algorithm 
and demonstrate its potential for quantum state preparation 
on near-term quantum computing platforms, 
while offering a promising route 
to reducing computational costs in future fault-tolerant architectures.

\section{variational quantum algorithms}\label{variational quantum algorithms}
In the last few years, variational quantum  algorithms (VQAs) 
have emerged as a promising framework to tackle the problem of eigenstates preparation, 
leveraging the hybrid quantum-classical paradigm of optimizing quantum states 
within a parameterized ansatz quantum circuit~\cite{cerezo2021variational, peruzzo2014variational, kuzmin2020variational}.
Variational quantum circuits employ a set of tunable parameters embedded within a given quantum circuit. These parameters are iteratively optimized 
by minimizing a cost function via classical optimization algorithms,
trying to obtain a circuit output that is as close as possible to the desired eigenstate.
This hybrid approach enables efficient exploration of the Hilbert space while mitigating hardware noise and quantum resource constraints. 
By tailoring the ansatz to the specific properties of the desired target state, VQAs can approximate complex quantum states with significantly reduced circuit depth 
as compared to exact state preparation methods,
which in general require exponential cost.
Despite these benefits, variational quantum circuits still encounter several challenges~\cite{sim2019expressibility, benedetti2019parameterized}. 
In fact, 
the initial choice for the variational ansatz circuit is crucial to achieve a faithful approximate solution, since poorly designed quantum circuits may ultimately fail to capture the structure of the target state~\cite{grimsley2019adaptive}.
On the other hand, an ansatz is essential because realizing the most generic parameterized unitary gate would require a number of tunable parameters that grows exponentially with system size, 
making the training computationally impractical. 
In addition, a significant limitation 
is the possible occurrence of ``barren plateaus,'' i.e., regions of the parameters space in which the gradient of the cost function decreases exponentially as the system size increases,
making the parameters optimization ineffective for large systems \cite{ragone2024lie,mcclean2018barren,holmes2022connecting,larocca2025barren}.
Finally, classical optimization bottlenecks, 
such as vanishing gradients or local minima, 
can ultimately hinder  convergence, in general~\cite{spall2005introduction, lavrijsen2020classical}.
\begin{figure*}[t]
\centering
    \begin{subfigure}[]{0.48\textwidth}
    \includegraphics[width = \textwidth]{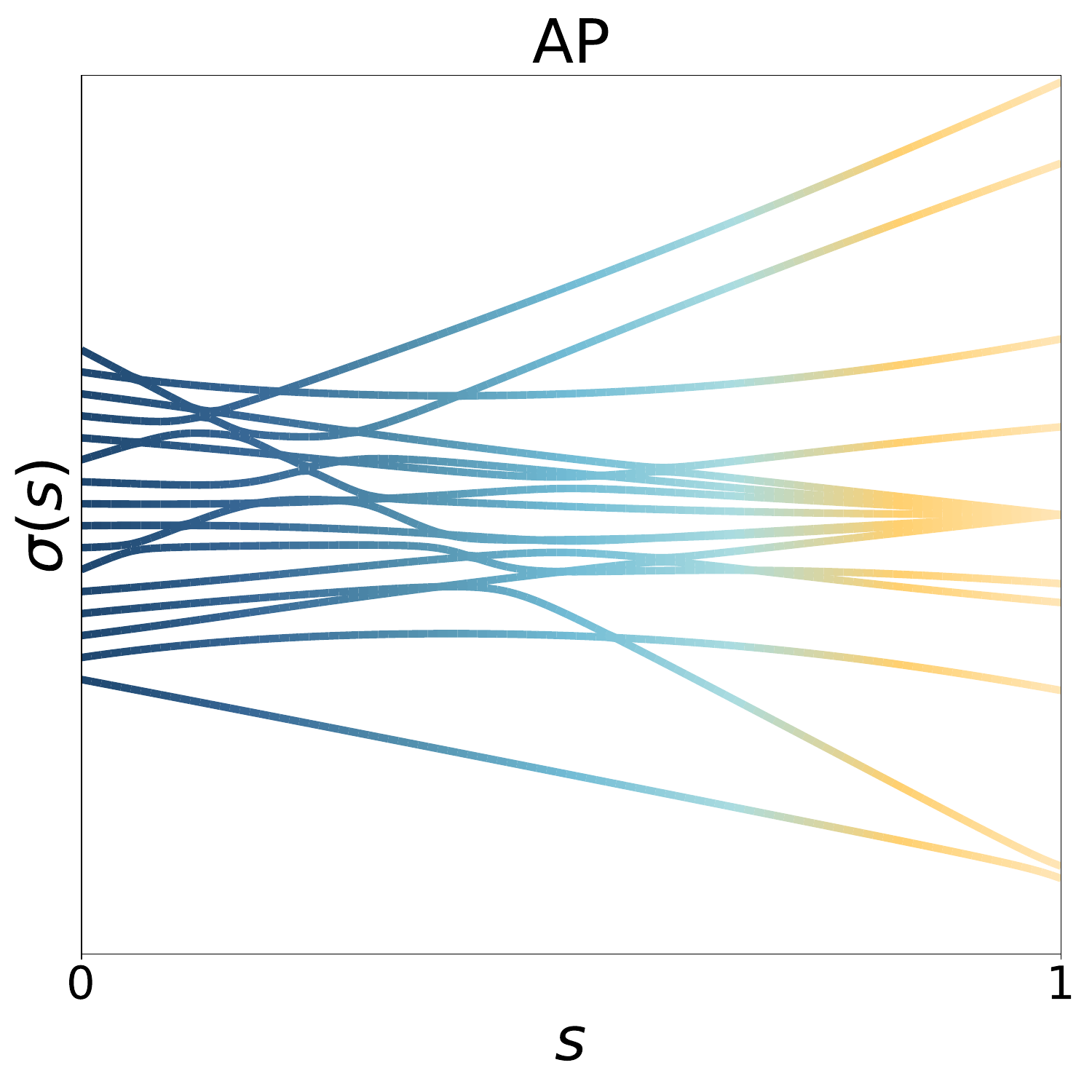}
    \caption{Conventional Adiabatic Preparation}
    \label{fig: AP spectrum}
    \end{subfigure}
    \hspace{0cm}
    \begin{subfigure}[]{0.48\textwidth}
    \includegraphics[width = \textwidth]{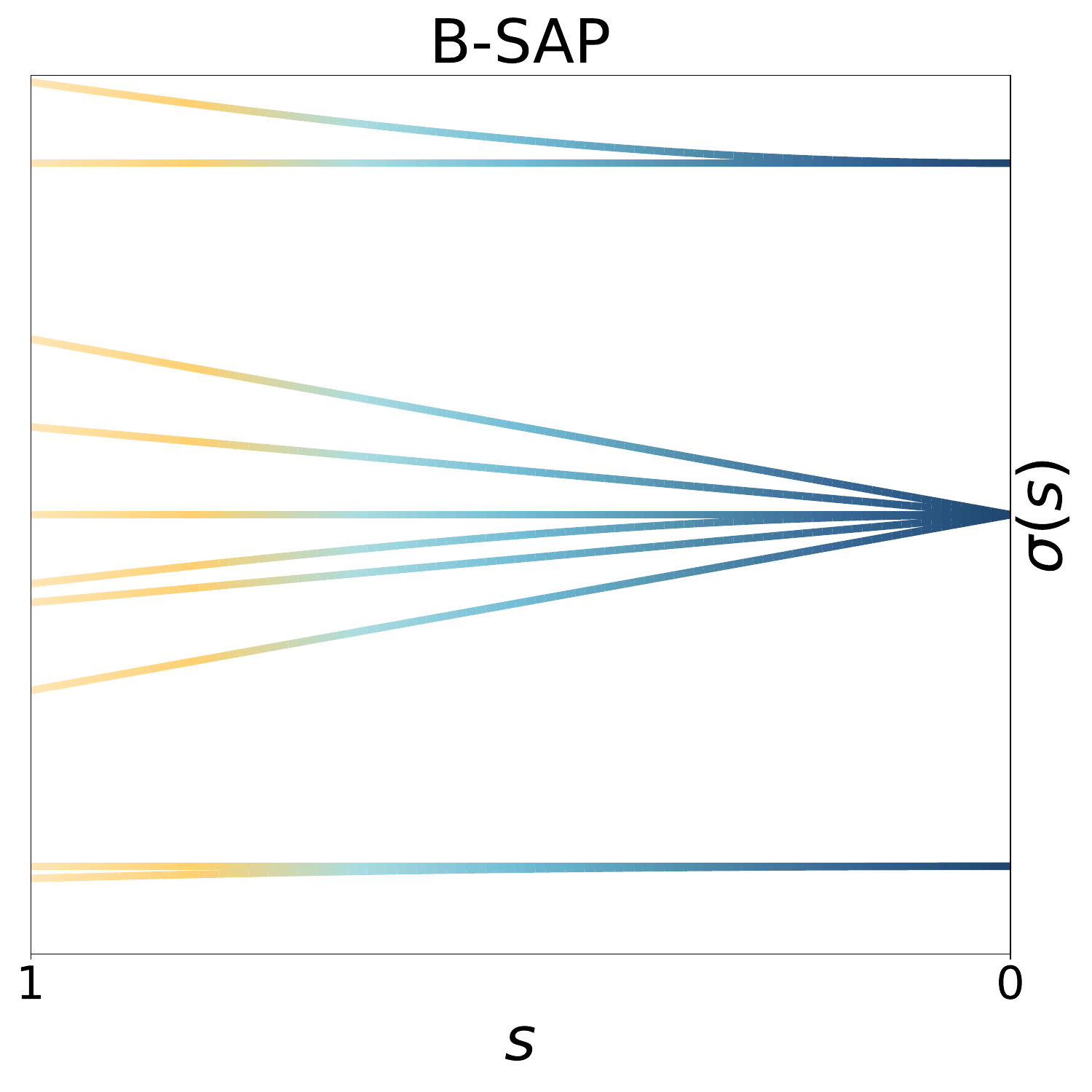}
    \caption{Branched-Subspace Adiabatic Preparation}
    \label{fig: BSAP spectrum}
    \end{subfigure}
    \caption{Illustrations of adiabatic energy spectra for two processes sharing the same final Hamiltonian but differing in their initial ones. 
    The left case~\ref{fig: AP spectrum}, related to conventional AP, starts with a non-degenerate spectrum, 
    but it features several level crossings. 
    In contrast, 
    the right panel illustrates the evolution performed through B-SAP~\ref{fig: BSAP spectrum}, which displays a forked structure with degeneracies that never increase throughout the evolution. Notice that the time evolution is plotted from 0 to 1 in~\ref{fig: AP spectrum}, and mirrored (i.e., from 1 to 0) in~\ref{fig: BSAP spectrum}, highlighting the differences in the final spectra. }
    \label{fig: spectra evolution}
\end{figure*}

\section{Adiabatic quantum-state preparation}\label{Adiabatic quantum state preparation}

We hereby summarize the Adiabatic Preparation (AP) algorithm \cite{farhi2001quantum,aspuru2005simulated,babbush2014adiabatic,albash2018adiabatic}
for the preparation of the eigenstates 
of a target Hamiltonian $H_T$. 
The pre-requisite is an auxiliary Hamiltonian $H_0$, whose eigenstates can be efficiently initialized on a quantum computer. From this, the algorithmic procedure consists of two main steps: 
(i)  the quantum system is prepared in an eigenstate of $H_0$, and then (ii) the system gets evolved under a time-dependent Hamiltonian of the form 
\begin{equation}
    H(s) = H_0 + f(s) [H_T - H_0]\,,
\end{equation}  
in which $s$ represents the time evolution parameter, and $f$ is a generic holomorphic function 
satisfying $f(0) = 0$ and $f(1) = 1$.
In principle, one is free to choose any auxiliary Hamiltonian $H_0$ and interpolating function $f$. 
This strategy is widely used, and it represents the basis for widespread algorithms, 
among which the Quantum Approximate Optimization Algorithm ~\cite{farhi2014quantum}. The evolution from $0$ to $s\in [0,1]$
is governed by the adiabatic evolution operator 
\begin{equation}
    U_\tau(s) = \mathcal{T} \exp\left[-i\tau \int_{0}^{s} H(s) ds \right]\,,
\end{equation}  
where $\mathcal{T}$ denotes the time-ordering operator, and $\tau > 0$ is the timescale of the adiabatic process.  

The efficiency of the AP approach can be formalized by considering the spectrum of $H(s)$, 
henceforth denoted as $\sigma(s)$, 
and by partitioning it into disjoint subsets
$\sigma(s) = \bigcup_n \sigma_n(s)$. 
Practically, 
such partitioning procedure is often based on physical considerations
such as an inherent coarse-graining of the energy measurement apparatus.
Let $P_n(s)$ be the projector into the subspace spanned by those eigenstates
whose eigenvalues belong to $\sigma_n(s)$. 
It is possible to quantify the AP error as 
\begin{equation}\label{eq: adiabatic error}
    \mathcal{E}_{AP} \left(s, \tau\right) = \left\| \left(\mathds{1}- P_n(s) \right)U_\tau P_n(0) \right\|, ,
\end{equation}
which is a generalization of the definition of ``infidelity'' ~\cite{nielsen2010quantum}.
The adiabatic theorem ~\cite{nenciu1993linear} ensures that if the spectral gap  
\begin{equation}\label{eq: gap condition}
    \Delta_n(s) = \min_{\substack{x \in \sigma_n(s) \\ y \in \;\sigma(s) \backslash \sigma_n(s)}} |x - y|
\end{equation}  
remains strictly positive for all $s \in [0, s^*]$, 
then, then the error decays exponentially fast with the protocol time $\tau$,
\begin{equation}\label{eq: exponential bound}
    \mathcal{E}_{AP}\left( s, \tau \right)  \sim \exp\left[-\tau/ \tilde{g_n}\right] \, ,
\end{equation}  
with $\tilde{g_n}$ being a positive constant 
that can be classically upper bounded a priori~\cite{cugini2025exponential}.
In particular, if the energy levels of $H(s)$ remain non-degenerate for all $s \in [0,1]$, one can take each $\sigma_n(s)$ to be the $n$-th eigenvalue. 
In this case, Eqs.~\eqref{eq: adiabatic error} and \eqref{eq: exponential bound} ensure that 
the $n$-th eigenstate of $H_0$ approximately evolves into the corresponding $n$-th eigenstate of $H_T$ at 
the final evolution step, $s = 1$, 
with an error that decays at least exponentially with $\tau$.  
If this is the case, 
the initialization of the n-th eigenstate of $H_0$,
followed by the application of $U_\tau \left( 1\right)$,
represents a simple way to approximate the n-th eigenstate of $H_T$
with arbitrary precision.
On the other hand, if any level crossing occurs at any point $s \in [0,1]$, Eq.~\eqref{eq: adiabatic error} no longer guarantees that a specific eigenstate of $H_0$ is uniquely mapped to its corresponding eigenstate of $H_T$.
In all these cases, which are very common in many-body systems, one needs to bundle all the crossing levels in the same partitioning $\sigma_n(s)$ in order to guarantee a strictly positive spectral gap $\Delta_n(s)$. Crucially, as a result the AP is only able to ensure that an initial eigenstate associated, e.g., to $\sigma_n(0)$, deterministically evolves into an unknown superposition of the eigenstates whose eigenvalues belong to $\sigma_n(1)$. 
From these general considerations, it is evident that level crossings stand as a fundamental limitation to the effectiveness of AP protocols, just as exponential closings of the energy gaps with system size $L$ 
can hinder the scaling in Eq.~\eqref{eq: exponential bound}, 
eventually requiring superpolynomial time $\tau$ for convergence.

Notwithstanding, a key general consideration can be made here to address this challenge, which will be central for the development of the B-SAP algorithm that we will present in the following Section. 
When multiple energy levels cross at a specific evolution parameter value \(s \in [0,1]\), the eigenvalues of the corresponding operator \(H(s)\) exhibit an increased degree of degeneracy. Degenerate eigenspaces correspond to irreducible representations (irreps) of the Hamiltonian symmetry group~\cite{landau2013quantum}, and the multiplicity of each energy level equals the dimension of the corresponding irrep. Therefore, a straightforward approach to hinder level crossings during adiabatic evolution is to design the auxiliary Hamiltonian, \(H_0\), such that the symmetry group of \(H(s)\) is as small as possible, thereby reducing the dimensions of the irreps, and thus limiting degenerate-eigenspace multiplicities that may give rise to crossings during the adiabatic evolution. 
This is typically achieved by adding terms to \(H_0\) that explicitly break the obvious symmetries of \(H_T\)~\cite{chakraborty2022classically, hamma2008adiabatic, ciavarella2023state}.  
However, care must be taken in this procedure, since the additional terms used to break symmetries can inadvertently introduce \emph{accidental symmetries}, i.e., hidden symmetries that are not immediately apparent from the geometric or physical structure of the Hamiltonian but arise from specific features of the model or the added terms. Such accidental symmetries can lead to unexpected degeneracies or level crossings in the instantaneous spectrum of \(H(s)\). As a classical example, we remind the Laplace--Runge--Lenz vector in the Kepler problem: although the obvious rotational symmetry is SO(3), the system has a hidden SO(4) symmetry arising from the \(1/r\) potential, which produces additional conserved quantities, and ensures that all bound orbits are closed ellipses. Its quantum analogue appears in the solution of eigenstates of the Hydrogen atom, in which the same SO(4) symmetry leads to degeneracies among energy levels with different angular momentum but the same principal quantum number. Similarly to the examples above, accidental symmetries can create undesired degeneracies during the adiabatic evolution in our setting, which may inevitably  complicate the adiabatic path.
These challenges highlight the need for a more refined strategy to enhance the robustness of adiabatic state preparation. In the following section, we introduce the B-SAP algorithm, which is specifically designed to mitigate issues arising from both obvious and accidental symmetries.

\section{Branched - subspaces adiabatic preparation}

\begin{figure}[t]
    \includegraphics[width = 0.48\textwidth]{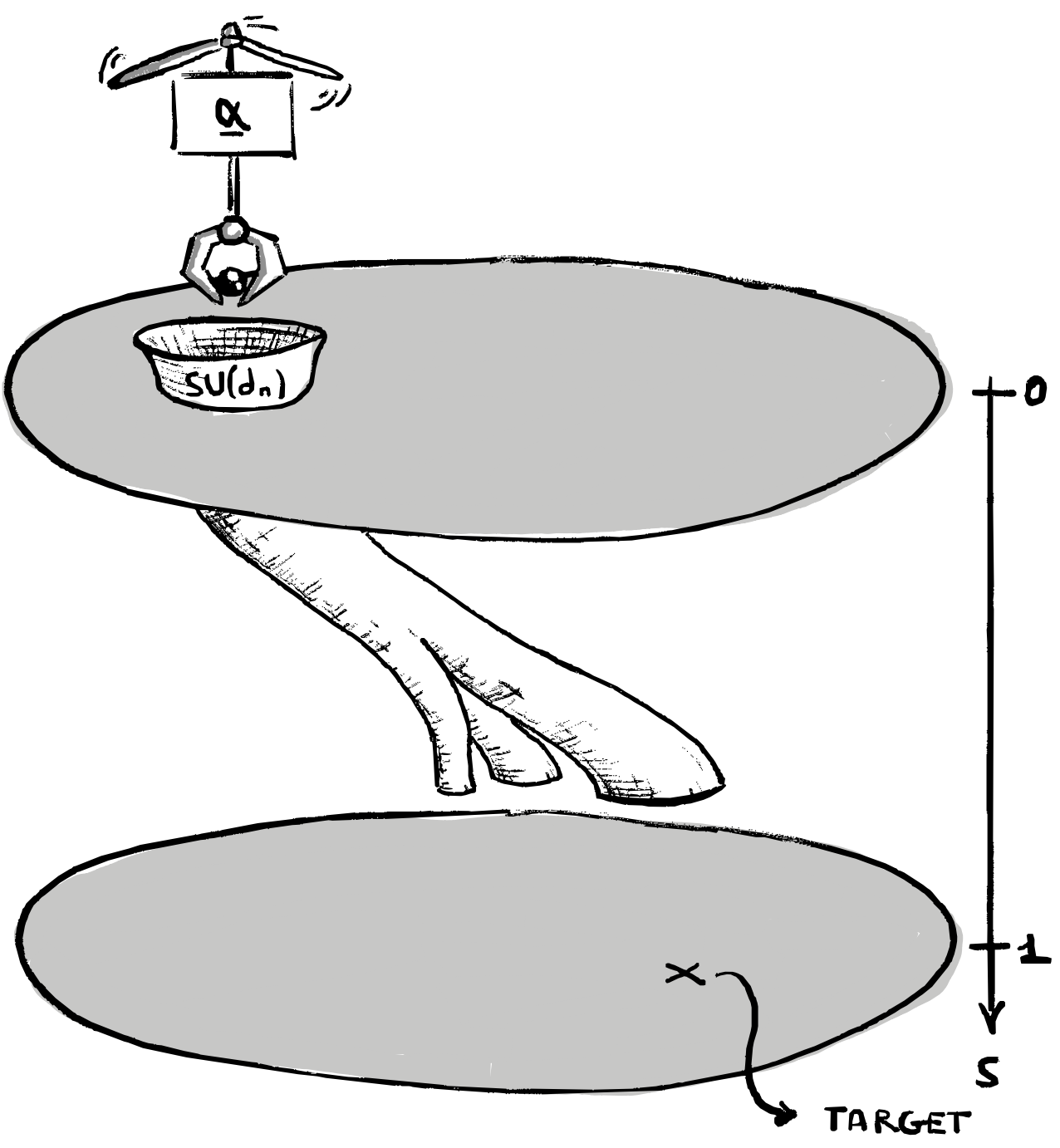}
    \caption{Schematic representation of the B-SAP protocol implemented on a quantum computer with $L$ qubits. 
    Two copies of the Hilbert space are depicted as gray manifolds.
    The hybrid algorithm trains a quantum circuit characterized by a polynomial number of tunable parameters $\{\underline{\alpha}\}$, enabling efficient exploration of a polynomially large subspace of the Hilbert space. 
    In contrast, a standard VQA would need to explore the exponentially large Hilbert space. 
    Finally, the adiabatic procedure, represented as a pipeline, bijectively maps the initial subspace onto a new one that includes the desired target state.}
    \label{fig : B-SAP}
\end{figure}

Here we introduce the ``Branched - Subspaces Adiabatic Preparation'' (B-SAP) method,  specifically designed to overcome the limitations of conventional AP and VQAs methods outlined in the previous Sections.
In fact, our strategy leverages the strengths of each technique, in order to mitigate their respective weaknesses.
We begin by reconsidering the framework 
for the adiabatic state preparation. In the previous Section we have outlined how, in the conventional AP approach,
an initial Hamiltonian is typically selected as possessing a symmetry group that is as small as possible. 
In complete contrast, we propose to choose the initial Hamiltonian $H_0$ such that its symmetry group is not only larger but, in fact, \textit{it contains the one related to the target Hamiltonian $H_T$ as a subgroup},
such that symmetries can only be broken during the AP process. 
As a result, symmetry does not cause distinct energy levels to merge. On the contrary, some of them may actually split along the adiabatic path, branching out
(see, e.g., Fig.~\ref{fig: spectra evolution}).
As a matter of fact, we realize that this choice may appear counterintuitive at first, 
as it results in energy levels at the beginning of the adiabatic process exhibiting equal or even larger degeneracy 
as compared to the final part. 
On the other hand,
the difficulty in discriminating various energy eigenstates of $H_0$ due to their high degeneracy
is now lifted by the fact 
that the initial energy spectrum is known, 
and the respective eigenstates can be efficiently initialized 
by assumption.
In practice, this allows to prepare well-characterized basis states with relatively shallow quantum circuits before the adiabatic sweep, rather than attempting to prepare complicated target eigenstates directly.
More precisely, let $\mathcal{B}$ denote an orthonormal basis of eigenstates of $H_0$,
and $\mathcal{B}_n \subset \mathcal{B}$ be
the subset of $d_n$ linearly independent eigenstates that span the degenerate subspace $\mathcal{S}\left( \mathcal{B}_n \right)$ corresponding to the n-th energy eigenvalue $E_n(s = 0)$. 
Our strategy is to exploit the knowledge of the basis $\mathcal{B}_n$ to construct the most general parametrized unitary gate, $\mathcal{G}(\underline{\alpha})$, acting as an endomorphism on $\mathcal{S}(\mathcal{B}_n)$.  
The action of $\mathcal{G}(\underline{\alpha})$ on a fixed initial state $\ket{\Psi} \in \mathcal{S}(\mathcal{B}_n)$,
while varying its $N_\alpha$ parameters $\underline{\alpha}$,
enables exploration of the entire subspace $\mathcal{S}(\mathcal{B}_n)$.  
Equivalently, the gate defines the mapping  
\begin{equation}\label{eq: G mapping}
    \mathcal{G}: \mathds{R}^{N_\alpha} \to \mathcal{S}(\mathcal{B}_n),
\end{equation}
which associates each parameter vector $\underline{\alpha}$ with the state $\mathcal{G}(\underline{\alpha})\ket{\Psi} \in \mathcal{S}(\mathcal{B}_n)$.    
Once $\mathcal{G}(\underline{\alpha})$ has been applied to $\ket{\Psi}$, the standard AP protocol can be executed.  
The output state is then measured, and the resulting outcomes are used to optimize the circuit parameters, as in any variational VQA.  
This procedure is illustrated schematically in Fig.~\ref{fig : B-SAP}.  
\begin{figure}[t]
    \includegraphics[width = 0.48\textwidth]{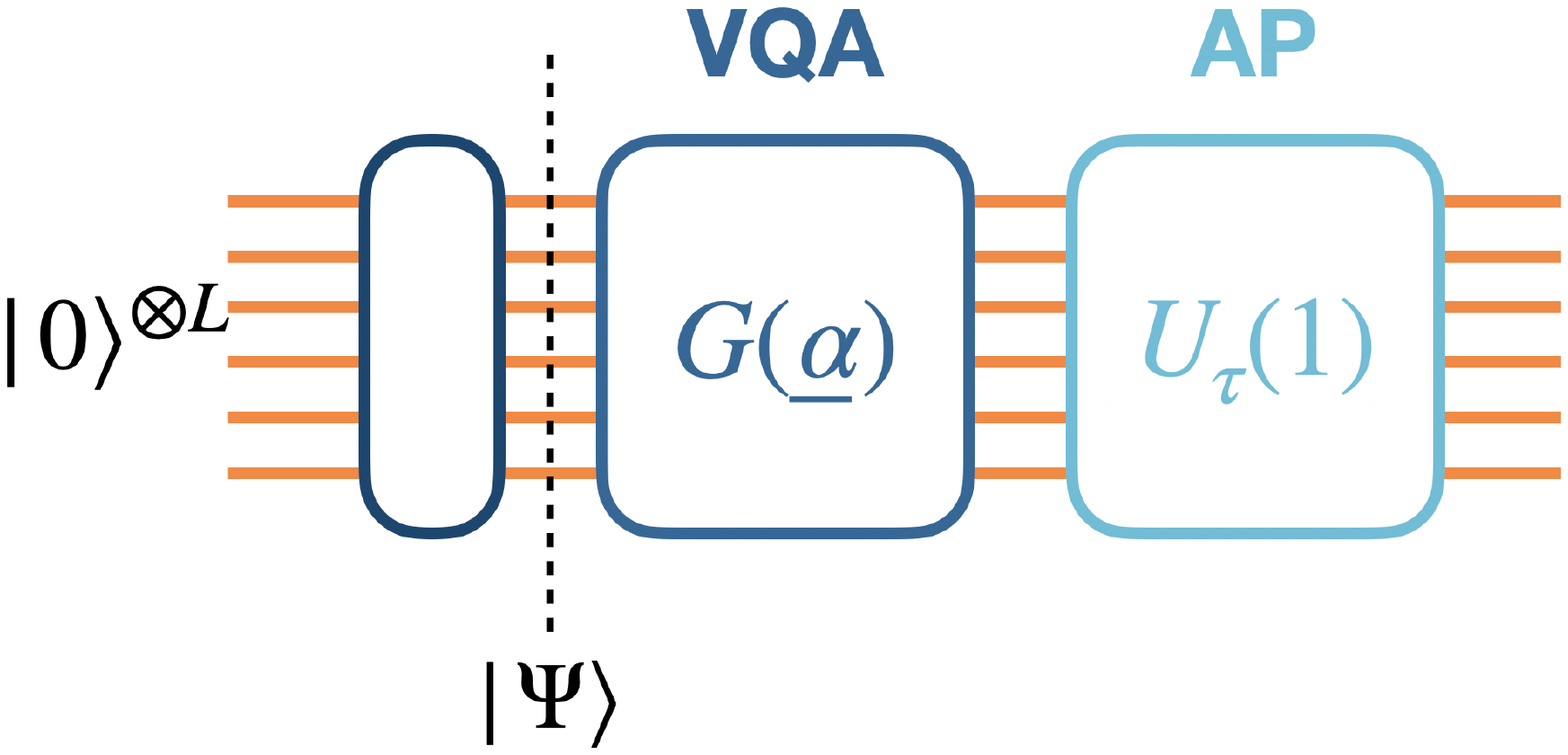}
    \caption{Schematic illustration of the quantum circuit implementing the B-SAP algorithm. The initial quantum state $\ket{\Psi} \in \mathcal{S}(\mathcal{B}_n)$ is first prepared by applying a low-depth circuit to the default $\ket{0}^{\otimes L}$ state. The parametrized unitary $G(\underline{\alpha})$ then allows exploration of the full subspace $\mathcal{S}(\mathcal{B}n)$, by varying the parameters $\underline{\alpha}$. Finally, the state undergoes adiabatic evolution via $U_\tau(1)$, after which the energy of the resulting state is measured to update the variational parameters.}
    \label{fig : circuit scheme}
\end{figure}
To formalize this idea, note that the set of unitary endomorphisms of $\mathcal{S}(\mathcal{B}_n)$ forms the unitary group $\mathrm{U}(d_n)$, whose Lie algebra $\mathfrak{u}(d_n)$ has dimension $d_n^2$.
Physically, the parametrized gate set performs all possible rotations and phase transformations within the degenerate subspace: each generator represents an infinitesimal rotation that mixes basis states or shifts their relative phases.  
We can then invoke the following result from Ref.~\cite{murnaghan1962unitary}.  
\begin{theorem}\label{theorem: decomposition}
Let $\{t_j\}_j$ be a set of generators of the Lie algebra $\mathfrak{u}(d_n)$.  
Then, for every $V \in \mathrm{U}(d_n)$, there exists a set of real parameters $\{\alpha_j\}_j$ such that
\begin{equation}\label{eq: inverse Lie theorem}
    V = \prod_{j=1}^{d_n^2} \exp\left[i \alpha_j t_j \right].
\end{equation}
\end{theorem}
\noindent
We stress that each generator appears only once in the product on the right-hand side of Eq.~\eqref{eq: inverse Lie theorem}.  
This theorem ensures that $\mathcal{G}$ can be implemented as a sequence of parametrized elementary gates
\begin{equation}\label{eq: parametrized gates}
    G_j(\alpha_j) = \exp\left[i\alpha_j t_j \right], \quad j \in \{1, 2, \ldots, d_n^2\}.
\end{equation}

Evidently, Theorem~\ref{theorem: decomposition} provides a method to construct the desired mapping $\mathcal{G}$.  
However, at this stage we notice that not all the elements of the group $\mathrm{U}(d_n)$ 
are actually necessary to reach other states in $\mathcal{S}(\mathcal{B}_n)$ when starting from a fixed state $\ket{\Psi}$.  
Indeed, the mapping $\mathcal{G}$ constructed so far is not a bijective operation.  
This is evident from the existence of a subgroup $\mathrm{U}(d_n-1) \subset \mathrm{U}(d_n)$ that leaves $\ket{\Psi}$ invariant, i.e., 
\[
V \ket{\Psi} = \ket{\Psi}, \quad \forall V \in \mathrm{U}(d_n-1) \, ,
\]  
including, essentially, all the unitaries that leave the direction of $\ket{\Psi}$ unchanged. From a group-theoretic perspective, the transformations that concretely change $\ket{\Psi}$ correspond to the coset space  
\(\mathrm{U}(d_n) / \mathrm{U}(d_n-1),\) 
whose dimension is  
\begin{align*}
    \dim \mathrm{U}(d_n) - \dim \mathrm{U}(d_n-1) 
    &= d_n^2 - (d_n-1)^2 \\
    &= 2 d_n - 1 \, .
\end{align*}  
Remarkably, this exactly matches the number of real parameters needed to uniquely describe normalized states in $\mathcal{S}(\mathcal{B}_n)$.  
Therefore, making the mapping $\mathcal{G}$ bijective would reduce 
both the number of trainable parameters and the number of $G_j$ gates in Eq.~\eqref{eq: parametrized gates} from $d_n^2$ to $2 d_n - 1$.  
To achieve this, we can order the generators $\{t_j\}_j$ in Theorem~\ref{theorem: decomposition} 
so that the $(d_n-1)^2$ generators
of the subalgebra $\mathfrak{u}(d_n-1) \subset \mathfrak{u}(d_n)$ act first on the state $\ket{\Psi}$. 
Since their effect on $\ket{\Psi}$ is trivial, they can be neglected,
leaving only those corresponding to the quotient $\mathrm{U}(d_n)/\mathrm{U}(d_n-1)$.
The action of $\mathcal{G}$ then reduces to  
\begin{equation}\label{eq: essential gates}
\begin{split}
\mathcal{G}(\alpha) \ket{\Psi} 
&= \underbrace{\prod_{j = 1}^{2d_n-1} G_j(\alpha_j)}_{U(d_n)/U(d_n-1)}
\underbrace{\prod_{j = 2d_n}^{d_n^2} G_j(\alpha_j)}_{\in U(d_n-1)} \ket{\Psi} \\
&= \prod_{j = 1}^{2 d_n-1} G_j(\alpha_j) \ket{\Psi},
\end{split}
\end{equation}
Intuitively, 
fixing the reference state $\ket{\Psi}$ removes redundant operations that leave  the state itself unchanged, so we only need the remaining ``active'' unitaries that truly change the observable amplitudes inside the subspace.
In conclusion, the number of generators (i.e., the number of real parameters) 
needed to prepare the generic state of $\mathcal{S}\left( \mathcal{B}_n\right)$ from $\ket{\Psi}$ is $2d_n-1$.
Moreover, 
if the target state in $\mathcal{S}\left( \mathcal{B}_n\right)$
only needs to be prepared up to a global phase, the generator that produces this phase shift can be omitted, thus reducing the required number of generators to $2(d_n-1)$.
As a result, for all those systems with  $d_n$ degeneracy of its initial n-th energy level that scales polynomially with the system size (i.e., $L$ in our case, coinciding with the qubits number in the register), \textit{also the number of required parameters showcase a polynomial growth}, avoiding the so called barren plateau scenario~\cite{mcclean2018barren,larocca2025barren}.
Moreover, the parametrized quantum circuit we proposed is not an ansatz, as it is commonly required in most applications involving VQA,
but a constructive set of operations determined 
by the degenerate subspace $\mathcal{S}\left(\mathcal{B}_n \right)$
that is being explored.

The following stage is the application
of the conventional adiabatic algorithm.
In what follows,
we assume that the hardware connectivity and native gate set allow to efficiently encode the time evolution, thereby maintaining scalability.
As previously discussed, the design of the B-SAP protocol ensures that the system’s symmetry group either remains constant or decreases throughout the adiabatic evolution. Consequently, its irreducible representations (irreps) either remain unchanged or split into smaller ones. Since each degenerate eigenspace corresponds to the representation of an irrep, this typically implies that the multiplicity of degenerate eigenspaces does not increase during the process, thereby suppressing level-crossing phenomena.
Nevertheless, it should be noted that, although symmetry is the dominant source of degeneracy, distinct irreps may coincidentally share the same energy eigenvalue. Hence, as with standard adiabatic passage, the B-SAP protocol is not universally guaranteed to succeed. However, it is worth emphasizing that the conditions under which B-SAP ensures accurate state preparation are less restrictive than those required for standard AP. 
Specifically, while AP demands that all energy gaps remain finite throughout the evolution for every $n$-th eigenstate, B-SAP only requires this condition for eigenstates that do not share the same eigenspace $\mathcal{S}\left(\mathcal{B}_n\right)$ 
at the beginning of the adiabatic process. 
This difference arises from the applicability of the finite-gap condition in Eq.~\eqref{eq: gap condition}.
Indeed, in the standard AP each eigenstate corresponds to a single $\sigma_n$, 
whereas in B-SAP an entire eigenspace $\mathcal{S}\left(\mathcal{B}_n\right)$ corresponds to a single $\sigma_n$. 
As a consequence, the finite-gap requirement, which in AP must hold between every pair of eigenstates, in B-SAP it is only needed between distinct eigenspaces.
In some cases, this condition can be guaranteed a priori. However, there exist scenarios in which this is not the case, 
yet it remains possible to demonstrate that the B-SAP protocol has successfully prepared the correct target state, a posteriori. 
These different situations, 
along with illustrative examples, 
are discussed in the Supplemental Material.
Building on this preparation, the procedure is then followed by a standard VQA workflow, in which the parametrized circuit serves as the starting point for classical optimization. In this stage, a loss function is evaluated from the circuit output, and it is subsequently minimized classically by adjusting the circuit parameters $\underline{\alpha}$ at each iteration step (see Fig.~\ref{fig : circuit scheme}).
Although the number of tunable parameters in the B-SAP circuit increases only polynomially with system size, 
the actual training process can become increasingly difficult as $L$ grows, making it challenging to determine whether the optimization converges to a local or a global minimum of the loss function.
For this reason, in the present work we bypass such conventional optimization procedure by making use of a different technique, called Multistate-Contracted Variational Quantum Eigensolver (MC-VQE)~\cite{parrish2019quantum}, 
which we now summarize for completeness.
Let $\{\ket{\Psi_m}\}_{m = 1}^{d_n}$ be the elements
of the basis $\mathcal{B}_{n}$.
Instead of preparing their generic superposition
via the application of the $G_j$ gates,
the main idea is to prepare on the quantum register the overlap states
\begin{equation}
    \ket{\Psi_{ml}^p} = \frac{1}{\sqrt{2}}\left(\ket{\Psi_m} + (i)^p \ket{\Psi_l} \right)\,, \qquad (p=0,1,2,3)
\end{equation}
for all the values $m$, $l \neq m$, and $p$, allowing to measure $\bra{\Psi_{ml}^p}U_\tau^\dagger(1) H_T U_\tau (1)\ket{\Psi_{ml}^p}$,
in order to obtain
\begin{equation}
\begin{split}
    \left[H_T\right]^{(n)}_{ml} = &\frac{1}{2}\big[\bra{\Psi_{ml}^0}U_\tau^\dagger(1) H_T U_\tau (1)\ket{\Psi_{ml}^0} \\
   &- \bra{\Psi_{ml}^2}U_\tau^\dagger(1) H_T U_\tau (1)\ket{\Psi_{ml}^2} \\
   &-i \bra{\Psi_{ml}^1}U_\tau^\dagger(1) H_T U_\tau (1)\ket{\Psi_{ml}^1} \\
   &+i \bra{\Psi_{ml}^3}U_\tau^\dagger(1) H_T U_\tau (1)\ket{\Psi_{ml}^3}\big] \,.
\end{split}
\end{equation}
Notice that $\left[H_T\right]^{(n)}_{ml}$ are the matrix elements of $H_T$ on the basis $U_\tau  \mathcal{B}_n$,
which we remind to have dimension $d_n \in \mathrm{poly}(L)$, and can be therefore diagonalized with classical techniques (see, e.g., Ref.~\cite{lanczos1950iteration}).
An additional crucial advantage for the applicability of this method to near-term quantum computing platforms is that the long circuit depth typically required to implement all the $G_j$ gates is traded with the repetition of much shallower circuits, which are required to prepare the states $\ket{\Psi^p_{ml}}$, for the different values of the indices $l,m,p$.
As a result of the classical diagonalization of the $d_n-$dimensional matrix $\left[H_T\right]^{(n)}_{ml}$, one immediately obtains an estimate of its $d_n$ eigenvalues $E^{(n)}_{m}(s=1)\in \sigma_{n}(1)$, together with the decomposition of their eigenvectors, $\ket{E^{(n)}_{m}(1)}$
on the basis $U_\tau(1)  \mathcal{B}_n$,
i.e.,
\begin{equation}
    \ket{E^{(n)}_{m}(1)} = U_\tau(1)\sum_{l}c_{ml}\ket{\Psi_l}\,.
\end{equation}
First, if the only goal is to characterize the energy eigenvalues,
e.g., to estimate spectral gaps in quantum matter~\cite{somma2013spectral, osterkorn2023gap, cubitt2015undecidability, gosset2016correlation, can2019spectral, arad2017rigorous, hastings2007area, landau2015polynomial,sachdev2011quantum,hastings2006spectral,kohn1964theory,szymanski2022universal,chung2023topological,rai2024spectral} or chemistry~\cite{tiwary2016spectral,bredas2014mind,griffith1957ligand,lupo2023two, deshpande2022importance},
no further steps are required.
Otherwise, when the goal is the effective preparation of a targeted eigenstate $\ket{E^{(n)}_{m}(1)}$
on the quantum register, one should finally proceed in finding the values of the parameters $\{\alpha_j\}$
such that the state $U_\tau(1)\prod_j G_j\left( \alpha_j\right)\ket{\Psi} = \sum_{l}a_{ml}\ket{\Psi_l}$ prepared on the quantum register coincides with $U_\tau(1)\sum_{l}c_{ml}\ket{\psi_l}$.
The dependence of the coefficients $a_{ml}$ 
on the parameters,
namely $a_{ml}\left( \underline{\alpha}\right)$ 
is defined by the specific choice of the $\{G_j\}_j$ operators and may be hard to invert analytically.
However, it is possible to optimize such parameters
by maximizing the overlap, i.e., by minimizing
\begin{equation}\label{eq: classical loss}
    \mathcal{L} = 1-\sum_l c_{ml}^* a_{ml}(\underline{\alpha})
\end{equation}
with classical methods.
Notice that such a minimization process is purely classical, since no quantum circuit is involved at this stage.
Moreover, the loss function, $\mathcal{L}$ in Eq.~\eqref{eq: classical loss}, has a known global minimum at the value $0$, which helps preventing any confusion between local and the global minima.

Here is a 6-steps summary of our proposed B-SAP algorithm: 
\begin{enumerate}
    \item Choose a simpler Hamiltonian $H_0$, whose eigenstates can be efficiently prepared on a quantum register, 
    and whose symmetry group contains that of $H_T$.
    \item Select the $n$-th energy level of $H_0$, whose associated eigenspace has dimension $d_n$ and is spanned by a known subset of eigenstates, $\mathcal{B}_n$.
    \item Prepare a state $\ket{\Psi} \in \mathcal{B}_n$ on the quantum register.
    \item Construct the parametrized quantum gates $\{G_j\}_j$
    defined in Eq.~\eqref{eq: parametrized gates}, 
    and then apply them as in the r.h.s. of Eq.~\eqref{eq: essential gates}.
    \item Perform the adiabatic evolution from $H_0$ to $H_T$.
    \item Tune the circuit parameters with MC-VQE
    to optimize the output state of the quantum circuit, namely $U_\tau(1)\prod_j G_j(\alpha_j)\ket{\Psi}$.
\end{enumerate}
In the following, we apply the above protocol to a prototypical spin model governed by the Heisenberg Hamiltonian.

\section{Application: the XYZ Heisenberg model}

The XYZ Heisenberg Hamiltonian is a prototype model of a many-body spin system, usually employed to study the different phases of many-body magnetic systems, or serving as a basis to simulate more complex quantum field theories~\cite{Tacchino2020AQT}. In particular, here we consider the case of a one-dimensional spin chain containing an even number of sites, $L$, with Periodic Boundary Conditions (PBC). 
Each spin site is labeled by an integer $i\in\{0,...,L-1\}$, such that the target  Hamiltonian is defined as
\begin{equation}\label{eq: Heisenberg Hamiltonian}
    H_T = -\sum_{i}\left[ J^x X_i X_{i+1}+ J^y Y_{i}Y_{i+1}+ J^z Z_{i}Z_{i+1}\right] \, ,
\end{equation}
where $\{X,Y,Z\}$ denotes the set of Pauli matrices defined in SU(2), and the $L$-th site of the chain
coincides with the $0$-th one according to PBC.
We will henceforth assume, without loss of generality, that $\abs{J^z} \geq \abs{J^x} \geq \abs{J^y}$. 
In fact, should this condition not be satisfied, a suitable rotation of the reference frame would be sufficient to restore it.
The target Hamiltonian showcases both topological and spin symmetries.
In particular, 
Eq.~\eqref{eq: Heisenberg Hamiltonian} is invariant both under translations and reflections,
which can be generated by mapping each spin site as $i \mapsto i+1$ and $i \mapsto L-i$,
respectively. 
Hence, the topological symmetry coincides with the Dihedral group $D_L$.
In the general case where the coupling constants are all distinct, 
i.e. $J^z \neq J^x \neq J^y$, the spin symmetry group is also a discrete one.
Specifically, 
the operators that commute with $H_T$ 
correspond to $\pi$ rotations of all the spins around one of the three coordinate axes. 
For instance, the operator $\mathcal{X} = \bigotimes_i X_i$, 
which we call \textit{parity}
and will play a relevant role in the following discussion, 
implements a global 
$\pi$-rotation around the $x$-axis.
In order to apply the B-SAP protocol, 
we hereby choose  
\begin{equation}\label{eq: initial Hamiltonian}
    H_0 = -J^z \sum_i Z_i Z_{i+1}
\end{equation}
to be the initial Hamiltonian.
Its spectrum, $\sigma(0)$, is highly degenerate, as clearly shown in Fig.~\ref{fig: spectra evolution} for $L=4$ (see Fig.~\ref{fig: BSAP spectrum} on the right end, corresponding to $s=0$). By definition, $H(0) = H_0$ is diagonal on the computational basis, hence we refer to its eigenstates as
\begin{equation}
    \mathcal{B} = \left\{ \ket{\underline{b}} =\bigotimes_{j = 0}^{L-1} \ket{b_j} , \; b_j  \in \{0,1\} \forall j\right\}\, .
\end{equation}
Their preparation on a quantum register is a trivial task.
Moreover, the $H_0$ symmetry group contains the one of $H_T$.
Indeed, while the topological symmetry group is still $D_L$, the spins symmetry group now includes all the possible rotations around the $z$-axis,
while maintaining the $\pi$-rotations around $x$ and $y$.
In particular, 
since $\mathcal{X}$ commutes with $H(s)$ for all values of $s$,
it is a constant of motion for the adiabatic process.
Consider now the subset $\mathcal{B}_n \subset \mathcal{B}$ of all the basis elements 
associated with the n-th energy level $E_n$ of $H(0)$. 
As explained in the previous Section, the goal is now to construct a parametrized quantum circuit that allows to ``explore'' only their span $\mathcal{S}\left(B_n\right)$
by simply varying the variational parameters.
We start by considering one term of $H_0$, explicitly written as
\begin{equation}
        \bra{\underline{b}}Z_iZ_{i+1}\ket{\underline{b}} = 1-2(b_i \oplus b_{i+1}) \,,
\end{equation}
in which $\oplus $ is the binary modulo operation. 
It is therefore natural to introduce the linear mapping 
$\Phi:~\mathcal{S}\left(\mathcal{B}\right)~\mapsto~\mathcal{S}\left(\mathcal{B}\right)$,
defined by its action on the computational basis as
\begin{equation}\label{eq: Phi definition}
     \Phi\ket{\underline{b}} = \bigotimes_{j=0}^{L-1} \ket{b_{j}\oplus b_{j+1}}.
\end{equation}
\begin{figure*}[t]
\centering
    \begin{subfigure}[]{0.4\textwidth}
        \includegraphics[width = \textwidth]{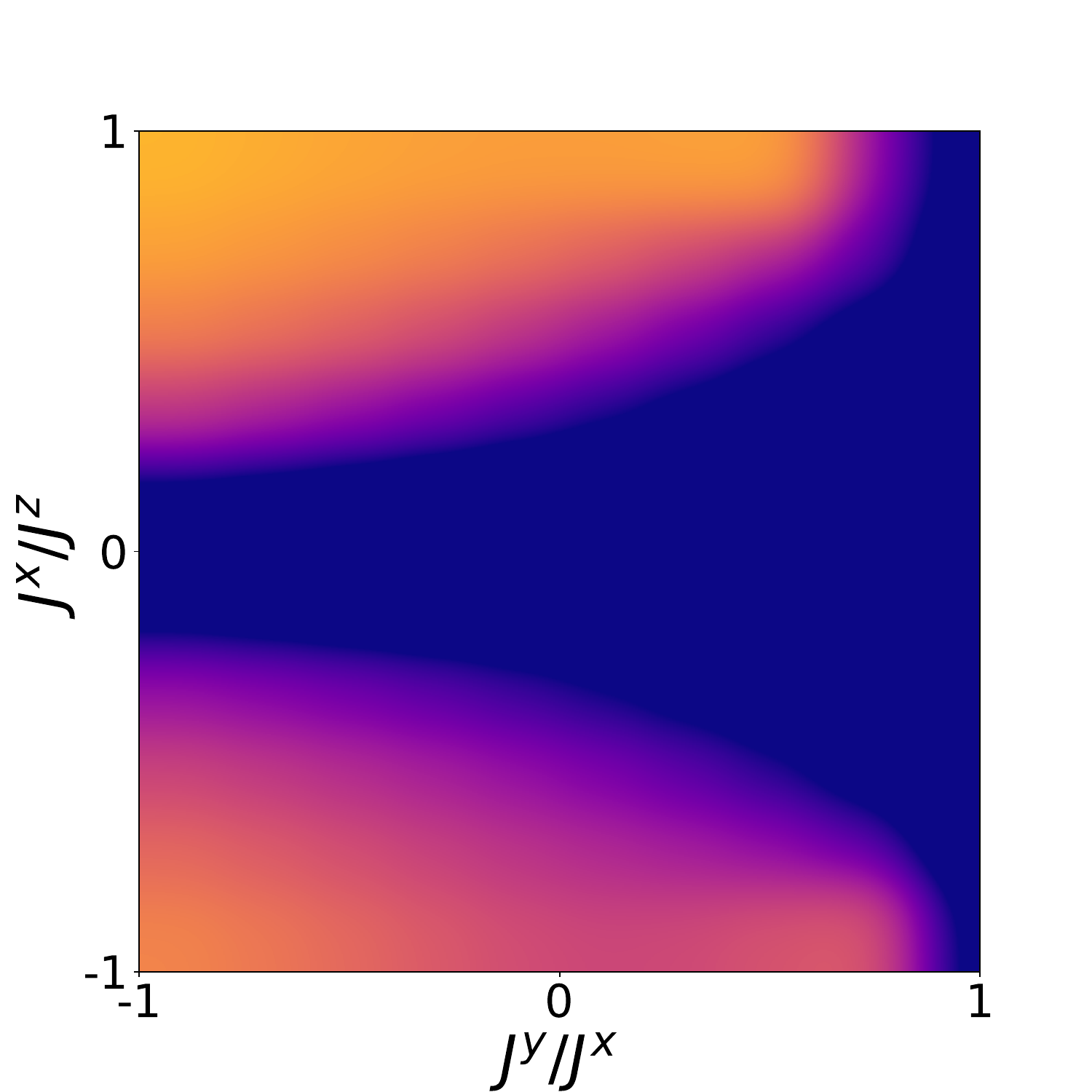}
        \caption{Ground state, $\mathcal{X} = +1$.}
    \end{subfigure}
    \begin{subfigure}[]{0.4\textwidth}
        \includegraphics[width = \textwidth]{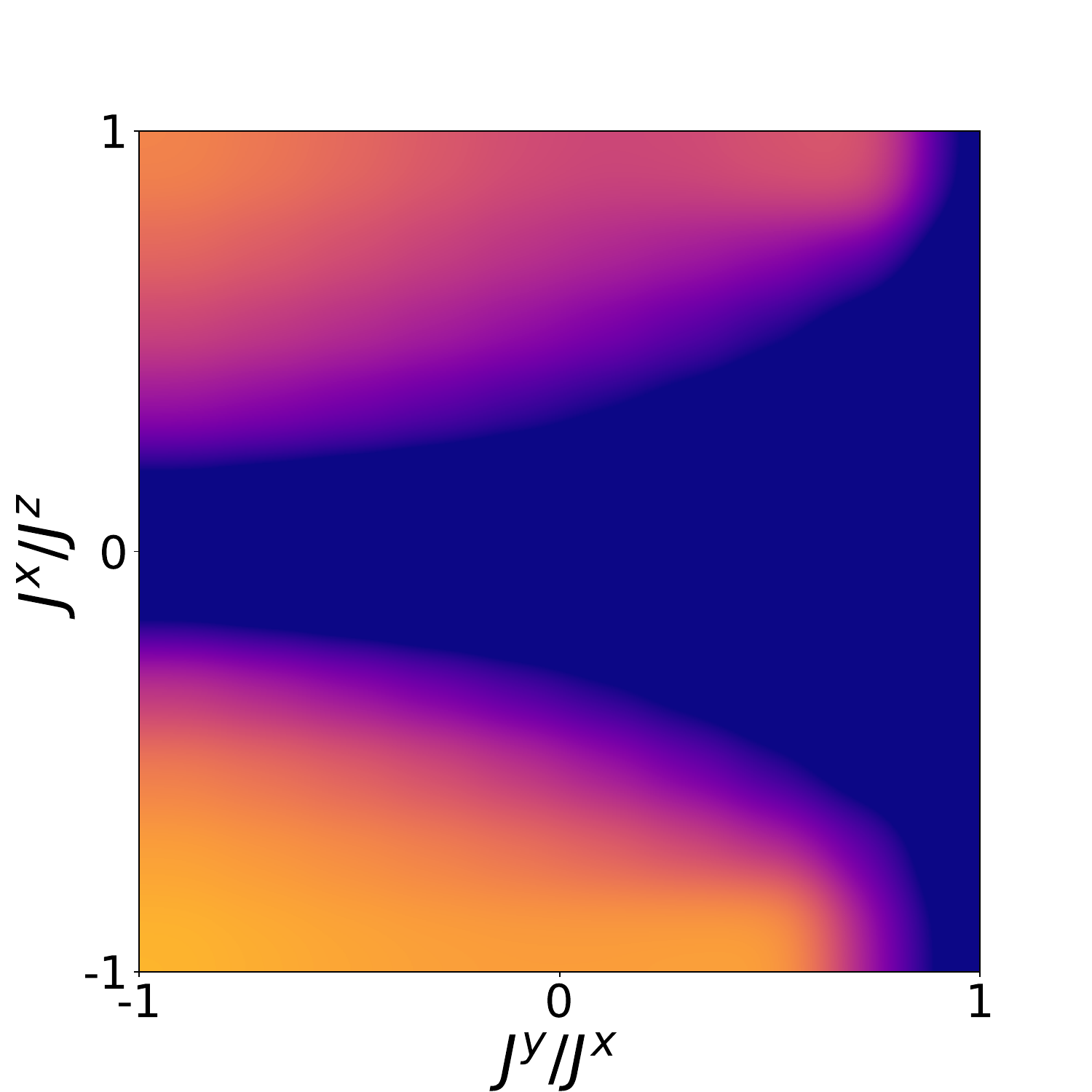}
        \caption{Ground state, $\mathcal{X} = -1$.}
    \end{subfigure}
    \begin{subfigure}[]{0.1\textwidth}
        \includegraphics[width = \textwidth]{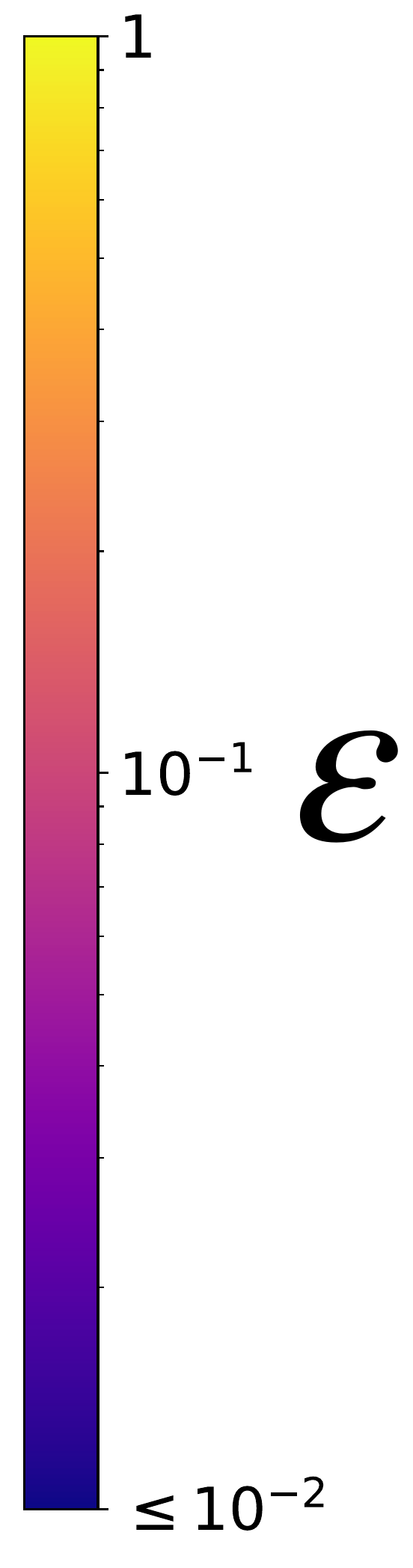}
    \end{subfigure}
    \caption{Error between the target ground state
    and the actually prepared state in the two parity sectors $\mathcal{X} = \pm 1$.
    Such an error is quantified by $\mathcal{E}$ (see Eq.~\eqref{eq: error B-SAP})
    for all the possible ratios 
    among the coupling constants of the XYZ Heisenberg Hamiltonian $H_T$
    in Eq.~\eqref{eq: Heisenberg Hamiltonian}.
    Both panels refer to the ferromagnetic model with $L = 10$.
    The adiabatic procedure has been performed 
    for $L/2 = 5$ Trotter steps, 
    equivalent to half of the chain length,
    each with $0.25\cdot[J^z]^{-1}$ duration.}
    \label{fig: numerical GS}
\end{figure*}
Notice that $\Phi$ is not bijective, as it stands.
Indeed, for any $\ket{\underline{b}}$ it holds
\begin{equation}
    \bigoplus_i \bra{\underline{b}}\Phi^\dagger \frac{\mathds{1}-Z_i}{2} \Phi \ket{\underline{b}} = 0\,,
\end{equation}
meaning that the Hamming weight 
\begin{equation}\label{eq:hamming}
    h\left(\Phi\ket{\underline{b}}\right) = \sum_i \bra{\underline{b}}\Phi^\dagger \frac{\mathds{1}-Z_i}{2} \Phi \ket{\underline{b}}
\end{equation}
is always even, and therefore the image of $\Phi$ is at most half of $\mathcal{S}\left(\mathcal{B}\right)$.
In order to make $\Phi$ invertible, 
we partition $\mathcal{B}$ into
\begin{align*}
    \mathcal{B}^{(0)} = \{ \ket{\underline{b}}\; | \; b_{0} = 0 \} \quad\quad
    \mathcal{B}^{(1)} = \{ \ket{\underline{b}} \; | \; b_{0} = 1 \} \, ,
\end{align*}
splitting the bit strings according to
their value on the first bit.
The restriction of the domain of $\Phi$
to either $\mathcal{S}\left(\mathcal{B}^{(0)}\right)$ or $\mathcal{S}\left(\mathcal{B}^{(1)}\right)$
leads to a bijective map.
For the case of $\Phi : \mathcal{S}\left(\mathcal{B}^{(0)}\right) \mapsto \mathcal{S}\left(\mathcal{B}\right)$,
the inverse operation is
\begin{equation}
     \Phi^{-1}_0\ket{\underline{b}} =  \bigotimes_{j=0}^{N-1} \ket{\bigoplus_{k = j}^{L-1} b_k} \, ,
\end{equation}
while for $\Phi : \mathcal{S}\left(\mathcal{B}^{(1)}\right) \mapsto \mathcal{S}\left(\mathcal{B}\right)$
its inverse is
\begin{equation}\label{eq: Inverse Phi1}
     \Phi^{-1}_1\ket{\underline{b}} = \mathcal{X}\Phi^{-1}_0\ket{\underline{b}}\,.
\end{equation}
Making use of these expressions, and reminding the Hamming weight defined in Eq.~\eqref{eq:hamming}, the eigenvalue equation for $H_0$ can now be recast into the more informative form
\begin{equation}
    H_0 \ket{\Phi \left( \underline{b} \right)} = -J^z \left[L-2h\left(\Phi\ket{\underline{b}}\right)\right]  \ket{\Phi \left( \underline{b} \right)}\,.
\end{equation}
This equation has important consequences.
First, we notice that $\Phi$ bijectively maps
the sets $\mathcal{B}^{(0,1)}_n = \mathcal{B}_n \cap \mathcal{B}^{(0,1)}$ into the set $\mathcal{W}_{2n} \subset \mathcal{B}$ 
of bit strings whose Hamming weight is equal to $2n$.
This in turn implies that there exist $L/2+1$ distinct energy levels, $\{E_n\}$,
all separated by the same gap $\Delta E = 4J^z$,
with a degeneracy equal to twice the size of $\mathcal{W}_{2n}$, i.e.,
\begin{equation}\label{eq:binom}
    d_n = 2 \,\binom{L}{2n} \, .
\end{equation}
Remarkably, Eq.~\eqref{eq:binom} allows to conclude that $d_n \in \mathrm{poly}\left( L^{2n}\right)$, such that the number of trainable parameters in the final circuit
will only grow polynomially with $L$.
If one manages to explore $\mathcal{S}\left( \mathcal{W}_{2n} \right)$,
one can then use $\Phi^{-1}_{0,1}$ 
to map the states back into $\mathcal{S}\left(\mathcal{B}^{(0,1)}_n\right)$.
However, this only allows to separately prepare the generic vector either in $\mathcal{S}\left(\mathcal{B}^{(0)}_n\right)$ 
or in $\mathcal{S}\left(\mathcal{B}^{(1)}_n\right)$,
while the goal is to prepare any $\ket{\psi_n}\in \mathcal{S}\left( \mathcal{B}_n\right)$ 
that is expressed as 
\begin{equation}\label{eq: B_0 B_1 overlap}
    \ket{\psi_n} = \mathrm{cos}\left(\theta\right) \ket{\psi^{(0)}_n} + \mathrm{sin}\left(\theta\right) e^{i \varphi} \ket{\psi^{(1)}_n},
\end{equation}
for proper $\ket{\psi_n^{0,1}}\in\mathcal{S}\left(\mathcal{B}^{(0,1)}_n\right)$, $\theta \in [0, \pi/2]$, and $\varphi \in [0, 2\pi)$.
The pivotal observation to overcome this problem is that $\mathcal{X}$, in light of Eq.~\eqref{eq: Inverse Phi1}, is an operator that bijectively maps $ \mathcal{B}_n^{(0)}$ into $\mathcal{B}_n^{(1)}$.
Moreover, $\mathcal{X}$ and $H(s)$ commute for any $s$, hence it is always possible to simultaneously diagonalize them such that the expectation value of $\mathcal{X}$ remains constant throughout the adiabatic procedure.
This is enough to have some insights about the form of Eq.~\eqref{eq: B_0 B_1 overlap}.
Let $\ket{\psi_n^\pm}$ be an eigenstate with energy $E_n$ and eigenvalue $\pm 1$ with respect to $\mathcal{X}$.
Then
\begin{equation}
\begin{split}
    \bra{\psi_n^\pm} \mathcal{X} \ket{\psi_n^\pm} &= \mathrm{sin}\left(2\theta\right) \mathrm{Re}\left[  e^{i \varphi} \bra{\psi^{(0)}_n}\mathcal{X}\ket{\psi^{(1)}_n}\right] = \pm1 \, ,
\end{split}
\end{equation}
in which the orthogonality between 
$\mathcal{S}\left(\mathcal{B}^{(0)}_n\right)$ and $\mathcal{S}\left(\mathcal{B}^{(1)}_n\right)$
has been used.

\noindent
It follows that
\begin{equation}
    \begin{cases}
        \theta = \pi/4\\
        e^{i\varphi}\ket{\psi^{(1)}_n} = \pm \mathcal{X}\ket{\psi^{(0)}_n} \, .
    \end{cases}
\end{equation}
Therefore
\begin{equation}
    \ket{\psi_n^\pm} = \frac{1}{\sqrt{2}}\left[ \mathds{1} \pm \mathcal{X}\right] \ket{\psi^{(0)}_n}\,,
\end{equation}
thus ensuring that, once we are able of exploring $\mathcal{S}\left(\mathcal{B}_n^{(0)}\right)$, also the exploration of $\mathcal{S}\left(\mathcal{B}_n\right)$ automatically follows.
Finally, we can constructively determine 
the unitary gates $G_{j}\left(\alpha_j \right)$ of Eq.~\eqref{eq: parametrized gates}
to explore $\mathcal{S}\left(\mathcal{B}_n^{(0)}\right)$, specifically focussing on the $n = 0$ and $n = 1$ cases as examples,
up to global phases.
\begin{figure*}[t]
\centering
    \begin{subfigure}[h!]{0.28\textwidth}
        \includegraphics[width = \textwidth]{Numerical_results/JZpos_n0_L10_Xpos.pdf}
        \caption{$L/2$ Trotter steps}
        \label{fig: L/2}
    \end{subfigure}
    \begin{subfigure}[h!]{0.28\textwidth}
        \includegraphics[width = \textwidth]{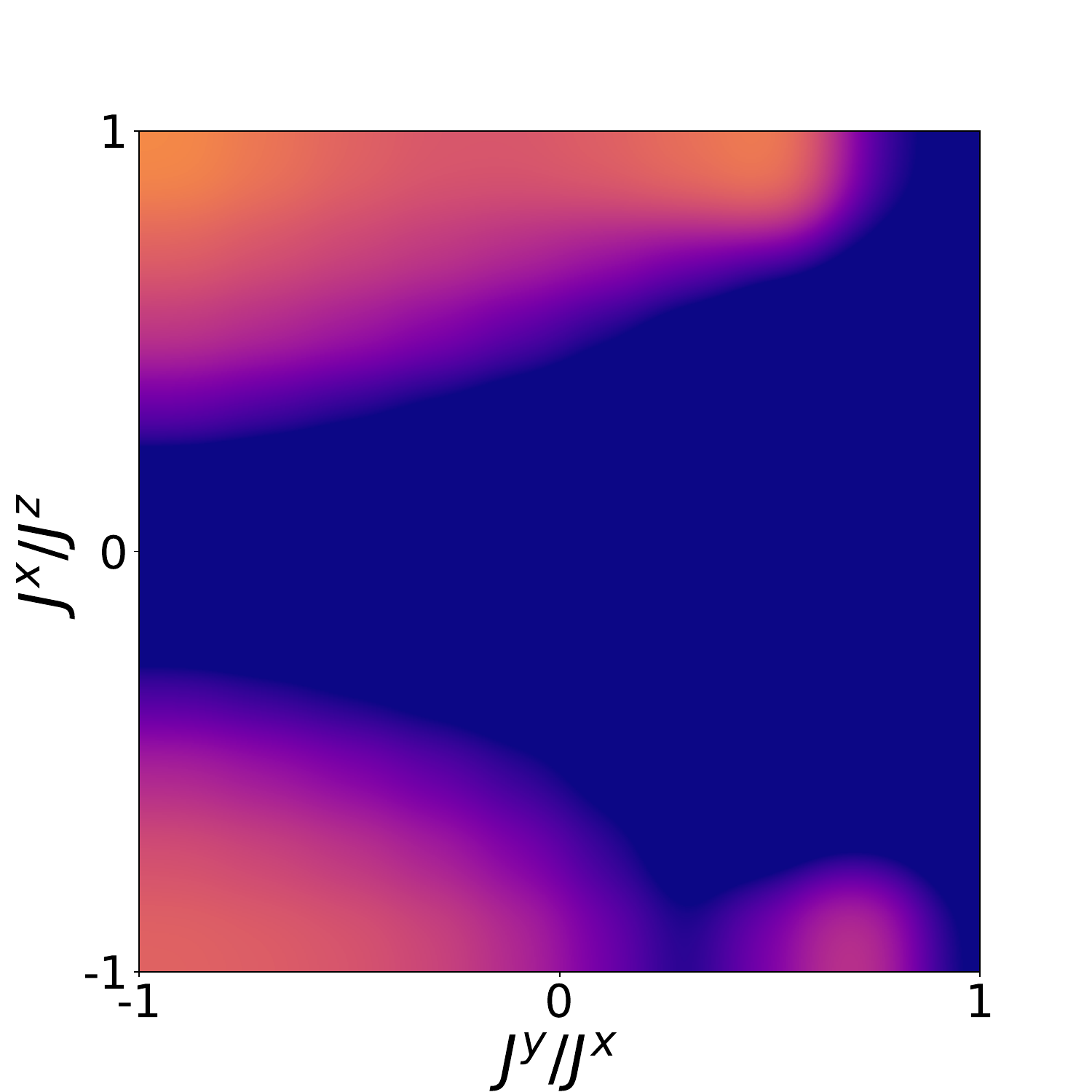}
        \caption{$L$ Trotter steps}
        \label{fig: L}
    \end{subfigure}
    \begin{subfigure}[h!]{0.28\textwidth}
        \includegraphics[width = \textwidth]{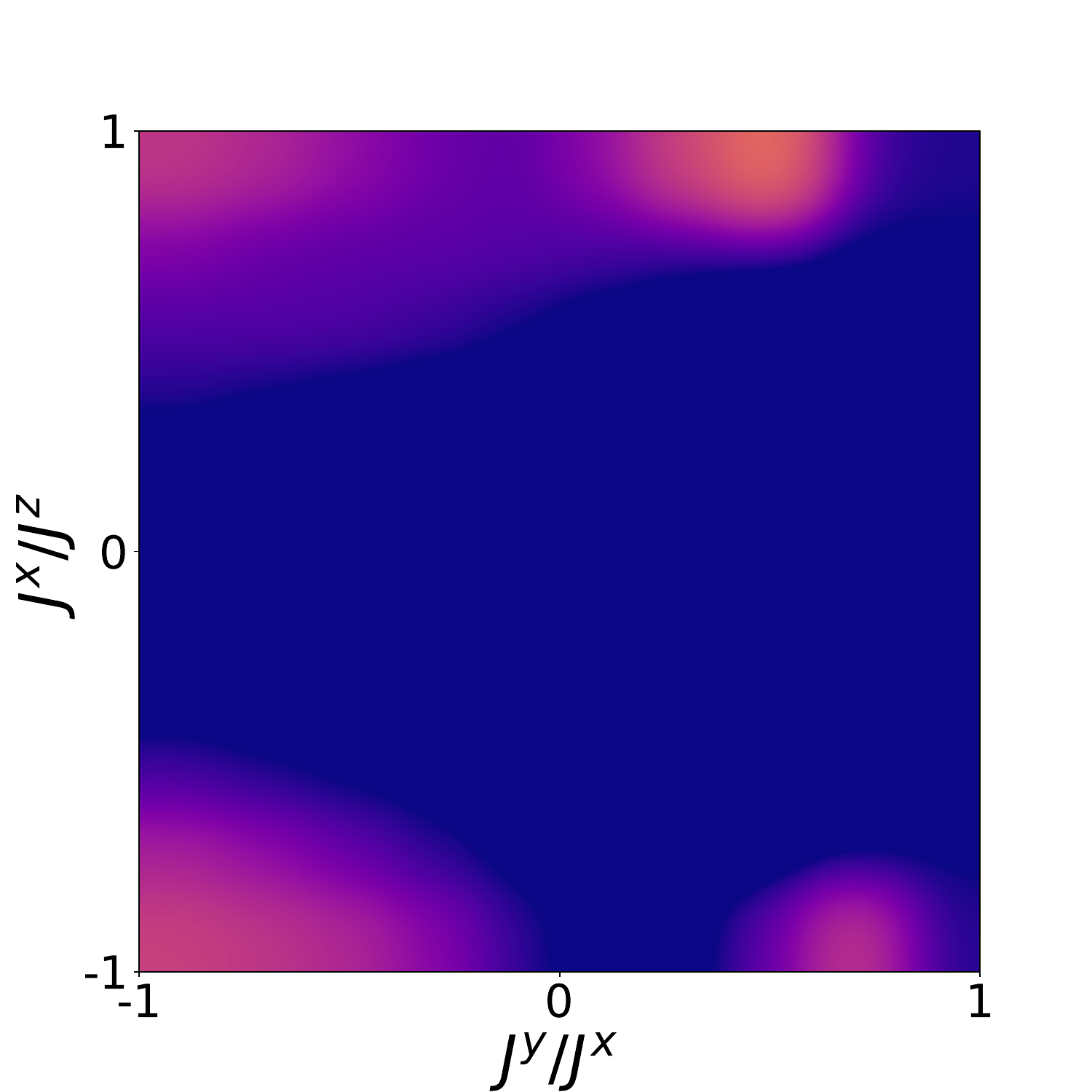}
        \caption{$2L$ Trotter steps}
        \label{fig: 2L}
    \end{subfigure}
    \begin{subfigure}[h!]{0.09\textwidth}
        \includegraphics[width = \textwidth]{Figures/cbar.pdf}
    \end{subfigure}
    \caption{Error between the target ground state
    and the actually prepared state in the $\mathcal{X} = +1$ sector, calculated for different numbers of Trotter steps in the adiabatic protocol of the B-SAP method.
    The error, $\mathcal{E}$ (see Eq.~\eqref{eq: error B-SAP}), is reported for possible ratios 
    among the coupling constants of the ferromagnetic XYZ Heisenberg Hamiltonian, see $H_T$
    in Eq.~\eqref{eq: Heisenberg Hamiltonian},
    with $L = 10$.
    The adiabatic procedures have been performed with (a) 5, (b) 10, and (c) 20 Trotter steps of duration $0.25\cdot [J^z]^{-1}$.}
    \label{fig: Trotter dependence}
\end{figure*}

In the following, we show numerical simulations of the B-SAP protocol performed by using the Qiskit software \cite{Qiskit}, with the aim of  determining several eigenstates of the Heisenberg Hamiltonian, Eq.~\eqref{eq: Heisenberg Hamiltonian}, defined on a register containing up to $L = 10$ qubits.
Specifically, we quantify the error in the preparation of the m-th energy eigenstate, $\ket{E_m}$, on the quantum register through
\begin{equation}\label{eq: error B-SAP}
    \mathcal{E} = \norm{\left(1-P_{E_m}\right)\left(U_\tau (1) \prod_j G_j(\alpha_j)\ket{\Psi}\right)}\,,
\end{equation}
in which $P_{E_m}$ is the projector on the subspace associated to the eigenvalue $E_m$.
Notice that if $E_m$ is non-degenerate,
$\mathcal{E}$ coincides with the infidelity ~\cite{nielsen2010quantum}.
Notice that all error sources, including finite-time adiabatic preparation and its Trotterization, contribute to $\mathcal{E}$.

\subsection{Example: target state for n=0}
\noindent
In this case, $d_0 = 1$
and the only state in $\mathcal{W}_{0}$
is $\ket{\Psi} = \ket{0}^{\otimes N}$, i.e., 
the default state of a digital quantum computer. 
Hence, no parameterization is required, 
and $\Phi^{-1}$, $\frac{1}{\sqrt{2}}\left[ \mathds{1} \pm \mathcal{X}\right]$ and $U_\tau(1)$ can be straightforwardly applied.
The error on the preparation of the ground state of $H_T$, defined in Eq.~\eqref{eq: error B-SAP}, as obtained by
using a simulator of a digital quantum computer for an Heisenberg chain of length $L = 10$, is reported in Fig.~\ref{fig: numerical GS}, in logarithmic scale for both $\mathcal{X} = 1$ and $\mathcal{X} = -1$.
In both cases, the numerical results obtained using only $L/2 = 5$ Trotter steps, the minimal number required to enable entanglement between any pair of sites in the periodic chain, show that $\mathcal{E}$ is negligible compared to $2^{-L}$, 
i.e., the average overlap between a random state in the Hilbert space and the target one.
Not surprisingly,
this is especially true for $\abs{J^x/J^z}, \abs{J^y/J^x} \ll 1$,
i.e. in the regions where the target Hamiltonian $H_T$ is not too different from $H_0$.
However, even for values close to critical points such as, e.g., $J_x=J_z$ or $J_y=J_x$,
the error achieved in state preparation by using the B-SAP protocol is never larger than $0.3$.
Finally, it is also crucial to stress that this error can be further significantly reduced,
for all the parameters,
by simply increasing the number of Trotter steps employed in the digital evolution for the adiabatic procedure. In Fig.~\ref{fig: Trotter dependence} this effect is explicitly shown for a number of Trotter steps increased from $L/2$ up to $L$, or $2L$, respectively.

\begin{figure*}[t]
    \begin{subfigure}[]{0.1\textwidth}
        \includegraphics[width = \textwidth]{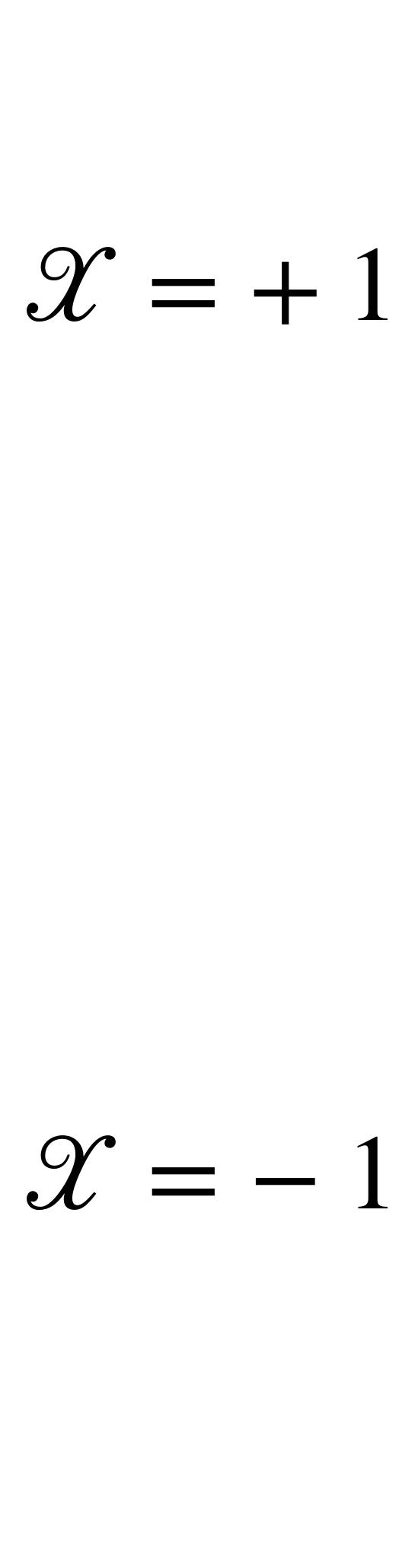}
    \end{subfigure}
    \begin{subfigure}[]{0.19\textwidth}
        \includegraphics[width = \textwidth]{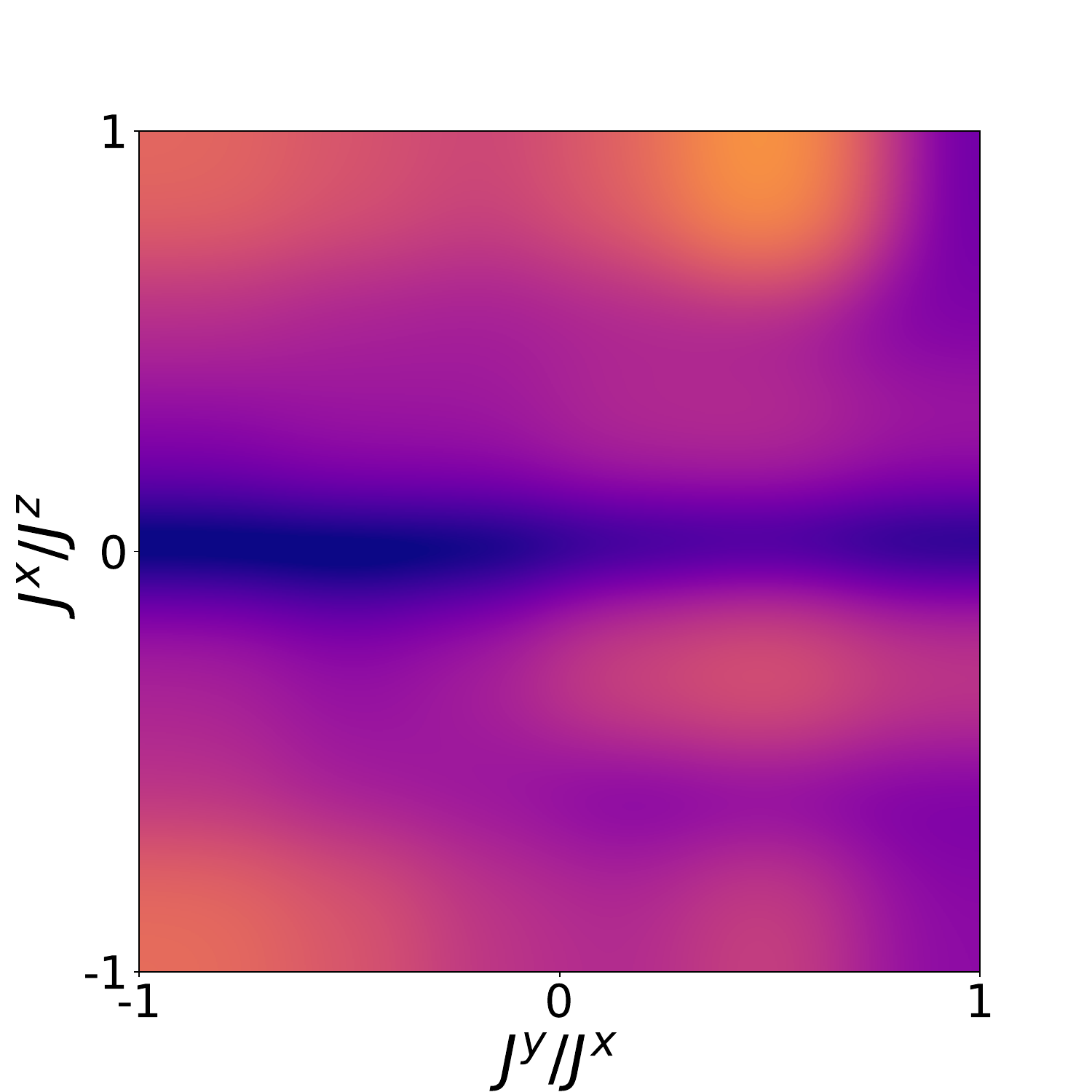}
        % \caption{\Romannum{1} exc, $\mathcal{X} = 1$ .}
        \includegraphics[width = \textwidth]{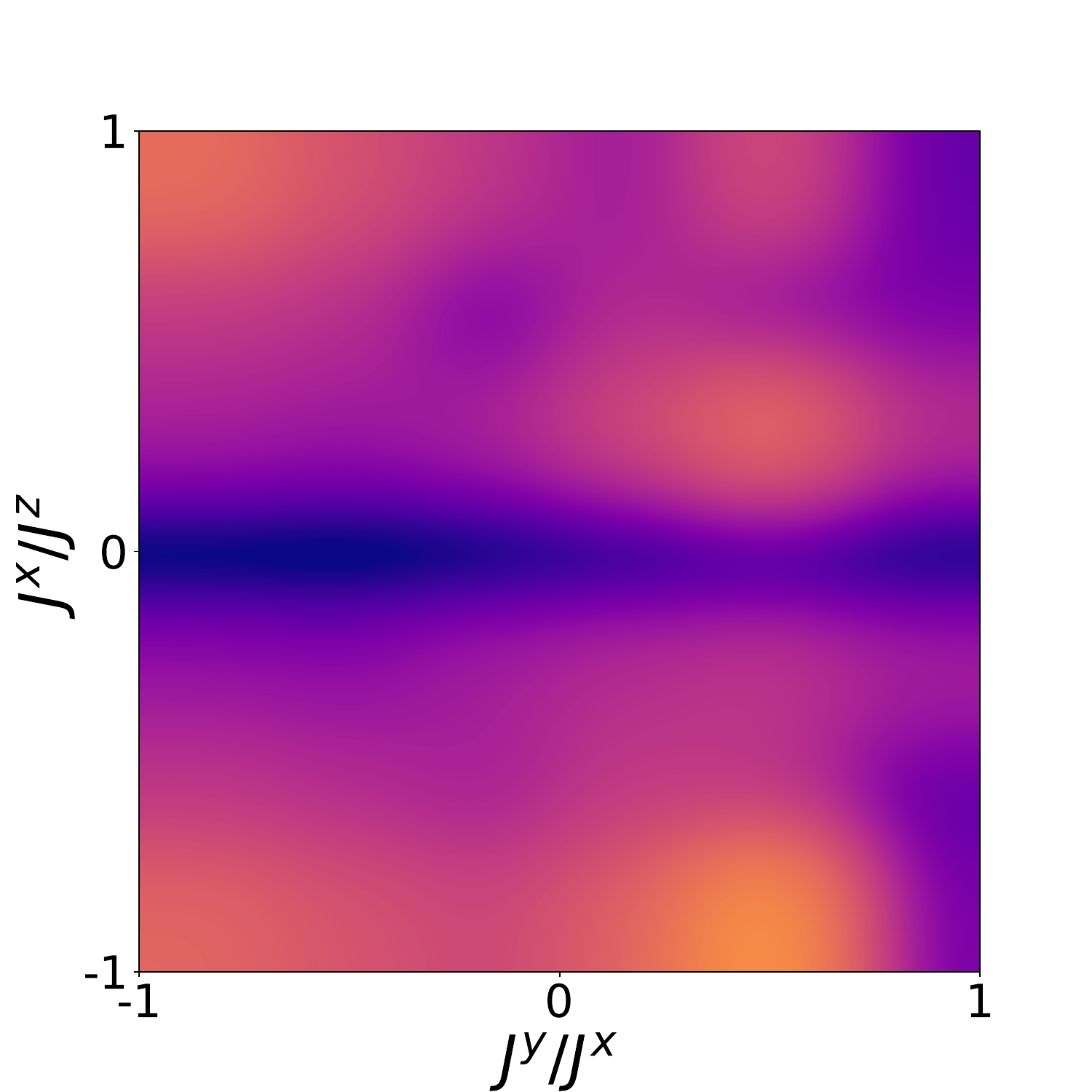}
        \caption{\Romannum{1} exc, $\mathcal{X} = \pm1$ .}
    \end{subfigure}
    \begin{subfigure}[]{0.19\textwidth}
        \includegraphics[width = \textwidth]{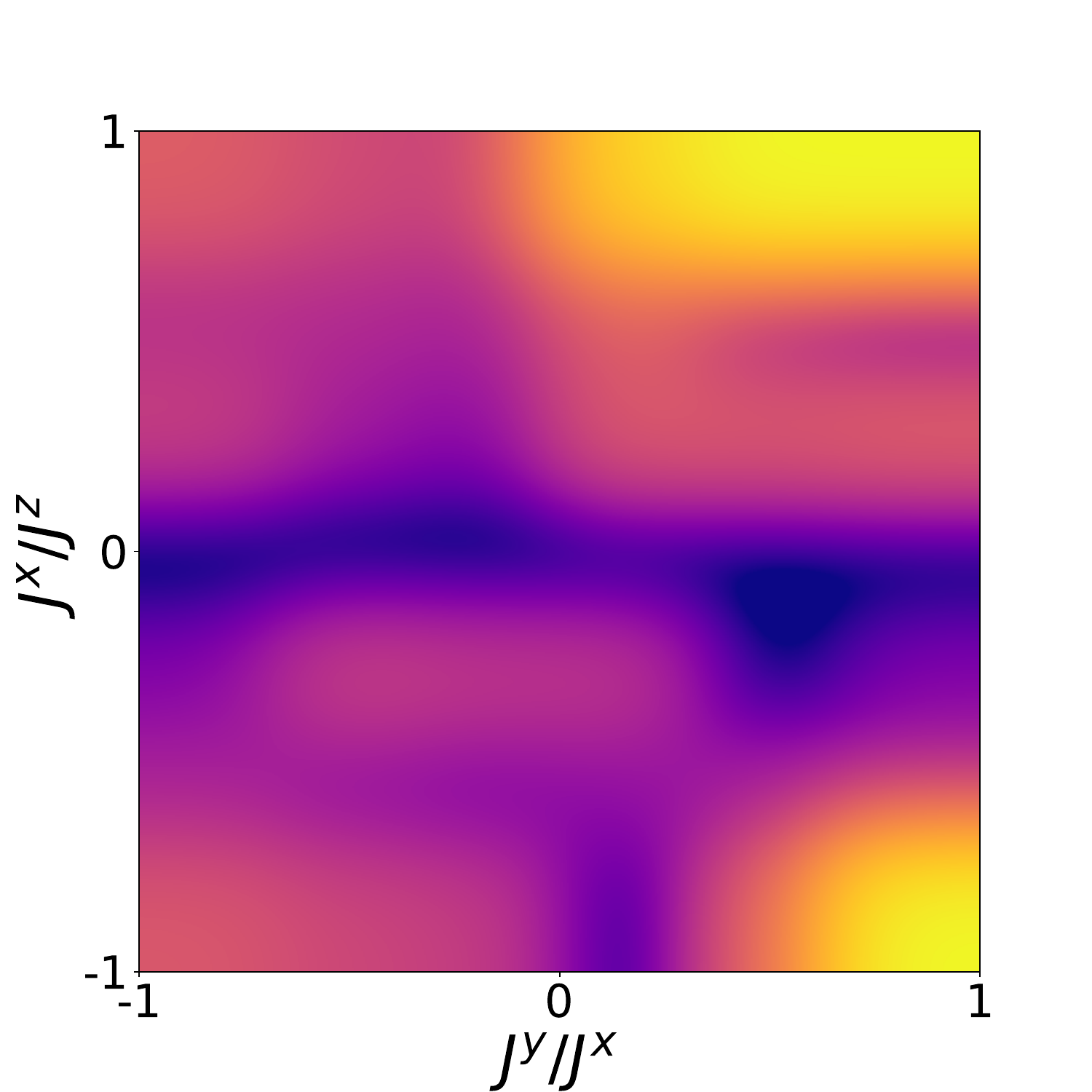}
        % \caption{\Romannum{2} exc, $\mathcal{X} = 1$ .}
        \includegraphics[width = \textwidth]{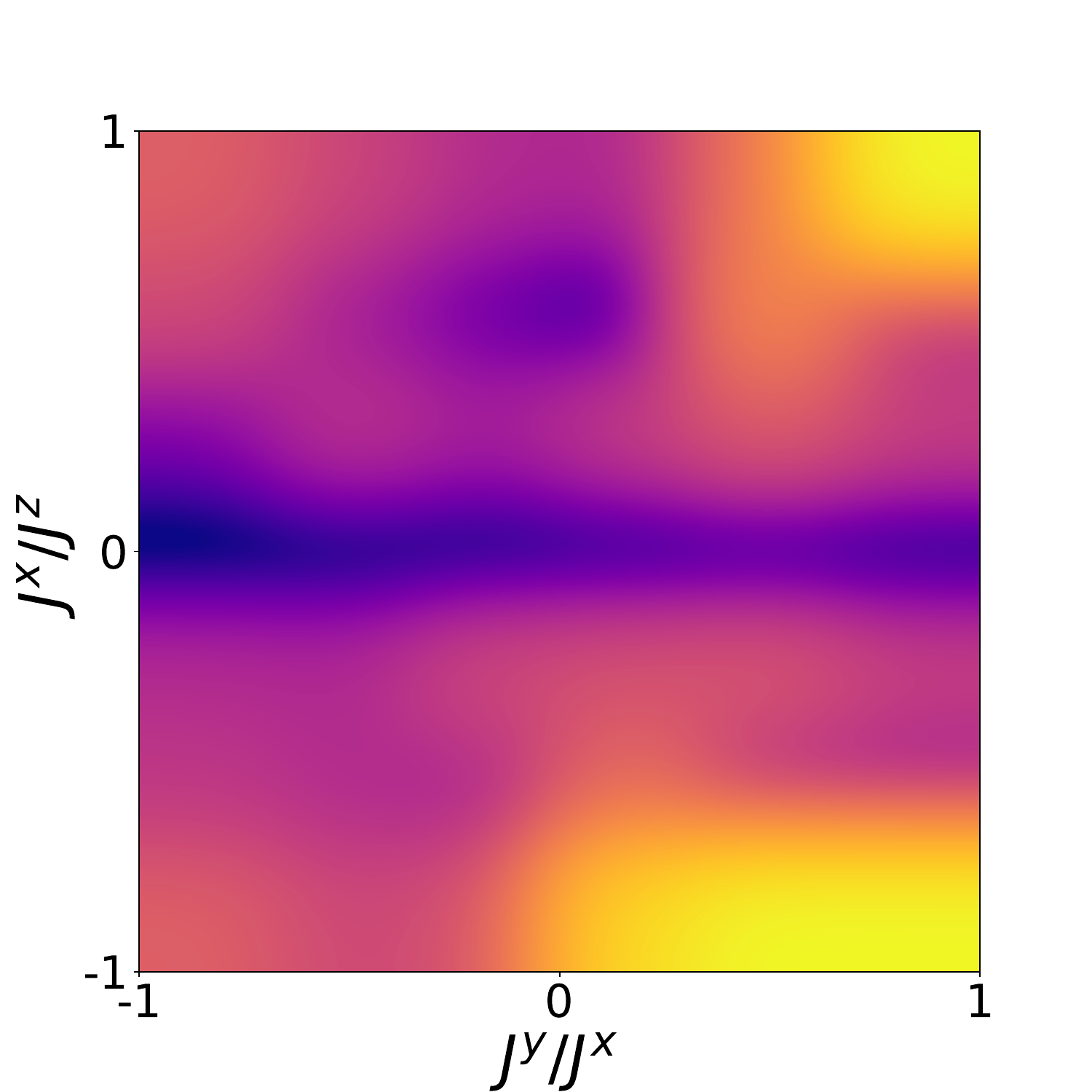}
        \caption{\Romannum{2} exc, $\mathcal{X} = \pm1$ .}
    \end{subfigure}
    \begin{subfigure}[]{0.19\textwidth}
        \includegraphics[width = \textwidth]{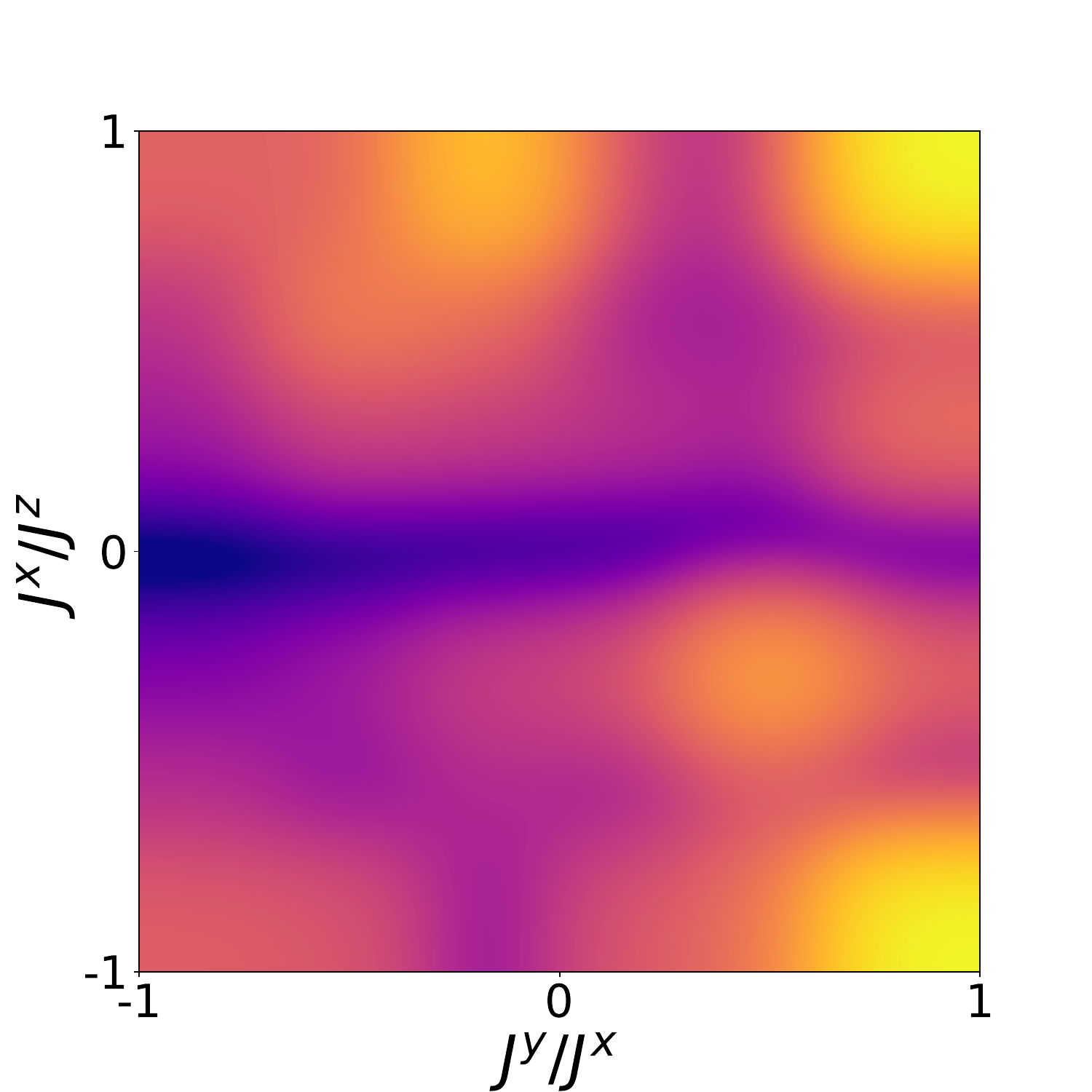}
        % \caption{\Romannum{3} exc, $\mathcal{X} = 1$ .}
        \includegraphics[width = \textwidth]{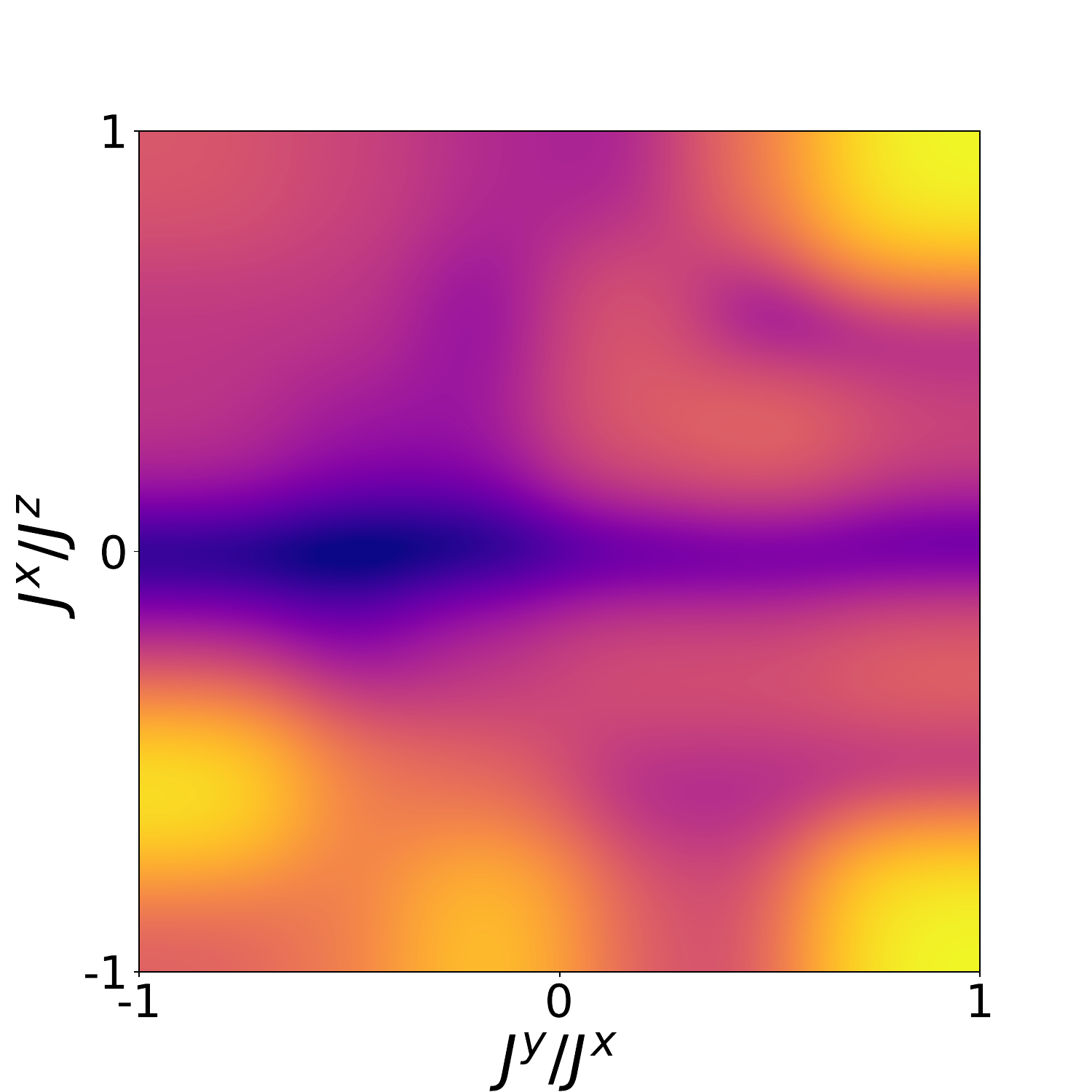}
        \caption{\Romannum{3} exc, $\mathcal{X} = \pm1$ .}
    \end{subfigure}
    \begin{subfigure}[]{0.19\textwidth}
        \includegraphics[width = \textwidth]{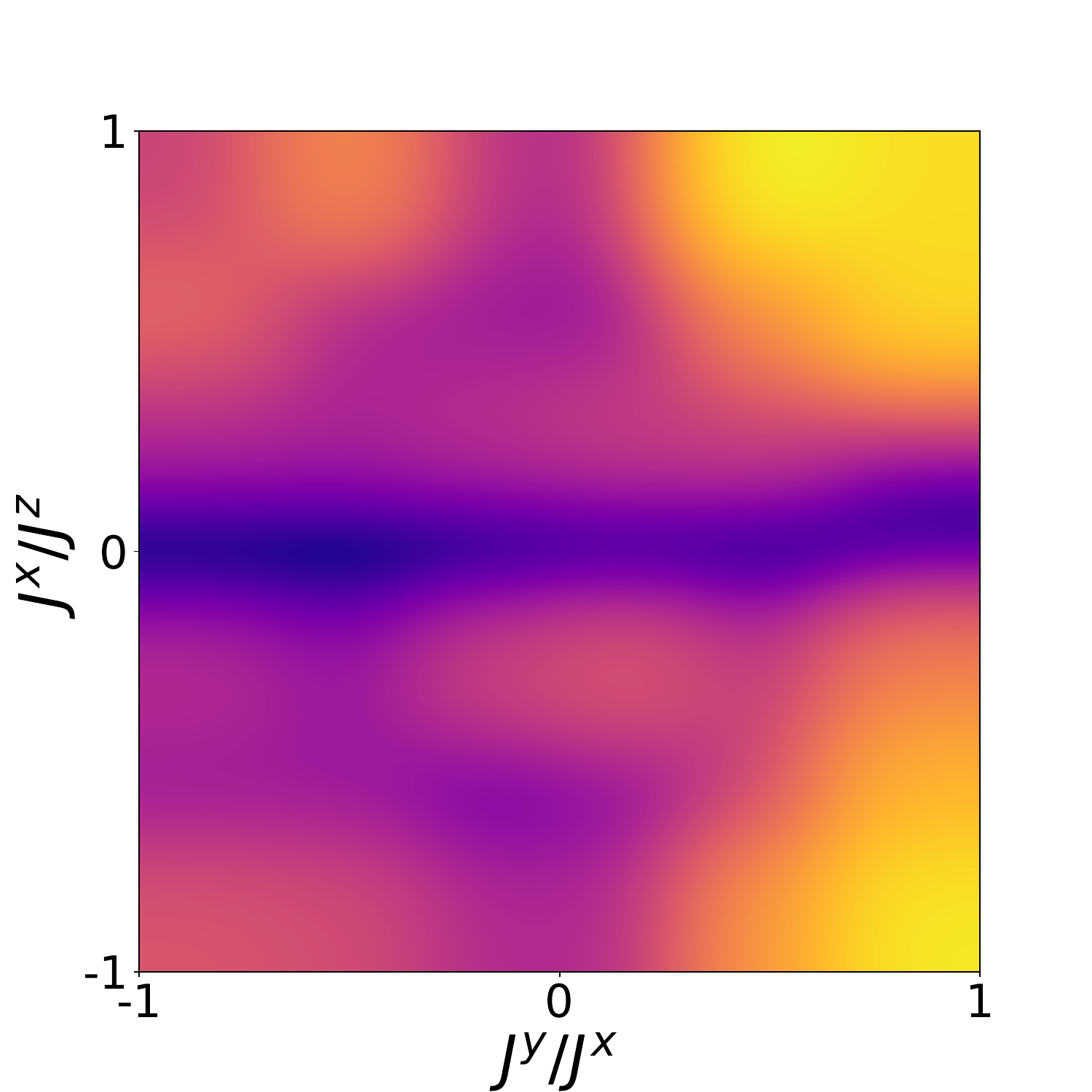}
        % \caption{\Romannum{4} exc, $\mathcal{X} = 1$ .}
        \includegraphics[width = \textwidth]{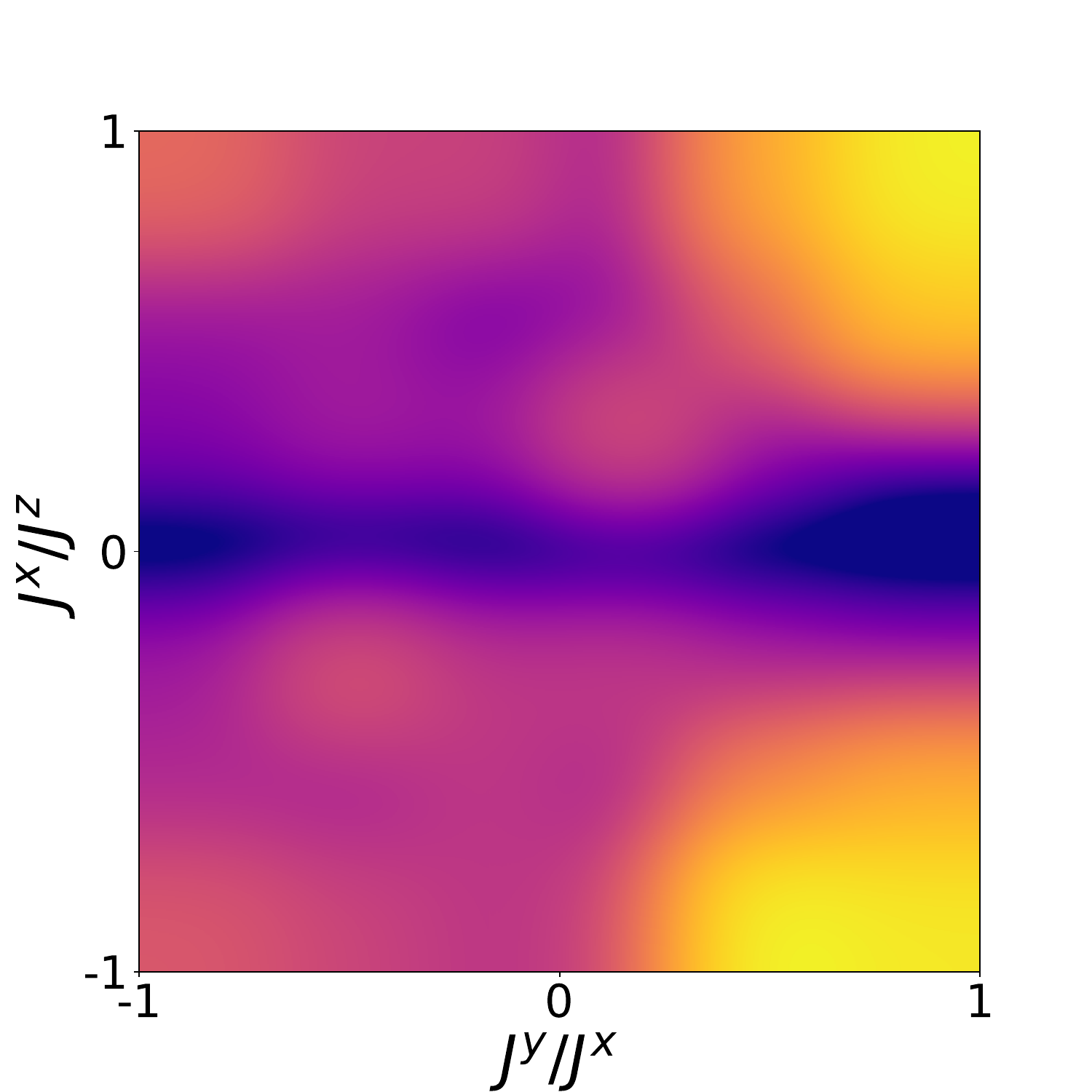}
        \caption{\Romannum{4} exc, $\mathcal{X} = \pm1$ .}
    \end{subfigure}
    \begin{subfigure}[]{0.1\textwidth}
        \includegraphics[width = \textwidth]{Figures/cbar.pdf}
    \end{subfigure}
    \caption{Error between the lowest-energy target excited states
    and the actually prepared states in both parity sectors, $\mathcal{X} = \pm 1$.
    The error, $\mathcal{E}$ (see Eq.~\eqref{eq: error B-SAP}), is reported for all the possible ratios among the coupling constants of the ferromagnetic XYZ Heisenberg Hamiltonian (see $H_T$
    in Eq.~\eqref{eq: Heisenberg Hamiltonian}).
    Calculations reported in this Figure have been performed for $L = 10$.
    The adiabatic procedure has been realized 
    with a number of Trotter steps that is 
    equivalent $L/2 = 5$, each with $0.25\cdot [J^z]^{-1}$ duration.}
    \label{fig: numerical Iexc}
\end{figure*}

\subsection{Example: target state for n = 1}
\noindent

In this case, the subspace dimension grows up to 
\begin{equation}
    d_1 = \frac{1}{2}L(L-1)
\end{equation}
and the Hamming weight is 2.
Since each state in $\mathcal{W}_{2} $ 
is completely defined by the positions
of its two non-zero terms in the bit string,
we will more conveniently denote them as $\ket{a, b}$, in which $a$ and $b $ label the positions 
of such terms in the chain,
and $0\leq a < b \leq L-1$ is assumed.
We hereby show the explicit implementation of the operators $G_j\left( \alpha_j\right)$
as quantum circuits, in which we assume $0 \leq i < l \leq L-1$ and $0 \leq k < r \leq L-1$ 
(see Supplemental Material for the detailed derivation): \\

\vspace{0.2 cm}
\Qcircuit @C=0.8em @R=0.5em {
q_i\quad & \multigate{2}{G_{Y}^{(i,l)}(\alpha)} &\qw  & & &\gate{H} & \ctrl{1} & \gate{\mathcal{R}_y(-\alpha)} & \ctrl{1} & \gate{H}  & \qw\\
& & & =& & & & &\\
q_l\quad & \ghost{G_{Y}^{(i,j)}(\alpha)} &\qw  & & &\gate{H} &\ctrl{-1} &  \gate{\mathcal{R}_y(\alpha)} & \ctrl{-1} & \gate{H} & 
\qw\\}
\vspace{0.5cm}
\Qcircuit @C=0.8em @R=0.5em {
q_i\quad & \multigate{2}{G_{X}^{(i,l)}(\alpha)} &\qw  & & &\gate{\mathcal{R}_z^{-1}(\frac{\pi}{4})} & \multigate{2}{G_{Y}^{(i,l)}(\alpha)}& \gate{\mathcal{R}_z(\frac{\pi}{4})}\\
& & & \hspace{0.1cm}=& & & &\\
q_l\quad & \ghost{G_{Y}^{(i,l)}(\alpha)} &\qw  & & &\gate{\mathcal{R}_z(\frac{\pi}{4})} &\ghost{G_{Y}^{(i,l)}(\alpha)} & \gate{\mathcal{R}_z^{-1}(\frac{\pi}{4})}\\}
\vspace{0.5cm}
\Qcircuit @C=0.8em @R=0.5em {
q_i\quad &  \multigate{4}{G_{Y}^{(i,l)(k,r)}(\alpha)} & \qw & & & \targ{} &  \qw&\ctrl{1} & \qw& \targ{}  & \qw\\
q_l\quad & \ghost{G_{Y}^{(i,l)(k,r)}(\alpha)} & \qw & & & \ctrlo{-1} & \qw& \multigate{2}{G_{Y}(\alpha)} & \qw& \ctrlo{-1} & 
\qw\\
& & & =& & & & &\\
q_k\quad &\ghost{G_{Y}^{(i,l)(k,r)}(\alpha)} & \qw & & & \ctrlo{1}  &\qw& \ghost{G_{Y}(\alpha)} &  \qw& \ctrlo{1} & \qw\\
q_r\quad &\ghost{G_{Y}^{(i,l)(k,r)}(\alpha)} & \qw & & & \targ{}  & \qw&\ctrl{-1} &  \qw&\targ{} & 
\qw\\}
\vspace{0.5cm}
\Qcircuit @C=0.8em @R=0.5em {
q_i\quad &  \multigate{4}{G_{X}^{(i,l)(k,r)}(\alpha)} & \qw & & & \targ{} &  \qw&\ctrl{1} & \qw& \targ{}  & \qw\\
q_l\quad & \ghost{G_{X}^{(i,l)(k,r)}(\alpha)} & \qw & & & \ctrlo{-1} & \qw& \multigate{2}{G_{X}(\alpha)} & \qw& \ctrlo{-1} & 
\qw\\
& & & =& & & & &\\
q_k\quad &\ghost{G_{X}^{(i,l)(k,r)}(\alpha)} & \qw & & & \ctrlo{1}  &\qw& \ghost{G_{X}(\alpha)} &  \qw& \ctrlo{1} & \qw\\
q_r\quad &\ghost{G_{X}^{(i,l)(k,r)}(\alpha)} & \qw & & & \targ{}  & \qw&\ctrl{-1} &  \qw&\targ{} & 
\qw\\}
\vspace*{0.5 cm}
\begin{figure*}[t]
\centering
    \begin{subfigure}[]{0.4\textwidth}
        \includegraphics[width = \textwidth]{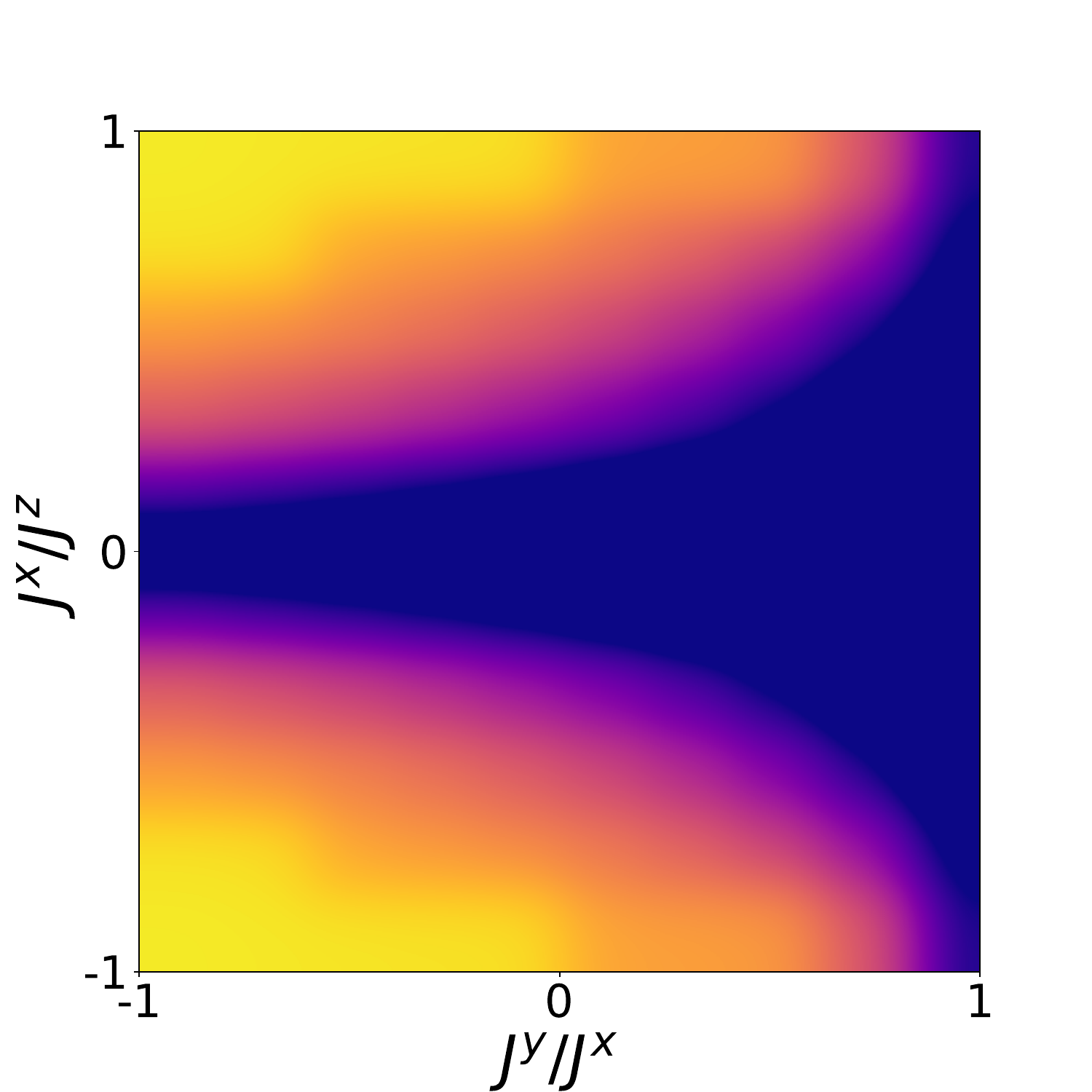}
        \caption{Ground state.}
    \end{subfigure}
    \begin{subfigure}[]{0.4\textwidth}
        \includegraphics[width = \textwidth]{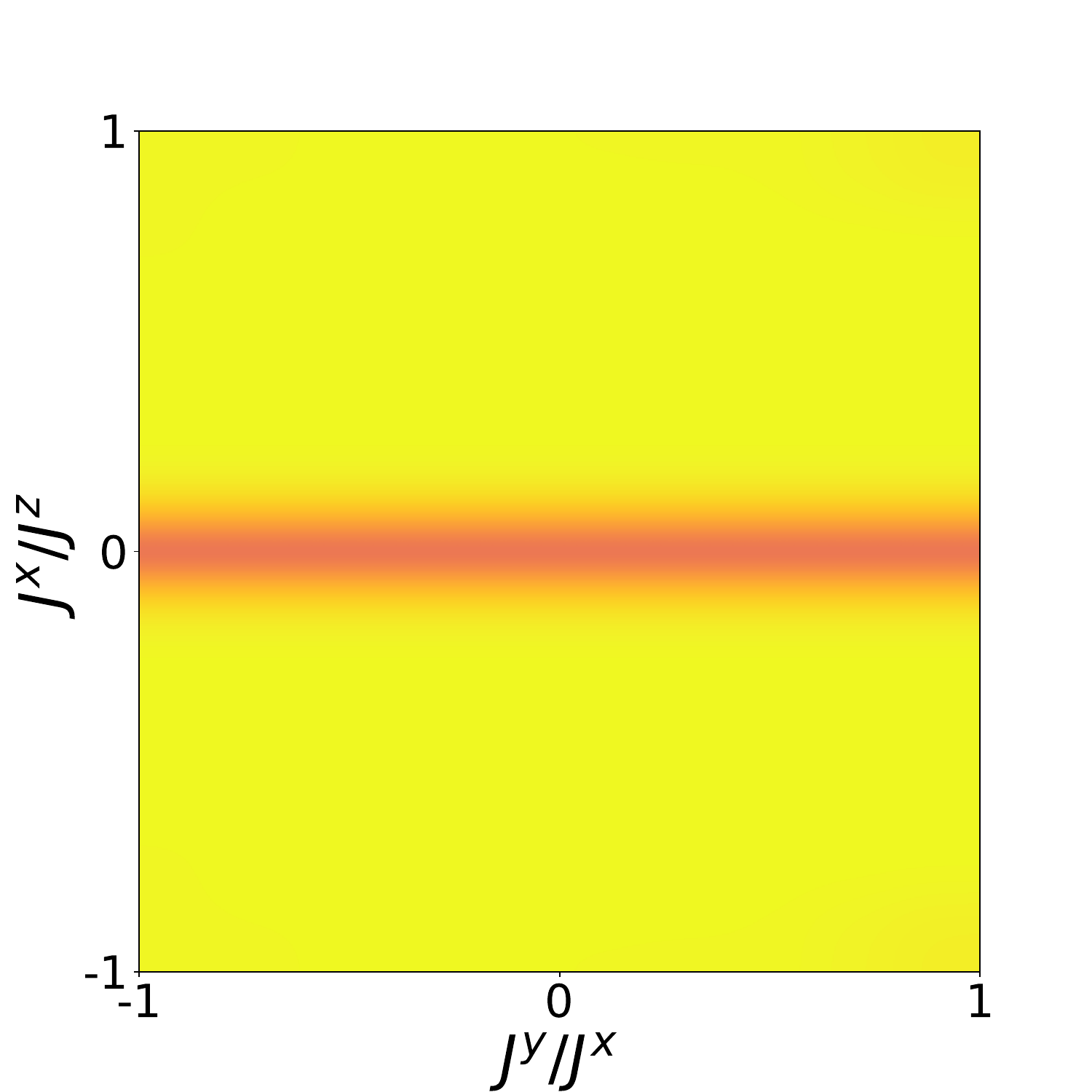}
        \caption{First excited state.}
    \end{subfigure}
    \begin{subfigure}[]{0.1\textwidth}
        \includegraphics[width = \textwidth]{Figures/cbar.pdf}
    \end{subfigure}
    \caption{Error in preparing the ground and first excited states of the ferromagnetic XYZ Heisenberg Hamiltonian $H_T$
    of Eq.~\eqref{eq: Heisenberg Hamiltonian}
    using standard adiabatic preparation. 
    Both panels show results for $L = 10$, 
    with $L/2 = 5$ Trotter steps of duration $0.25\cdot [J^z]^{-1}$ each.}
    \label{fig: AP numerical results}
\end{figure*}

\noindent
Fixing the state in Eq.~\eqref{eq: G mapping} to be $\ket{\Psi} = \ket{0,1}$,
one is left with $2(d_1-1) = L^2-L-2$ gates
whose action on $\ket{\Psi}$ is non-trivial.
Those are $\{G_Y^{(0,a)}\}$, $\{G_X^{(0,a)}\}$,
$\{G_Y^{(1,a)}\}$, $\{G_X^{(1,a)}\}$, $\{G_Y^{(0,1)(a,b)}\}$ and
$\{G_X^{(0,1)(a,b)}\}$, with $2 \leq a < b \leq L-1$.
The subsequent application of all the above mentioned gates,
irrespectively from the order,
allows us to move in the desired subspace $\mathcal{S}\left( \mathcal{W}_2\right)$.
Importantly, the structure of $\mathcal{G}\left(\underline{\alpha}\right)$ 
is independent from $H_T$, and it is only tailored on the initial Hamiltonian $H_0$ in Eq.~\eqref{eq: initial Hamiltonian}.
In what follows, we will further reduce the number of trainable parameters by exploiting some specific properties of the target Hamiltonian~\eqref{eq: Heisenberg Hamiltonian}.
First, we notice that the latter is purely real,
and hence there exists an orthogonal (rather than simply unitary) change of basis that diagonalizes it.
Hence, we do not need to reproduce the action of the full $\mathrm{U}(d_1)$ symmetry group,
but we can just focus on its subgroup $ \mathrm{O}(d_1)$. The Lie group $ \mathrm{O}(d_1)$ has two connected components, and the associated Lie algebra has dimension $\frac{1}{2}d_1 \left(d_1 -1 \right)$.  Its generators are those of $\mathrm{U}(d_1)$, whose associated exponentiation 
is of type $G_Y^{(i,j)}$ or $G_Y^{(i,j)(k,r)}$.
Restricting to those generators whose action on $\ket{\Psi}$ is not trivial, we are left with $d_1-1$ generators, i.e., $d_1-1$ real parameters in addition to a binary parameter indicating 
which of the two connected components of $ \mathrm{O}(d_1)$ is explored.
Hence, the total number of trainable parameters ultimately reduces to $d_1 = L(L-1)/2$.
Now that we have showcased how to explore $\mathcal{S}\left( \mathcal{W}_{2}\right)$,
we follow the steps for the B-SAP algorithm put forward above.
\Romannum{1}) 
The starting point is the preparation $\ket{\Psi}$ on the quantum register;
\Romannum{2})
It follows the application of all the parametrized gates required 
to enable the exploration of the real-valued span $\mathcal{S}_\mathds{R}\left( \mathcal{W}_{2}\right)$;
\Romannum{3}) the subsequent application 
of $\Phi_0^{-1}$ (or $\Phi_1^{-1}$) via
the following circuit
\vspace{0.2cm}
\begin{align*}
\Qcircuit @C=0.8em @R=0.5em {
q_{0}\quad & \multigate{6}{\Phi_0^{-1}}  & \qw & &&\qw & \qw & \qw & \qw & \targ{}& \qw  \\
&\ghost{\Phi_0^{-1}} &\qw & & &\qw & \qw & \qw &  \targ{} & \ctrl{-1} & \qw \\
&\ghost{\Phi_0^{-1}} &\qw  & &&\qw & \qw & \targ{} & \ctrl{-1}  \qw & \qw & \qw \\
& & & =& & & & &\\
&\ghost{\Phi_0^{-1}} &\qw  & &&\qw & \targ{} & \ctrl{-2} & \qw & \qw & \qw  \\
q_{L-2}\quad\quad & \ghost{\Phi_0^{-1}} & \qw& & &\targ{} & \ctrl{-1} &  \qw & \qw & \qw  &  \qw\\
q_{L-1}\quad\quad  &  \ghost{\Phi_0^{-1}} & \qw& & &\ctrl{-1}& \qw &  \qw & \qw & \qw   & \qw\\}    
\end{align*}
enables to access the generic state in $\mathcal{S}_\mathds{R}\left( B_{2}\right)$.
\Romannum{4}) At this point one applies
either $\frac{1}{\sqrt{2}}\left( \mathds{1} + \mathcal{X} \right)$ or $\frac{1}{\sqrt{2}}\left( \mathds{1} - \mathcal{X} \right)$,
depending on the targeted sector of $\mathcal{X}$. 
Notice that steps \Romannum{1} - \Romannum{4} together
correspond to points 3. and 4. of the B-SAP algorithm outlined above. 
\Romannum{5}) At this point, a (trotterized version) of the adiabatic protocol
is applied to the quantum register;
\Romannum{6}) finally, the circuit parameters are optimized
through the MC-VQE method,
i.e. buy using quantum resources to construct the  square matrix $E^{(1)}_{ml} \in \mathrm{Mat}_{d_1\times d_1 }\left( \mathds{R}\right)$
that can be therefore diagonalized classiacally 
(see Supplemental Material for details).

We hereby present the results 
of this procedure for the quantum state preparation of the first four excited eigenstates of the Heisenberg Hamiltonian, 
for each parity sector of the $\mathcal{X}$ operator, by using a quantum register with up to $L = 10$ qubits 
and $L/2 = 5$ Trotter steps for the adiabatic-protocol encoding. 
Figure~\ref{fig: numerical Iexc} 
shows the calculated error, $\mathcal{E}$ from Eq.~\eqref{eq: error B-SAP}, in a colored logarithmic scale,
as a function of the Hamiltonian parameters.
As evident from the plots, 
the error in states preparation remains either very small (i.e., below the percent level) 
or limited to values below $10\%$, in most of the parameters space. As it can be visually appreciated, the regions of larger error tend to spread on increasing the order of the excited energy level, at fixed Trotter step size. 
This is because such eigenvalues lie in close proximity to one another, making the classical diagonalization of the $E_{mn}$ matrix less effective 
in distinguishing the corresponding eigenstates.
These results fully confirm the effectiveness of the B-SAP approach even for preparation of excited energy eigenstates, which is a major outcome of the present work. 

Finally, we conclude this work by showcasing a relevant point: the comparison between the B-SAP and the conventional AP. In Fig.~\ref{fig: AP numerical results} we report the results of the standard AP performed on either ground or first excited state of the very same model Hamiltonian considered above. In these simulations, in order to guarantee a fair comparison, we keep the same number of qubits, $L=10$, and the same Trotter-steps number. The initial Hamiltonian for the AP has been chosen as 
\begin{equation}
    H_0 = - J^z \sum_i \frac{Z_i}{2^i} \, ,
\end{equation}
motivated by the fact that it is already diagonal on the computational basis and it does not present any degenerate level.
In particular, ground state errors remain limited to relatively small values in a large fraction of the parameters space. In fact, the results of Fig.~\ref{fig: AP numerical results}(a) should be compared to the B-SAP results of Fig.~\ref{fig: Trotter dependence}(a), which were obtained for the same number of qubits and Trotter steps: the overall behavior of the error is similar and quantitatively comparable, although the B-SAP performs slightly better. On the other hand, crucially, the results on the error for the preparation of the first excited state displayed in Fig.~\ref{fig: Trotter dependence}(b) confirm that, apart from the trivial parameter region $J^x/J^z=0$, the AP completely fails, due to the inherent energy level crossing during the adiabatic evolution (see, e.g., the plotted spectrum of the Hamiltonian in Fig.~\ref{fig: AP spectrum}). \\

\section{Conclusions}

We have presented a quantum algorithm
that leverages the group-theoretical notion of irreducible representations
to combine in a fresh way the strengths of both variational quantum algorithms and adiabatic state preparation. 
Our approach, which we call B-SAP, 
does not rely on ansatz circuits, 
avoids the problem of barren plateaus, 
and relaxes the strict energy-gap conditions required in conventional adiabatic protocols.
The algorithm leverages structured quantum queries followed by classical postprocessing, following a philosophy similar to that of Ref.~\cite{cerezo2023does}.

The algorithm has been validated on the XYZ Heisenberg-ring model, preparing its lowest-energy eigenstates across a broad parameter space, by using a quantum computer simulator. Our results show that B-SAP achieves the desired state preparation with very low error using a circuit whose depth scales polynomially in the system size, making it well-suited for direct implementations into near-term quantum hardware and appealing for fault-tolerant quantum simulations.
Looking forward, the flexibility and efficiency of our approach open pathways for broader applications, including quantum simulation of complex materials, quantum chemistry, and field-theoretical systems. Future developments will focus on integrating quantum error mitigation techniques to enable deployment on real quantum devices.\\\\

\section{acknowledgments}
This research was supported by the Italian Ministry of Research (MUR): D.G. acknowledges the PNRR project PE0000023 - National Quantum Science Technology Institute (NQSTI), G.G. acknowledges the MUR grant “Rita Levi-Montalcini”. The authors warmly acknowledge C. Dappiaggi, F. Scala and F. Tacchino for stimulating scientific discussions.

\clearpage
\onecolumngrid
\begin{center}
\textbf{Supplementary Material}
\end{center}

\begin{figure*}[b]
\centering
    \begin{subfigure}[t]{0.32\textwidth}
    \includegraphics[width=\textwidth]{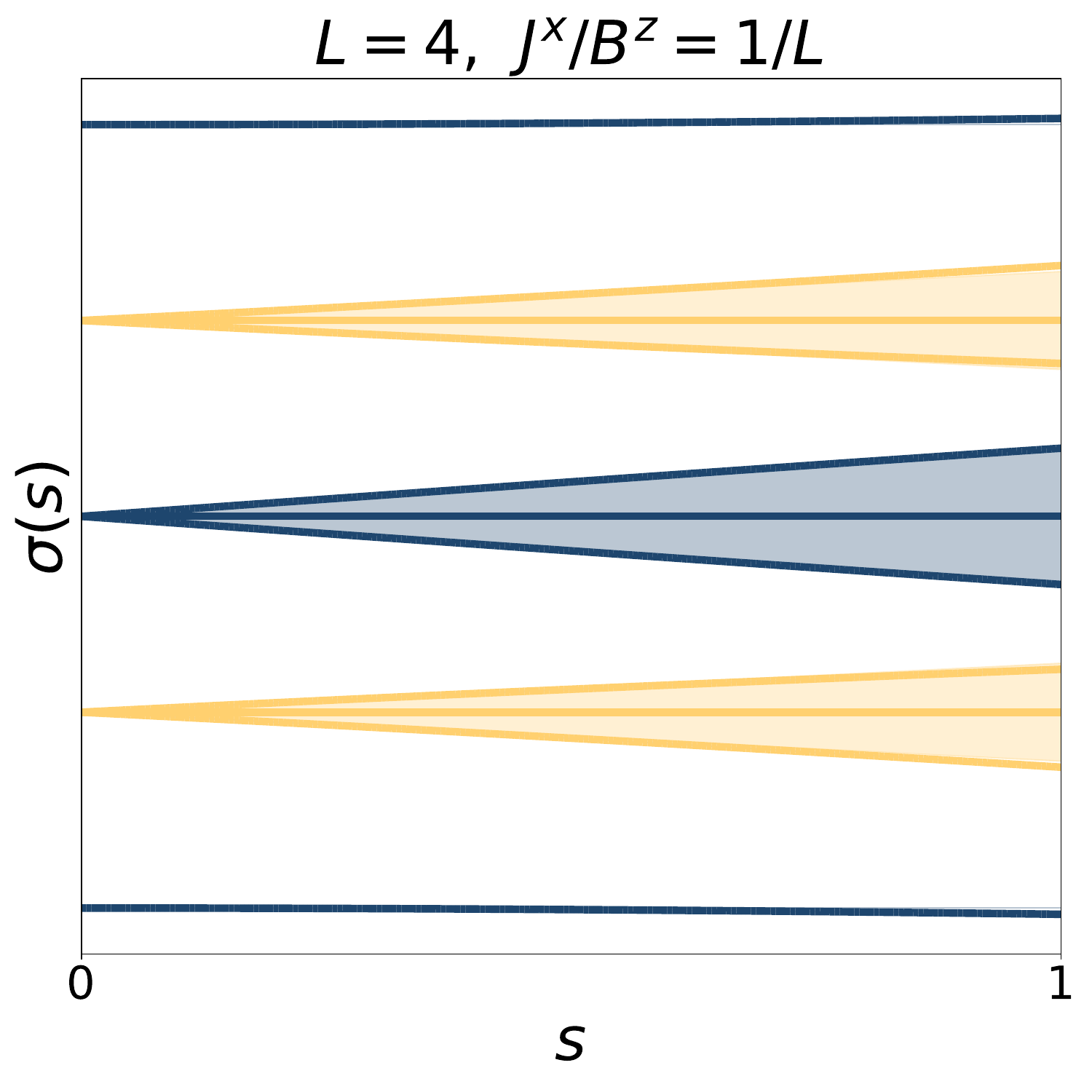}
    \caption{Small ratio $J^x/B^z = 1/L$}
    \label{fig:eigenvalues_small_ratio}
    \end{subfigure}
    \hspace{0cm}
    \begin{subfigure}[t]{0.32\textwidth}
    \includegraphics[width=\textwidth]{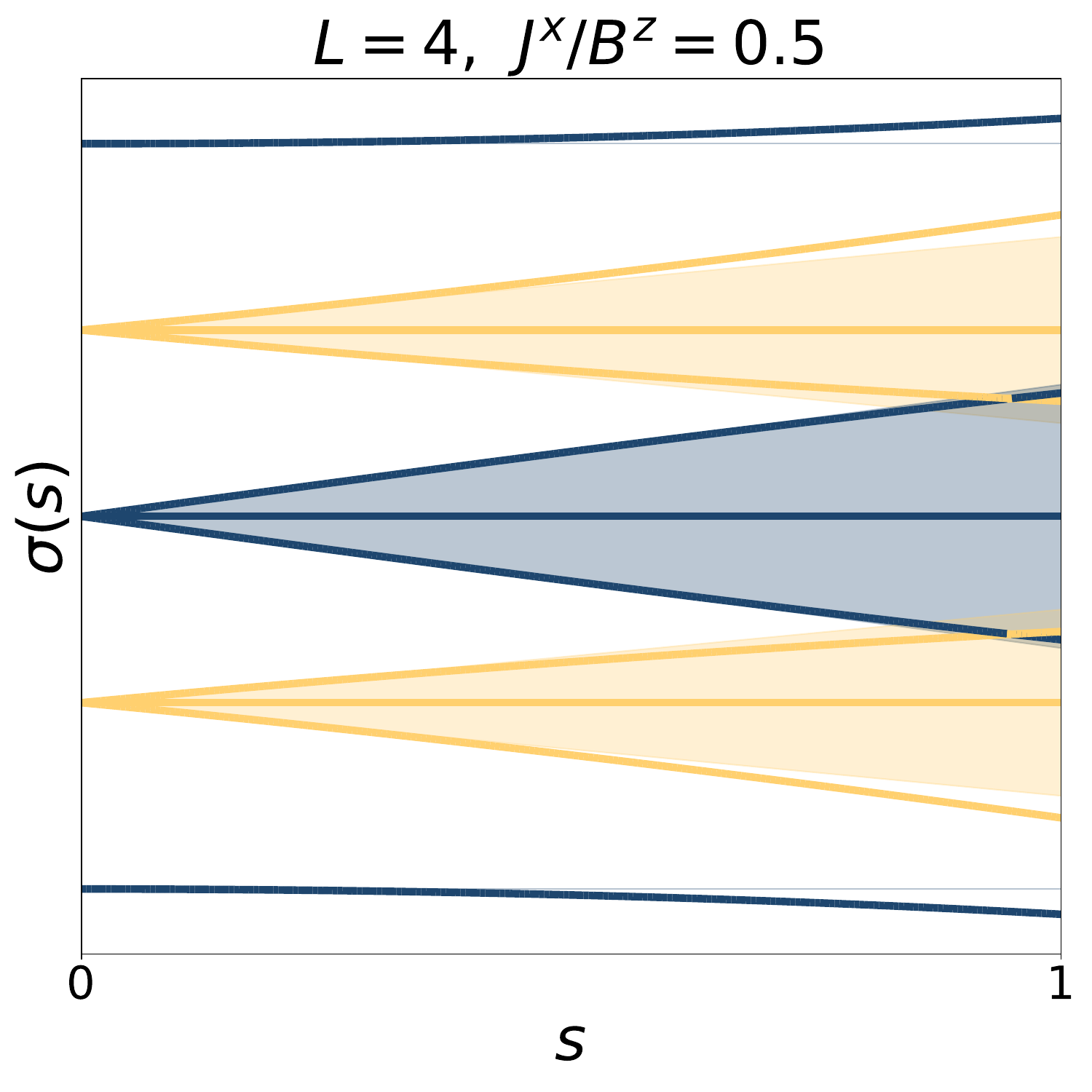}
    \caption{Larger ratio $J^x/B^z$ with possible near-crossings}
    \label{fig:eigenvalues_large_ratio}
    \end{subfigure}
    \hspace{0cm}
    \begin{subfigure}[t]{0.32\textwidth}
    \includegraphics[width=\textwidth]{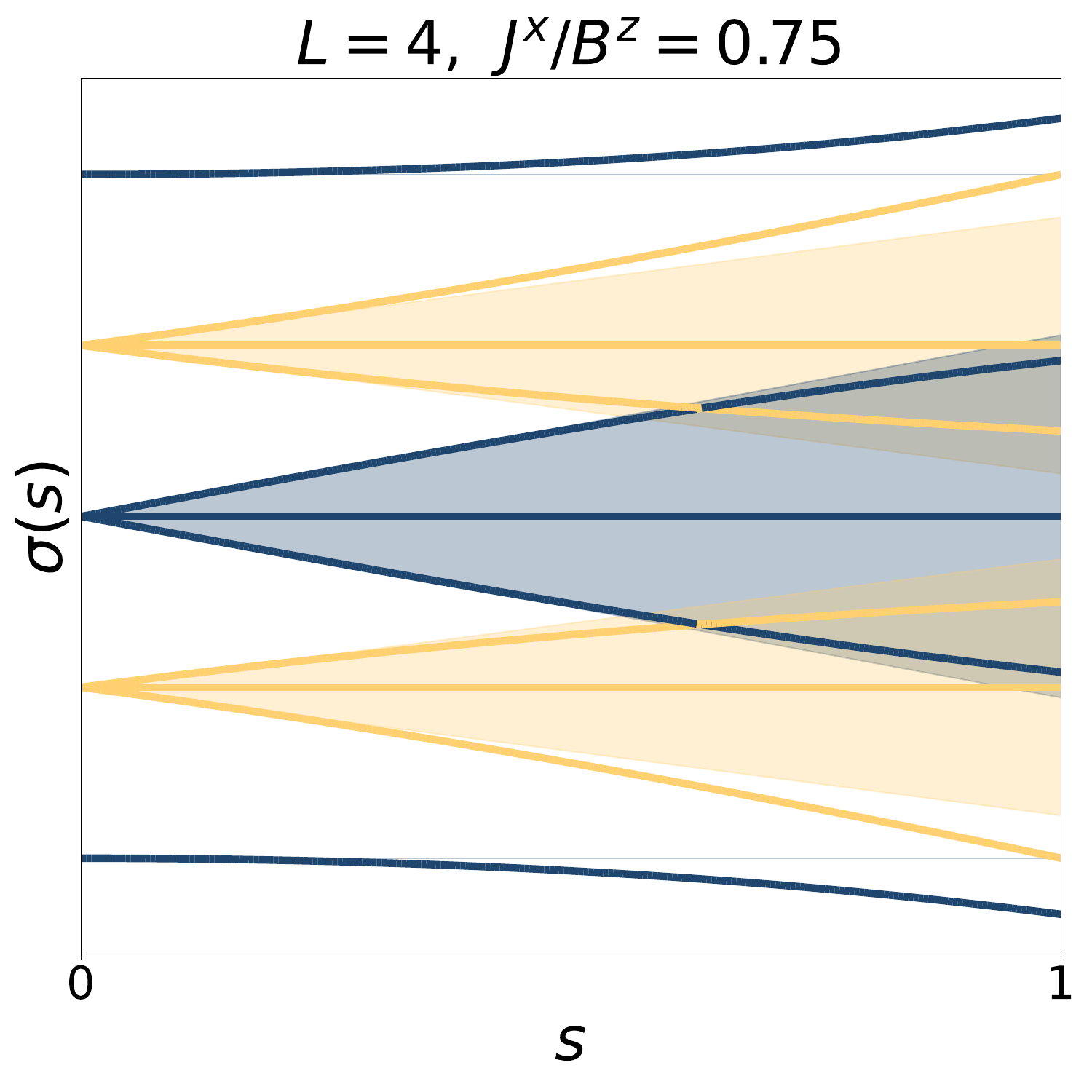}
    \caption{Even larger ratio $J^x/B^z$ with more pronounced near-crossings}
    \label{fig:eigenvalues_largest_ratio}
    \end{subfigure}
    \caption{Spectrum of the transverse-field Ising chain with $L=4$ sites as a function of the interpolation parameter $s$. The thick lines show the exact eigenvalues of $H(s)$. The shaded areas around the initial Hamming-weight energies $\sigma_n(0)$ represent the estimated maximal shifts $\pm s \, \|\Pi_n V \Pi_n\|$ for each Hamming-weight sector $n$. Panel (a) corresponds to a small ratio $J^x/B^z = 1/L$, where the gaps between sectors are large and the bound provides a clear separation. Panels (b) and (c) correspond to larger ratios $J^x/B^z$, in which case one should also evolve the closest Hamming-weight eigenspaces that may cross or become nearly degenerate, with an additional polynomial overhead, to ensure the correct ordering of eigenstates.}
    \label{fig:eigenvalues_comparison}
\end{figure*}

\section{Guarantee criteria for B-SAP conditions satisfiability}
\noindent
In this section we examine different situations that may arise when attempting to guarantee that the gap conditions required by B-SAP are satisfied.  As a representative testbed we consider the transverse-field Ising model on a closed chain of \(L\) sites.  The target Hamiltonian is
\begin{align}
    H_T &= -J^x \sum_{i=1}^L X_i X_{i+1} + B^z \sum_{i=1}^L Z_i ,
\end{align}
with periodic boundary conditions \(X_{L+1}\equiv X_1\).  \(H_T\) is invariant under site translations, reflections of the chain, and \(\pi\)-rotations of every spin about the \(z\)-axis.

Two natural choices for the initial Hamiltonian \(H_0\) are
\begin{align}
    H_0 &= -J^x \sum_{i=1}^L X_i X_{i+1}, \qquad \text{if } |J^x| \ge |B^z|, \\
\end{align}
and
\begin{align}
    H_0 &= B^z \sum_{i=1}^L Z_i, \qquad \text{if } |B^z| \ge |J^x|\,.
\end{align}
In the following we focus on the latter choice and use \(|B^z|\) as the unit of energy.  By a suitable rotation of the reference frame we may take \(B^z\ge 0\) and we assume this in what follows.

We adopt the standard convention \(Z\ket{0}=+1\ket{0}\), \(Z\ket{1}=-1\ket{1}\).  With this convention the eigenstates of
\begin{align}
    H_0 = B^z \sum_{i=1}^L Z_i
\end{align}
are the computational basis states \(\ket{\underline{b}}\) and their energies depend only on the Hamming weight \(h(\ket{\underline{b}})=n\) (the number of ones) according to
\begin{align}
    H_0 \ket{\underline{b}} = B^z\,(L-2n)\,\ket{\underline{b}}\,.
\end{align}
Consequently, the eigenspace \(\mathcal{S}(\mathcal{B}_n)\) corresponding to Hamming weight \(n\) has degeneracy
\begin{align}
    d_n = \binom{L}{n},
\end{align}
and adjacent Hamming-weight eigenspaces are separated at \(s=0\) by an energy gap \(2B^z\).

We consider the usual linear interpolation
\begin{align}
    H(s) = H_0 + s\bigl(H_T - H_0\bigr) , \qquad s\in[0,1],
\end{align}
so that the perturbation that moves us away from the unperturbed spectrum is
\begin{align}
    V := H_T - H_0 = J^x \sum_{i=1}^L X_i X_{i+1}.
\end{align}
B-SAP will evolve the initial eigenspace \(\mathcal{S}(\mathcal{B}_n)\) with exponentially small mixing if, for all \(s\in[0,1]\), the spectrum \(\sigma_n(s)\) originating from \(\mathcal{S}(\mathcal{B}_n)\) remains separated from all other spectra \(\sigma_m(s)\) by a finite gap.  
At this scope, we can require a bound on the maximal energy shift of levels 
inside a fixed Hamming-weight \(n\) sector under the perturbation \(sV\).
For clarity: projecting onto the different Hamming-weight $n$ sectors provides an approximate, 
sectorwise account of level shifts. 
A strictly rigorous alternative is to use Weyl’s theorem to bound the motion of the full spectrum, 
but that approach produces considerably looser (non-tight) bounds.
We project the perturbation to the Hamming-weight \(n\) subspace by defining
\begin{align}
    H_{\mathrm{eff}}^{(n)} := \Pi_n H_T \Pi_n = \Pi_n \bigl(H_0 + V\bigr) \Pi_n = H_0|_{\mathcal{S}(\mathcal{B}_n)} + \Pi_n V \Pi_n ,
\end{align}
where \(\Pi_n\) is the projector onto \(\mathcal{S}(\mathcal{B}_n)\).  Note that \(X_i X_{i+1}\) flips the pair of bits \((b_i,b_{i+1})\).  Such a local flip preserves the Hamming weight only when it acts on a boundary of the form \(01\) or \(10\); flips of \(00\) or \(11\) change the Hamming weight and are removed by the projection.
For a bitstring \(\underline{b}\) of weight \(n\) define \(q(\underline{b})\) as the number of boundaries of type \(01\) or \(10\).  The maximal number of such boundaries over all bitstrings of weight \(n\) (with periodic boundary conditions) is obtained by alternating bits as much as possible, giving
\begin{align}
    q_{\max}(n) = 2\min(n,\,L-n).
\end{align}

Within \(\mathcal{S}(\mathcal{B}_n)\) each allowed boundary flip contributes an off-diagonal matrix element of magnitude \(|J^x|\).  Hence every row of \(\Pi_n V \Pi_n\) contains at most \(q_{\max}(n)\) nonzero entries of magnitude \(|J^x|\).  By the Gershgorin circle theorem (or a simple row-sum bound) the operator norm satisfies
\begin{align}
    \bigl\|\Pi_n V \Pi_n\bigr\| \le |J^x|\, q_{\max}(n) = 2|J^x| \min(n,\,L-n) .
\end{align}
Therefore the projected perturbation \(s\,\Pi_n V \Pi_n\) causes at most the following maximal shift of any eigenvalue originating from the \(n\)-th subspace:
\begin{align}\label{eq:delta_sigma_general}
    \Delta\sigma_n(s) \;:=\; \max_{x\in \sigma_n(s),\,y\in\sigma_n(0)} |x-y|
    \;\le\; 2 s |J^x| \min(n,\,L-n) .
\end{align}

Two spectra \(\sigma_n(s)\) and \(\sigma_m(s)\) (with \(n\neq m\)) cannot cross for all \(s\in[0,1]\) if the sum of their maximal possible shifts does not exceed their initial separation:
\begin{align}
    \Delta\sigma_n(s) + \Delta\sigma_m(s) \le \bigl|\sigma_n(0)-\sigma_m(0)\bigr|
    = 2B^z\,|n-m| .
\end{align}
Using Eq.~\eqref{eq:delta_sigma_general} and taking the worst case \(s=1\) (which gives the largest possible shift over the interpolation), a sufficient condition to avoid any crossings for all \(s\in[0,1]\) is
\begin{align}\label{eq:avoid_crossing_general}
    2|J^x|\bigl(\min(n,L-n)+\min(m,L-m)\bigr) \le 2B^z\,|n-m|.
\end{align}
If one restricts to sectors with \(n,m \le L/2\) (the complementary cases follow by particle–hole symmetry), then \(\min(n,L-n)=n\) and \(\min(m,L-m)=m\), and Eq.~\eqref{eq:avoid_crossing_general} simplifies to
\begin{align}\label{eq:avoid_crossing_small_n}
    |J^x|(n+m) \le B^z\,|n-m|.
\end{align}
Eq.~\eqref{eq:avoid_crossing_small_n} is a conservative sufficient condition: if it holds, then the spectra originating from the two Hamming-weight sectors cannot overlap for any \(s\in[0,1]\).  Note that the most stringent requirement for avoiding crossings between nearby sectors typically occurs for \(m=n\pm 1\).
Clearly, for 
\begin{align}
\frac{|J^x|}{B^z} \le \frac{1}{L},
\end{align}
no crossing occurs and the B-SAP protocol is guaranteed \textit{a priori} to succeed for all Hamming-weight sectors $n$.  
However, even outside this strict regime, it is still possible to efficiently prepare the target state with an additional cost that remains polynomial in $L$.  

Indeed, for generic $|J^x| \le B^z$ and fixed $n$, the only spectrum partitions that may potentially cross $\sigma_n(s)$ are those corresponding to Hamming weights 
\begin{align}
m \in [\, n - \ell_\text{max}, \, n + k_\text{max} \,],
\end{align} 
with 
\begin{align}
\ell_\text{max} &= \left\lfloor \frac{2 n |J^x|}{B^z + |J^x|} \right\rfloor, \\
k_\text{max} &= \left\lfloor \frac{2 n |J^x|}{B^z - |J^x|} \right\rfloor.
\end{align} 
Notice that the number of such Hamming-weight sectors is independent of $L$.  

Each of the corresponding eigenspaces $\mathcal{S}(\mathcal{B}_m)$ has dimension
\begin{align}
d_m = \binom{L}{m}.
\end{align}
For fixed $m$, this scales polynomially with $L$:
\begin{align}
d_m \sim \frac{L^m}{m!}.
\end{align}

Therefore, the total number of states that may need to be adiabatically evolved or checked is at most
$\mathcal{O}\left(L^{\,n+k_\text{max}\,}\right)$
up to factors that depend only on $n$ and $k_\text{max}$ but not on $L$.  
This makes the protocol efficiently implementable, since both the number of potentially crossing sectors and the size of each sector grow at most polynomially in $L$.  
Adiabatically evolving all of them allows one to verify if any crossings occurred and, if necessary, determine \textit{a posteriori} the correct ordering of the eigenstates.

\clearpage 
\section{ Construction of quantum gates $G_j\left(\alpha_j\right)$ for $n=1$ }
\noindent
Here we present the construction of a gate set designed to explore the first excited eigenspace of the Hamiltonian
\begin{equation}
    H_0 = - \sum_i J^z Z_i Z_{i+1}\,,
\end{equation}
which we denote by $\mathcal{S}\left( \mathcal{W}_2 \right)$.
Each state in the subspace $\mathcal{W}_{2}$ 
is uniquely specified by the positions of the two non-zero entries 
(i.e., spin flips) in the bit string $a,b \in \{0,..., L-1\}$ (with $L$ being the number of qubits and $ \; a<b \,$ by convention). 
We therefore label these states by $\ket{a, b}$.
To efficiently parameterize this subspace, we construct a family of quantum gates 
\begin{equation}
    G_j\left( \alpha_j \right) = \mathrm{exp}\left[ -i \alpha_j t_j\right]
\end{equation}
for all the integer values $j \in \{1, .., d_1^2-1\}$, with $d_1=L(L-1)/2$ being the dimension of the first excited eigenspace, and where $\{ t_j \}$ is a chosen basis of generators of the associated Lie algebra $\mathfrak{su}(d_1)$.\\
In particular, we choose the basis to be the set of all generators of the type
\begin{align*}
        &\bra{c,d}t_X^{(i,l)}\ket{a,b} = 
     \left(\delta_{a,c} \delta_{i,d}\delta_{l,b}
    +\delta_{b,d}  \delta_{i,c}\delta_{l,a} 
    +\delta_{a,d} \delta_{i,c}\delta_{l,b}
    +\delta_{a,c} \delta_{i,b}\delta_{l,d}
    +\delta_{b,d}\delta_{i,a}\delta_{l,c}
    +\delta_{b,c}\delta_{i,a}\delta_{l,d}\right)\;,\\
    &\bra{c,d}t_X^{(i,l)(k,r)}\ket{a,b} =  \delta_{a,i}\delta_{b,l}\delta_{c,k}\delta_{d,r} +\delta_{a,k}\delta_{b,r}\delta_{c,i}\delta_{d,l}\;,\\
    &\bra{c,d}t_Y^{(i,l)}\ket{a,b} = 
     i\left(\delta_{a,c} \delta_{i,d}\delta_{l,b}
    +\delta_{b,d}  \delta_{i,c}\delta_{l,a} 
    +\delta_{a,d} \delta_{i,c}\delta_{l,b}
    -\delta_{a,c} \delta_{i,b}\delta_{l,d}
    -\delta_{b,d}\delta_{i,a}\delta_{l,c}
    -\delta_{b,c}\delta_{i,a}\delta_{l,d}\right)\;,\\
    &\bra{c,d}t_Y^{(i,l)(k,r)}\ket{a,b} =  i\left(\delta_{a,i}\delta_{b,l}\delta_{c,k}\delta_{d,r} -\delta_{a,k}\delta_{b,r}\delta_{c,i}\delta_{d,l}\right)\;,\\
    &\bra{c,d}t_Z^{(i,l)}\ket{a,b} =  \delta_{a,c}\delta_{b,d}\left[\delta_{i,a}+\delta_{i,b}-\delta_{l,a}-\delta_{l,b}\right]\,,
\end{align*}
with $i,\,l,\,k,\,r \in \{0,\,...,\, L-1 \}$.
The motivation for this choice is that the operators $t_X^{(i,l)}$, $t_Y^{(i,l)}$ and $t_Z^{(i,l)}$
only couple elements of $\mathcal{W}_2$ 
that differ solely in the values of the $i$-th and $l$-th bits, 
by acting as Pauli $X$, $Y$ and $Z$ operators, respectively.
On the other hand,  
$t_X^{(i,l)(k,r)}$ and $t_Y^{(i,l)(k,r)}$
act as Pauli $X$ and $Y$ 
on the elements $ \ket{i,l},\,\ket{k,r} \in \mathcal{W}_2$,
while leaving all the others unchanged.
Overall, those are $L^2-L+1$ generators 
but they need to be linearly independent in order to form a basis.
We now prove that the selected generators
are orthogonal with respect to the Hilbert-Schmidt scalar product,
which is defined as
\begin{equation}
    \langle A, B \rangle \equiv \mathrm{Tr}\left( A^\dagger B\right) =\sum_{a}\sum_{b>a} \bra{a,b}A^\dagger B \ket{a,b} = \sum_{a,c}\sum_{b>a,d>c} \bra{a,b}A^\dagger \ket{c,d}\bra{c,d} B \ket{a,b} \,,
\end{equation}
and hence they form an orthogonal basis.
We start by explicitly computing
\begin{equation}
\begin{split}
        \langle t_Y^{(i,l)(k,r)}, t_Y^{(q,m)(n,p)} \rangle 
        &= \sum_{a,c}\sum_{b>a,d>c}  \bra{c,d} t_Y^{(i,l)(k,r)}\ket{a,b}^*\bra{c,d}t_Y^{(q,m)(n,p)} \ket{a,b}\\
        &= \sum_{a,c}\sum_{b>a,d>c} \left( \delta_{a,i}\delta_{b,l}\delta_{c,k}\delta_{d,r}
        - \delta_{c,i}\delta_{d,l}\delta_{a,k}\delta_{b,r}\right)
        \left( \delta_{a,q}\delta_{b,m}\delta_{c,n}\delta_{d,p}
        - \delta_{c,q}\delta_{d,m}\delta_{a,n}\delta_{b,p}\right)\\
        &= 2\delta_{i,q}\delta_{l,m}\delta_{k,n}\delta_{r,p}-2\delta_{i,n}\delta_{l,p}\delta_{k,q}\delta_{r,m
        }\\
        &= 2\delta_{i,q}\delta_{l,m}\delta_{k,n}\delta_{r,p}\, ,
\end{split}
\end{equation}
where we have used $\delta_{i,n}\delta_{l,p}\delta_{k,q}\delta_{r,m} = 0$ 
since, by construction $i < k$ and $q < n$.
Similarly
\begin{equation}
\begin{split}
        \langle t_X^{(i,l)(k,r)}, t_X^{(q,m)(n,p)} \rangle 
        &= \sum_{a,c}\sum_{b>a,d>c}  \bra{c,d} t_Y^{(i,l)(k,r)}\ket{a,b}^*\bra{c,d}t_Y^{(q,m)(n,p)} \ket{a,b}\\
        &= \sum_{a,c}\sum_{b>a,d>c} \left( \delta_{a,i}\delta_{b,l}\delta_{c,k}\delta_{d,r}
        + \delta_{c,i}\delta_{d,l}\delta_{a,k}\delta_{b,r}\right)
        \left( \delta_{a,q}\delta_{b,m}\delta_{c,n}\delta_{d,p}
        + \delta_{c,q}\delta_{d,m}\delta_{a,n}\delta_{b,p}\right)\\
        &= 2\delta_{i,q}\delta_{l,m}\delta_{k,n}\delta_{r,p}+2\delta_{i,n}\delta_{l,p}\delta_{k,q}\delta_{r,m
        }\\
        &= 2\delta_{i,q}\delta_{l,m}\delta_{k,n}\delta_{r,p}\, ,
\end{split}
\end{equation}
while
\begin{equation}
\begin{split}
        \langle t_Y^{(i,l)(k,r)}, t_X^{(q,m)(n,p)} \rangle 
        &= \sum_{a,c}\sum_{b>a,d>c}  \bra{c,d} t_Y^{(i,l)(k,r)}\ket{a,b}^*\bra{c,d}t_Y^{(q,m)(n,p)} \ket{a,b}\\
        &= -i\sum_{a,c}\sum_{b>a,d>c} \left( \delta_{a,i}\delta_{b,l}\delta_{c,k}\delta_{d,r}
        -\delta_{c,i}\delta_{d,l}\delta_{a,k}\delta_{b,r}\right)
        \left( \delta_{a,q}\delta_{b,m}\delta_{c,n}\delta_{d,p}
        + \delta_{c,q}\delta_{d,m}\delta_{a,n}\delta_{b,p}\right)\\
        &= 0 \, .
\end{split}
\end{equation}
Moving to the remaining generators
\begin{equation}
\begin{split}
        \langle t_Y^{(i,l)}, t_Y^{(k,r)} \rangle 
        &= \sum_{a,c}\sum_{b>a,d>c}  \bra{c,d} t_Y^{(i,l)}\ket{a,b}^*\bra{c,d}t_Y^{(k,r)} \ket{a,b}\\
        &= \sum_{a,c}\sum_{b>a,d>c}  \left(\delta_{a,c} \delta_{i,d}\delta_{l,b}
        +\delta_{b,d}  \delta_{i,c}\delta_{l,a} 
        +\delta_{a,d} \delta_{i,c}\delta_{l,b}
        -\delta_{a,c} \delta_{i,b}\delta_{l,d}
        -\delta_{b,d}\delta_{i,a}\delta_{l,c}
        -\delta_{b,c}\delta_{i,a}\delta_{l,d}\right)\\
        &\quad\left(\delta_{a,c} \delta_{k,d}\delta_{r,b}
        +\delta_{b,d}  \delta_{k,c}\delta_{r,a} 
        +\delta_{a,d} \delta_{k,c}\delta_{r,b}
        -\delta_{a,c} \delta_{k,b}\delta_{r,d}
        -\delta_{b,d}\delta_{k,a}\delta_{r,c}
        -\delta_{b,c}\delta_{k,a}\delta_{r,d}\right)\\
        &= \sum_{a,c}\sum_{b>a,d>c}  
        \big(
        \delta_{a,c} \delta_{i,d}\delta_{k,d}\delta_{l,b}\delta_{r,b}
        +\delta_{b,d} \delta_{i,c}\delta_{k,c}\delta_{l,a}\delta_{r,a}
        +\delta_{a,d} \delta_{i,c}\delta_{k,c}\delta_{l,b}\delta_{r,b}\\
        & \quad\quad \delta_{a,c} \delta_{i,b}\delta_{k,b}\delta_{l,d}\delta_{r,d}
        +\delta_{b,d} \delta_{i,a}\delta_{k,a   }\delta_{l,c}\delta_{r,c}
        +\delta_{b,c} \delta_{i,a}\delta_{k,a   }\delta_{l,d}\delta_{r,d}
        \big)\\
        &= \mathcal{N}\delta_{i,k}\delta_{l,r} \, ,
\end{split}
\end{equation}
where $\mathcal{N}$ is a proportionality factor. Similarly, we have
\begin{equation}
\begin{split}
        \langle t_X^{(i,l)}, t_X^{(k,r)} \rangle 
        &= \sum_{a,c}\sum_{b>a,d>c}  \bra{c,d} t_Y^{(i,l)}\ket{a,b}^*\bra{c,d}t_Y^{(k,r)} \ket{a,b}\\
        &= \sum_{a,c}\sum_{b>a,d>c}  \left(\delta_{a,c} \delta_{i,d}\delta_{l,b}
        +\delta_{b,d}  \delta_{i,c}\delta_{l,a} 
        +\delta_{a,d} \delta_{i,c}\delta_{l,b}
        +\delta_{a,c} \delta_{i,b}\delta_{l,d}
        +\delta_{b,d}\delta_{i,a}\delta_{l,c}
        +\delta_{b,c}\delta_{i,a}\delta_{l,d}\right)\\
        &\quad\left(\delta_{a,c} \delta_{k,d}\delta_{r,b}
        +\delta_{b,d}  \delta_{k,c}\delta_{r,a} 
        +\delta_{a,d} \delta_{k,c}\delta_{r,b}
        +\delta_{a,c} \delta_{k,b}\delta_{r,d}
        +\delta_{b,d}\delta_{k,a}\delta_{r,c}
        +\delta_{b,c}\delta_{k,a}\delta_{r,d}\right)\\
        &= \sum_{a,c}\sum_{b>a,d>c}  
        \big(
        \delta_{a,c} \delta_{i,d}\delta_{k,d}\delta_{l,b}\delta_{r,b}
        +\delta_{b,d} \delta_{i,c}\delta_{k,c}\delta_{l,a}\delta_{r,a}
        +\delta_{a,d} \delta_{i,c}\delta_{k,c}\delta_{l,b}\delta_{r,b}\\
        & \quad\quad \delta_{a,c} \delta_{i,b}\delta_{k,b}\delta_{l,d}\delta_{r,d}
        +\delta_{b,d} \delta_{i,a}\delta_{k,a   }\delta_{l,c}\delta_{r,c}
        +\delta_{b,c} \delta_{i,a}\delta_{k,a   }\delta_{l,d}\delta_{r,d}
        \big)\\
        &= \mathcal{N}\delta_{i,k}\delta_{l,r} \, ,
\end{split}
\end{equation}
and
\begin{equation}
\begin{split}
        \langle t_Y^{(i,l)}, t_X^{(k,r)} \rangle 
        &= \sum_{a,c}\sum_{b>a,d>c}  \bra{c,d} t_Y^{(i,l)}\ket{a,b}^*\bra{c,d}t_Y^{(k,r)} \ket{a,b}\\
        &= -i \sum_{a,c}\sum_{b>a,d>c}  \left(\delta_{a,c} \delta_{i,d}\delta_{l,b}
        +\delta_{b,d}  \delta_{i,c}\delta_{l,a} 
        +\delta_{a,d} \delta_{i,c}\delta_{l,b}
        -\delta_{a,c} \delta_{i,b}\delta_{l,d}
        -\delta_{b,d}\delta_{i,a}\delta_{l,c}
        -\delta_{b,c}\delta_{i,a}\delta_{l,d}\right)\\
        &\quad\left(\delta_{a,c} \delta_{k,d}\delta_{r,b}
        +\delta_{b,d}  \delta_{k,c}\delta_{r,a} 
        +\delta_{a,d} \delta_{k,c}\delta_{r,b}
        +\delta_{a,c} \delta_{k,b}\delta_{r,d}
        +\delta_{b,d}\delta_{k,a}\delta_{r,c}
        +\delta_{b,c}\delta_{k,a}\delta_{r,d}\right)\\
        &= \sum_{a,c}\sum_{b>a,d>c}  
        \big(
        \delta_{a,c} \delta_{i,d}\delta_{k,d}\delta_{l,b}\delta_{r,b}
        +\delta_{b,d} \delta_{i,c}\delta_{k,c}\delta_{l,a}\delta_{r,a}
        +\delta_{a,d} \delta_{i,c}\delta_{k,c}\delta_{l,b}\delta_{r,b}\\
        & \quad\quad -\delta_{a,c} \delta_{i,b}\delta_{k,b}\delta_{l,d}\delta_{r,d}
        -\delta_{b,d} \delta_{i,a}\delta_{k,a   }\delta_{l,c}\delta_{r,c}
        -\delta_{b,c} \delta_{i,a}\delta_{k,a   }\delta_{l,d}\delta_{r,d}
        \big)\\
        &= 0 \, .
\end{split}
\end{equation}
Finally, one should compute the mixed products, such as
\begin{equation}
\begin{split}
        \langle t_Y^{(i,l)}, t_Y^{(q,m)(n,p)} \rangle 
        &= \sum_{a,c}\sum_{b>a,d>c}  \bra{c,d} t_Y^{(i,l)}\ket{a,b}^*\bra{c,d}t_Y^{(q,m)(n,p)}  \ket{a,b}\\
        &= \sum_{a,c}\sum_{b>a,d>c}  \left(\delta_{a,c} \delta_{i,d}\delta_{l,b}
        +\delta_{b,d}  \delta_{i,c}\delta_{l,a} 
        +\delta_{a,d} \delta_{i,c}\delta_{l,b}
        -\delta_{a,c} \delta_{i,b}\delta_{l,d}
        -\delta_{b,d}\delta_{i,a}\delta_{l,c}
        -\delta_{b,c}\delta_{i,a}\delta_{l,d}\right)\\
        &\quad \left( \delta_{a,q}\delta_{b,m}\delta_{c,n}\delta_{d,p}
        - \delta_{c,q}\delta_{d,m}\delta_{a,n}\delta_{b,p}\right)\\
        &= 0 \, .
\end{split}
\end{equation}
In the same way we obtain
\begin{equation}
    \langle t_X^{(i,l)}, t_X^{(q,m)(n,p)} \rangle = \langle t_Y^{(i,l)}, t_X^{(q,m)(n,p)} \rangle = \langle t_X^{(i,l)}, t_Y^{(q,m)(n,p)} \rangle = 0 \, ,
\end{equation}
thus proving that the chosen set is an orthogonal basis.
It is now possible to proceed with the exponentiation of each of the generators in the basis.
Since $t_j ^2 = \mathds{1}$ holds for each of these generators, then we have
\begin{equation}
    G_j\left( \alpha_j \right) = \mathrm{exp}\left[ -i \alpha_j t_j\right] = \mathrm{cos}\left( \alpha_j\right)\mathds{1}-i\,\mathrm{sin}\left( \alpha_j\right) t_j\,,
\end{equation}
from which the matrix representation of the action of $G_j(\alpha_j)$ 
follows straightforwardly.
In particular, by changing the notation to $G_{X,Y,Z}^{(i,l)}(\alpha) = \mathrm{exp}\left[ -i\, \alpha \,t_{X,Y,Z}^{(i,l)}\right]$
and $G_{X,Y}^{(i,l)(k,r)}(\alpha) = \mathrm{exp}\left[ -i \,\alpha \,t_{X,Y}^{(i,l)(k,r)}\right]$,
one obtains that the $G$ operators act trivially on all the qubits,
except for qubits $(i,l)$ or $(i,l,k,r)$, respectively.
Tracing out the qubits on which the action is trivial, the matrices on the computational basis read:
\begin{equation}\label{eq: G matrices}
G_X^{(i,l)}(\alpha) = \begin{bmatrix}
        1 & 0 &0 &0\\
        0 & \mathrm{cos}\alpha & -i\,\mathrm{sin}\alpha & 0\\
        0 & -i\,\mathrm{sin}\alpha  & \mathrm{cos}\alpha  & 0\\
        0 &0 & 0 & 1
    \end{bmatrix}
\quad\quad
    G_Y^{(i,l)}(\alpha) = \begin{bmatrix}
        1 & 0 &0 &0\\
        0 & \mathrm{cos}\alpha & -\mathrm{sin}\alpha & 0\\
        0 & \mathrm{sin}\alpha  & \mathrm{cos}\alpha  & 0\\
        0 &0 & 0 & 1
    \end{bmatrix}
\quad\quad G_Z^{(i,j)}(\alpha) = \begin{bmatrix}
        1 & 0 &0 &0\\
        0 & e^{i\alpha} & 0 & 0\\
        0 & 0 & e^{-i\alpha}  & 0\\
        0 &0 & 0 & 1
    \end{bmatrix} \, .
\end{equation}
Similarly, 
those for $G_{X,Y}^{(i,l)(k,r)}(\alpha)$ are equal to the identity, except for two terms of the diagonal, in correspondence of the binary elements $2^i+2^j+1$ and $2^k+2^r+1$ of the computational basis,
and two terms in row $2^i+2^j+1$ and column $2^k+2^r+1$, and vice versa,
which are given as $-i\,\mathrm{sin}\alpha$, or $\pm \mathrm{sin}\alpha$ as in Eq.~\eqref{eq: G matrices}.
Inspired by $G_Y^{(i,l)}(\alpha)$ in Eq.~\eqref{eq: G matrices},
which is known as the generic Givens rotation in the Quantum Computing literature, it can be implemented as the following quantum circuit~\cite{arrazola2022universal},
\begin{align*}
    \Qcircuit @C=1em @R=0.7em {
    q_i\quad\quad\quad & \gate{H} & \ctrl{1} & \gate{\mathcal{R}_y(-\alpha)} & \ctrl{1} & \gate{H}  & \qw\\
    q_l\quad\quad\quad & \gate{H} & \ctrl{-1} &  \gate{\mathcal{R}_y(\alpha)} & \ctrl{-1} & \gate{H} & 
    \qw\\} \, .
\end{align*}

\clearpage
\section{State preparation for MC-VQE}
We hereby describe how to prepare the quantum states 
\begin{equation}
    \ket{\Psi_{ml}^p} = \frac{1}{\sqrt{2}}\left(\ket{\Psi_m} + (-1)^p \ket{\Psi_l} \right)\,, \quad (p=0,1)
\end{equation}
on the quantum register,
where $\ket{\Psi_m}$ are the elements of the basis $\mathcal{B}_1$.
In the main text, it is shown that 
$\mathcal{B}_1 = \Phi^{-1}\mathcal{W}_2$ is a bijective mapping.
We denote the elements of $\mathcal{W}_2$ as $\ket{a,b}$, with $a, b \in \{0, \, ...,\, L-1\}$ and $a < b$.
Hence our aim is to construct the generic superposition
\begin{equation}
    \ket{\Psi_{(a,b)(c,d)}^p} = \frac{1}{\sqrt{2}}\Phi^{-1}\left(\ket{a,b} + (-1)^p \ket{c,d} \right)\,, \quad (p=0,1)\,.
\end{equation}
For $(a,b) = (c,d)$, this is trivial.
If this is not the case, we are either in the case of $a \neq c$, or $b \neq d$.
If $a \neq c \neq b$, it is always possible to prepare $\ket{\Psi_{(a,b)(c,d)}^p}$ by applying the following circuit to the default state $\ket{0}^{\otimes L}$.\\
\begin{align*}
\Qcircuit @C=0.8em @R=0.5em {
q_c\quad &  \gate{R_Y\left[(-1)^p \pi/2 \right]} & \ctrl{1} & \ctrlo{2} & \ctrlo{3} &  \qw\\
q_d\quad &  \qw & \targ{} & \qw& \qw& \qw \\
q_a\quad &  \qw & \qw & \targ{}& \qw& \qw \\
q_b\quad &  \qw & \qw & \qw& \targ{}& \qw \\}
\end{align*}
Otherwise, it is enough to use the circuit
\begin{align*}
\Qcircuit @C=0.8em @R=0.5em {
q_d\quad &  \gate{R_Y\left[(-1)^p \pi/2 \right]} & \ctrl{1} & \ctrlo{2} & \ctrlo{3} &  \qw\\
q_c\quad &  \qw & \targ{} & \qw& \qw& \qw \\
q_a\quad &  \qw & \qw & \targ{}& \qw& \qw \\
q_b\quad &  \qw & \qw & \qw& \targ{}& \qw \\}
\end{align*}

\bibliography{main}

\begin{thebibliography}{67}%
\makeatletter
\providecommand \@ifxundefined [1]{%
 \@ifx{#1\undefined}
}%
\providecommand \@ifnum [1]{%
 \ifnum #1\expandafter \@firstoftwo
 \else \expandafter \@secondoftwo
 \fi
}%
\providecommand \@ifx [1]{%
 \ifx #1\expandafter \@firstoftwo
 \else \expandafter \@secondoftwo
 \fi
}%
\providecommand \natexlab [1]{#1}%
\providecommand \enquote  [1]{``#1''}%
\providecommand \bibnamefont  [1]{#1}%
\providecommand \bibfnamefont [1]{#1}%
\providecommand \citenamefont [1]{#1}%
\providecommand \href@noop [0]{\@secondoftwo}%
\providecommand \href [0]{\begingroup \@sanitize@url \@href}%
\providecommand \@href[1]{\@@startlink{#1}\@@href}%
\providecommand \@@href[1]{\endgroup#1\@@endlink}%
\providecommand \@sanitize@url [0]{\catcode `\\12\catcode `\$12\catcode `\&12\catcode `\#12\catcode `\^12\catcode `\_12\catcode `\%12\relax}%
\providecommand \@@startlink[1]{}%
\providecommand \@@endlink[0]{}%
\providecommand \url  [0]{\begingroup\@sanitize@url \@url }%
\providecommand \@url [1]{\endgroup\@href {#1}{\urlprefix }}%
\providecommand \urlprefix  [0]{URL }%
\providecommand \Eprint [0]{\href }%
\providecommand \doibase [0]{https://doi.org/}%
\providecommand \selectlanguage [0]{\@gobble}%
\providecommand \bibinfo  [0]{\@secondoftwo}%
\providecommand \bibfield  [0]{\@secondoftwo}%
\providecommand \translation [1]{[#1]}%
\providecommand \BibitemOpen [0]{}%
\providecommand \bibitemStop [0]{}%
\providecommand \bibitemNoStop [0]{.\EOS\space}%
\providecommand \EOS [0]{\spacefactor3000\relax}%
\providecommand \BibitemShut  [1]{\csname bibitem#1\endcsname}%
\let\auto@bib@innerbib\@empty
%</preamble>
\bibitem [{\citenamefont {Aolita}\ \emph {et~al.}(2015)\citenamefont {Aolita}, \citenamefont {Gogolin}, \citenamefont {Kliesch},\ and\ \citenamefont {Eisert}}]{aolita2015reliable}%
  \BibitemOpen
  \bibfield  {author} {\bibinfo {author} {\bibfnamefont {L.}~\bibnamefont {Aolita}}, \bibinfo {author} {\bibfnamefont {C.}~\bibnamefont {Gogolin}}, \bibinfo {author} {\bibfnamefont {M.}~\bibnamefont {Kliesch}},\ and\ \bibinfo {author} {\bibfnamefont {J.}~\bibnamefont {Eisert}},\ }\bibfield  {title} {\bibinfo {title} {Reliable quantum certification of photonic state preparations},\ }\href@noop {} {\bibfield  {journal} {\bibinfo  {journal} {Nature communications}\ }\textbf {\bibinfo {volume} {6}},\ \bibinfo {pages} {8498} (\bibinfo {year} {2015})}\BibitemShut {NoStop}%
\bibitem [{\citenamefont {Eisert}\ \emph {et~al.}(2020)\citenamefont {Eisert}, \citenamefont {Hangleiter}, \citenamefont {Walk}, \citenamefont {Roth}, \citenamefont {Markham}, \citenamefont {Parekh}, \citenamefont {Chabaud},\ and\ \citenamefont {Kashefi}}]{eisert2020quantum}%
  \BibitemOpen
  \bibfield  {author} {\bibinfo {author} {\bibfnamefont {J.}~\bibnamefont {Eisert}}, \bibinfo {author} {\bibfnamefont {D.}~\bibnamefont {Hangleiter}}, \bibinfo {author} {\bibfnamefont {N.}~\bibnamefont {Walk}}, \bibinfo {author} {\bibfnamefont {I.}~\bibnamefont {Roth}}, \bibinfo {author} {\bibfnamefont {D.}~\bibnamefont {Markham}}, \bibinfo {author} {\bibfnamefont {R.}~\bibnamefont {Parekh}}, \bibinfo {author} {\bibfnamefont {U.}~\bibnamefont {Chabaud}},\ and\ \bibinfo {author} {\bibfnamefont {E.}~\bibnamefont {Kashefi}},\ }\bibfield  {title} {\bibinfo {title} {Quantum certification and benchmarking},\ }\href@noop {} {\bibfield  {journal} {\bibinfo  {journal} {Nature Reviews Physics}\ }\textbf {\bibinfo {volume} {2}},\ \bibinfo {pages} {382} (\bibinfo {year} {2020})}\BibitemShut {NoStop}%
\bibitem [{\citenamefont {Tacchino}\ \emph {et~al.}(2020)\citenamefont {Tacchino}, \citenamefont {Chiesa}, \citenamefont {Carretta},\ and\ \citenamefont {Gerace}}]{Tacchino2020AQT}%
  \BibitemOpen
  \bibfield  {author} {\bibinfo {author} {\bibfnamefont {F.}~\bibnamefont {Tacchino}}, \bibinfo {author} {\bibfnamefont {A.}~\bibnamefont {Chiesa}}, \bibinfo {author} {\bibfnamefont {S.}~\bibnamefont {Carretta}},\ and\ \bibinfo {author} {\bibfnamefont {D.}~\bibnamefont {Gerace}},\ }\bibfield  {title} {\bibinfo {title} {Quantum computers as universal quantum simulators: State-of-the-art and perspectives},\ }\href {https://doi.org/https://doi.org/10.1002/qute.201900052} {\bibfield  {journal} {\bibinfo  {journal} {Advanced Quantum Technologies}\ }\textbf {\bibinfo {volume} {3}},\ \bibinfo {pages} {1900052} (\bibinfo {year} {2020})}\BibitemShut {NoStop}%
\bibitem [{\citenamefont {Farhi}\ \emph {et~al.}(2000)\citenamefont {Farhi}, \citenamefont {Goldstone}, \citenamefont {Gutmann},\ and\ \citenamefont {Sipser}}]{farhi2000quantum}%
  \BibitemOpen
  \bibfield  {author} {\bibinfo {author} {\bibfnamefont {E.}~\bibnamefont {Farhi}}, \bibinfo {author} {\bibfnamefont {J.}~\bibnamefont {Goldstone}}, \bibinfo {author} {\bibfnamefont {S.}~\bibnamefont {Gutmann}},\ and\ \bibinfo {author} {\bibfnamefont {M.}~\bibnamefont {Sipser}},\ }\bibfield  {title} {\bibinfo {title} {Quantum computation by adiabatic evolution},\ }\href@noop {} {\bibfield  {journal} {\bibinfo  {journal} {arXiv preprint quant-ph/0001106}\ } (\bibinfo {year} {2000})}\BibitemShut {NoStop}%
\bibitem [{\citenamefont {Hogg}(2003)}]{hogg2003adiabatic}%
  \BibitemOpen
  \bibfield  {author} {\bibinfo {author} {\bibfnamefont {T.}~\bibnamefont {Hogg}},\ }\bibfield  {title} {\bibinfo {title} {Adiabatic quantum computing for random satisfiability problems},\ }\href@noop {} {\bibfield  {journal} {\bibinfo  {journal} {Physical Review A}\ }\textbf {\bibinfo {volume} {67}},\ \bibinfo {pages} {022314} (\bibinfo {year} {2003})}\BibitemShut {NoStop}%
\bibitem [{\citenamefont {Preskill}(2018)}]{preskill2018simulating}%
  \BibitemOpen
  \bibfield  {author} {\bibinfo {author} {\bibfnamefont {J.}~\bibnamefont {Preskill}},\ }\bibfield  {title} {\bibinfo {title} {Simulating quantum field theory with a quantum computer},\ }\href@noop {} {\bibfield  {journal} {\bibinfo  {journal} {arXiv preprint arXiv:1811.10085}\ } (\bibinfo {year} {2018})}\BibitemShut {NoStop}%
\bibitem [{\citenamefont {Funcke}\ \emph {et~al.}(2023)\citenamefont {Funcke}, \citenamefont {Hartung}, \citenamefont {Jansen},\ and\ \citenamefont {K{\"u}hn}}]{funcke2023review}%
  \BibitemOpen
  \bibfield  {author} {\bibinfo {author} {\bibfnamefont {L.}~\bibnamefont {Funcke}}, \bibinfo {author} {\bibfnamefont {T.}~\bibnamefont {Hartung}}, \bibinfo {author} {\bibfnamefont {K.}~\bibnamefont {Jansen}},\ and\ \bibinfo {author} {\bibfnamefont {S.}~\bibnamefont {K{\"u}hn}},\ }\bibfield  {title} {\bibinfo {title} {Review on quantum computing for lattice field theory},\ }\href@noop {} {\bibfield  {journal} {\bibinfo  {journal} {arXiv preprint arXiv:2302.00467}\ } (\bibinfo {year} {2023})}\BibitemShut {NoStop}%
\bibitem [{\citenamefont {Martinez}\ \emph {et~al.}(2016)\citenamefont {Martinez}, \citenamefont {Muschik}, \citenamefont {Schindler}, \citenamefont {Nigg}, \citenamefont {Erhard}, \citenamefont {Heyl}, \citenamefont {Hauke}, \citenamefont {Dalmonte}, \citenamefont {Monz}, \citenamefont {Zoller},\ and\ \citenamefont {Blatt}}]{martinez2016}%
  \BibitemOpen
  \bibfield  {author} {\bibinfo {author} {\bibfnamefont {E.~A.}\ \bibnamefont {Martinez}}, \bibinfo {author} {\bibfnamefont {C.~A.}\ \bibnamefont {Muschik}}, \bibinfo {author} {\bibfnamefont {P.}~\bibnamefont {Schindler}}, \bibinfo {author} {\bibfnamefont {D.}~\bibnamefont {Nigg}}, \bibinfo {author} {\bibfnamefont {A.}~\bibnamefont {Erhard}}, \bibinfo {author} {\bibfnamefont {M.}~\bibnamefont {Heyl}}, \bibinfo {author} {\bibfnamefont {P.}~\bibnamefont {Hauke}}, \bibinfo {author} {\bibfnamefont {M.}~\bibnamefont {Dalmonte}}, \bibinfo {author} {\bibfnamefont {T.}~\bibnamefont {Monz}}, \bibinfo {author} {\bibfnamefont {P.}~\bibnamefont {Zoller}},\ and\ \bibinfo {author} {\bibfnamefont {R.}~\bibnamefont {Blatt}},\ }\bibfield  {title} {\bibinfo {title} {Real-time dynamics of lattice gauge theories with a few-qubit quantum computer},\ }\href@noop {} {\bibfield  {journal} {\bibinfo  {journal} {Nature}\ }\textbf {\bibinfo {volume} {534}},\ \bibinfo {pages} {516} (\bibinfo {year} {2016})}\BibitemShut {NoStop}%
\bibitem [{\citenamefont {Klco}\ \emph {et~al.}(2018)\citenamefont {Klco}, \citenamefont {Dumitrescu}, \citenamefont {McCaskey}, \citenamefont {Morris}, \citenamefont {Pooser}, \citenamefont {Sanz}, \citenamefont {Solano}, \citenamefont {Lougovski},\ and\ \citenamefont {Savage}}]{klco2018quantum}%
  \BibitemOpen
  \bibfield  {author} {\bibinfo {author} {\bibfnamefont {N.}~\bibnamefont {Klco}}, \bibinfo {author} {\bibfnamefont {E.~F.}\ \bibnamefont {Dumitrescu}}, \bibinfo {author} {\bibfnamefont {A.~J.}\ \bibnamefont {McCaskey}}, \bibinfo {author} {\bibfnamefont {T.~D.}\ \bibnamefont {Morris}}, \bibinfo {author} {\bibfnamefont {R.~C.}\ \bibnamefont {Pooser}}, \bibinfo {author} {\bibfnamefont {M.}~\bibnamefont {Sanz}}, \bibinfo {author} {\bibfnamefont {E.}~\bibnamefont {Solano}}, \bibinfo {author} {\bibfnamefont {P.}~\bibnamefont {Lougovski}},\ and\ \bibinfo {author} {\bibfnamefont {M.~J.}\ \bibnamefont {Savage}},\ }\bibfield  {title} {\bibinfo {title} {Quantum-classical computation of schwinger model dynamics using quantum computers},\ }\href@noop {} {\bibfield  {journal} {\bibinfo  {journal} {Physical Review A}\ }\textbf {\bibinfo {volume} {98}},\ \bibinfo {pages} {032331} (\bibinfo {year} {2018})}\BibitemShut {NoStop}%
\bibitem [{\citenamefont {Eisert}\ \emph {et~al.}(2015)\citenamefont {Eisert}, \citenamefont {Friesdorf},\ and\ \citenamefont {Gogolin}}]{eisert2015quantum}%
  \BibitemOpen
  \bibfield  {author} {\bibinfo {author} {\bibfnamefont {J.}~\bibnamefont {Eisert}}, \bibinfo {author} {\bibfnamefont {M.}~\bibnamefont {Friesdorf}},\ and\ \bibinfo {author} {\bibfnamefont {C.}~\bibnamefont {Gogolin}},\ }\bibfield  {title} {\bibinfo {title} {Quantum many-body systems out of equilibrium},\ }\href@noop {} {\bibfield  {journal} {\bibinfo  {journal} {Nature Physics}\ }\textbf {\bibinfo {volume} {11}},\ \bibinfo {pages} {124} (\bibinfo {year} {2015})}\BibitemShut {NoStop}%
\bibitem [{\citenamefont {Bertoni}\ \emph {et~al.}(2025)\citenamefont {Bertoni}, \citenamefont {Wassner}, \citenamefont {Guarnieri},\ and\ \citenamefont {Eisert}}]{bertoni2025typical}%
  \BibitemOpen
  \bibfield  {author} {\bibinfo {author} {\bibfnamefont {C.}~\bibnamefont {Bertoni}}, \bibinfo {author} {\bibfnamefont {C.}~\bibnamefont {Wassner}}, \bibinfo {author} {\bibfnamefont {G.}~\bibnamefont {Guarnieri}},\ and\ \bibinfo {author} {\bibfnamefont {J.}~\bibnamefont {Eisert}},\ }\bibfield  {title} {\bibinfo {title} {Typical thermalization of low-entanglement states},\ }\href@noop {} {\bibfield  {journal} {\bibinfo  {journal} {Communications Physics}\ }\textbf {\bibinfo {volume} {8}},\ \bibinfo {pages} {301} (\bibinfo {year} {2025})}\BibitemShut {NoStop}%
\bibitem [{\citenamefont {Daley}\ \emph {et~al.}(2022)\citenamefont {Daley}, \citenamefont {Bloch}, \citenamefont {Kokail}, \citenamefont {Flannigan}, \citenamefont {Pearson}, \citenamefont {Troyer},\ and\ \citenamefont {Zoller}}]{daley2022practical}%
  \BibitemOpen
  \bibfield  {author} {\bibinfo {author} {\bibfnamefont {A.~J.}\ \bibnamefont {Daley}}, \bibinfo {author} {\bibfnamefont {I.}~\bibnamefont {Bloch}}, \bibinfo {author} {\bibfnamefont {C.}~\bibnamefont {Kokail}}, \bibinfo {author} {\bibfnamefont {S.}~\bibnamefont {Flannigan}}, \bibinfo {author} {\bibfnamefont {N.}~\bibnamefont {Pearson}}, \bibinfo {author} {\bibfnamefont {M.}~\bibnamefont {Troyer}},\ and\ \bibinfo {author} {\bibfnamefont {P.}~\bibnamefont {Zoller}},\ }\bibfield  {title} {\bibinfo {title} {Practical quantum advantage in quantum simulation},\ }\href@noop {} {\bibfield  {journal} {\bibinfo  {journal} {Nature}\ }\textbf {\bibinfo {volume} {607}},\ \bibinfo {pages} {667} (\bibinfo {year} {2022})}\BibitemShut {NoStop}%
\bibitem [{\citenamefont {Julia}(2025)}]{julia2025catalysis}%
  \BibitemOpen
  \bibfield  {author} {\bibinfo {author} {\bibfnamefont {F.}~\bibnamefont {Julia}},\ }\bibfield  {title} {\bibinfo {title} {Catalysis in the excited state: Bringing innate transition metal photochemistry into play},\ }\href@noop {} {\bibfield  {journal} {\bibinfo  {journal} {ACS catalysis}\ }\textbf {\bibinfo {volume} {15}},\ \bibinfo {pages} {4665} (\bibinfo {year} {2025})}\BibitemShut {NoStop}%
\bibitem [{\citenamefont {Bauer}\ \emph {et~al.}(2020)\citenamefont {Bauer}, \citenamefont {Bravyi}, \citenamefont {Motta},\ and\ \citenamefont {Chan}}]{bauer2020quantum}%
  \BibitemOpen
  \bibfield  {author} {\bibinfo {author} {\bibfnamefont {B.}~\bibnamefont {Bauer}}, \bibinfo {author} {\bibfnamefont {S.}~\bibnamefont {Bravyi}}, \bibinfo {author} {\bibfnamefont {M.}~\bibnamefont {Motta}},\ and\ \bibinfo {author} {\bibfnamefont {G.~K.-L.}\ \bibnamefont {Chan}},\ }\bibfield  {title} {\bibinfo {title} {Quantum algorithms for quantum chemistry and quantum materials science},\ }\href@noop {} {\bibfield  {journal} {\bibinfo  {journal} {Chemical reviews}\ }\textbf {\bibinfo {volume} {120}},\ \bibinfo {pages} {12685} (\bibinfo {year} {2020})}\BibitemShut {NoStop}%
\bibitem [{\citenamefont {Gonz{\'a}lez}\ and\ \citenamefont {Lindh}(2020)}]{gonzalez2020quantum}%
  \BibitemOpen
  \bibfield  {author} {\bibinfo {author} {\bibfnamefont {L.}~\bibnamefont {Gonz{\'a}lez}}\ and\ \bibinfo {author} {\bibfnamefont {R.}~\bibnamefont {Lindh}},\ }\href@noop {} {\emph {\bibinfo {title} {Quantum chemistry and dynamics of excited states: methods and applications}}}\ (\bibinfo  {publisher} {John Wiley \& Sons},\ \bibinfo {year} {2020})\BibitemShut {NoStop}%
\bibitem [{\citenamefont {Roldao}\ \emph {et~al.}(2022)\citenamefont {Roldao}, \citenamefont {Oliveira}, \citenamefont {Mili{\'a}n-Medina}, \citenamefont {Gierschner},\ and\ \citenamefont {Roca-Sanjuan}}]{roldao2022quantum}%
  \BibitemOpen
  \bibfield  {author} {\bibinfo {author} {\bibfnamefont {J.~C.}\ \bibnamefont {Roldao}}, \bibinfo {author} {\bibfnamefont {E.~F.}\ \bibnamefont {Oliveira}}, \bibinfo {author} {\bibfnamefont {B.}~\bibnamefont {Mili{\'a}n-Medina}}, \bibinfo {author} {\bibfnamefont {J.}~\bibnamefont {Gierschner}},\ and\ \bibinfo {author} {\bibfnamefont {D.}~\bibnamefont {Roca-Sanjuan}},\ }\bibfield  {title} {\bibinfo {title} {Quantum-chemistry study of the ground and excited state absorption of distyrylbenzene: Multi vs single reference methods},\ }\href@noop {} {\bibfield  {journal} {\bibinfo  {journal} {The Journal of Chemical Physics}\ }\textbf {\bibinfo {volume} {156}} (\bibinfo {year} {2022})}\BibitemShut {NoStop}%
\bibitem [{\citenamefont {Levine}(2017)}]{levine2017photochemistry}%
  \BibitemOpen
  \bibfield  {author} {\bibinfo {author} {\bibfnamefont {R.}~\bibnamefont {Levine}},\ }\bibfield  {title} {\bibinfo {title} {Photochemistry of highly excited states},\ }\href@noop {} {\bibfield  {journal} {\bibinfo  {journal} {Proceedings of the National Academy of Sciences}\ }\textbf {\bibinfo {volume} {114}},\ \bibinfo {pages} {13594} (\bibinfo {year} {2017})}\BibitemShut {NoStop}%
\bibitem [{\citenamefont {Cerezo}\ \emph {et~al.}(2021)\citenamefont {Cerezo}, \citenamefont {Arrasmith}, \citenamefont {Babbush}, \citenamefont {Benjamin}, \citenamefont {Endo}, \citenamefont {Fujii}, \citenamefont {McClean}, \citenamefont {Mitarai}, \citenamefont {Yuan}, \citenamefont {Cincio} \emph {et~al.}}]{cerezo2021variational}%
  \BibitemOpen
  \bibfield  {author} {\bibinfo {author} {\bibfnamefont {M.}~\bibnamefont {Cerezo}}, \bibinfo {author} {\bibfnamefont {A.}~\bibnamefont {Arrasmith}}, \bibinfo {author} {\bibfnamefont {R.}~\bibnamefont {Babbush}}, \bibinfo {author} {\bibfnamefont {S.~C.}\ \bibnamefont {Benjamin}}, \bibinfo {author} {\bibfnamefont {S.}~\bibnamefont {Endo}}, \bibinfo {author} {\bibfnamefont {K.}~\bibnamefont {Fujii}}, \bibinfo {author} {\bibfnamefont {J.~R.}\ \bibnamefont {McClean}}, \bibinfo {author} {\bibfnamefont {K.}~\bibnamefont {Mitarai}}, \bibinfo {author} {\bibfnamefont {X.}~\bibnamefont {Yuan}}, \bibinfo {author} {\bibfnamefont {L.}~\bibnamefont {Cincio}}, \emph {et~al.},\ }\bibfield  {title} {\bibinfo {title} {Variational quantum algorithms},\ }\href@noop {} {\bibfield  {journal} {\bibinfo  {journal} {Nature Reviews Physics}\ }\textbf {\bibinfo {volume} {3}},\ \bibinfo {pages} {625} (\bibinfo {year} {2021})}\BibitemShut {NoStop}%
\bibitem [{\citenamefont {Born}\ and\ \citenamefont {Fock}(1928)}]{born1928beweis}%
  \BibitemOpen
  \bibfield  {author} {\bibinfo {author} {\bibfnamefont {M.}~\bibnamefont {Born}}\ and\ \bibinfo {author} {\bibfnamefont {V.}~\bibnamefont {Fock}},\ }\bibfield  {title} {\bibinfo {title} {Beweis des adiabatensatzes},\ }\href@noop {} {\bibfield  {journal} {\bibinfo  {journal} {Zeitschrift f{\"u}r Physik}\ }\textbf {\bibinfo {volume} {51}},\ \bibinfo {pages} {165} (\bibinfo {year} {1928})}\BibitemShut {NoStop}%
\bibitem [{\citenamefont {Albash}\ and\ \citenamefont {Lidar}(2018)}]{albash2018adiabatic}%
  \BibitemOpen
  \bibfield  {author} {\bibinfo {author} {\bibfnamefont {T.}~\bibnamefont {Albash}}\ and\ \bibinfo {author} {\bibfnamefont {D.~A.}\ \bibnamefont {Lidar}},\ }\bibfield  {title} {\bibinfo {title} {Adiabatic quantum computation},\ }\href@noop {} {\bibfield  {journal} {\bibinfo  {journal} {Reviews of Modern Physics}\ }\textbf {\bibinfo {volume} {90}},\ \bibinfo {pages} {015002} (\bibinfo {year} {2018})}\BibitemShut {NoStop}%
\bibitem [{\citenamefont {Babbush}\ \emph {et~al.}(2014)\citenamefont {Babbush}, \citenamefont {Love},\ and\ \citenamefont {Aspuru-Guzik}}]{babbush2014adiabatic}%
  \BibitemOpen
  \bibfield  {author} {\bibinfo {author} {\bibfnamefont {R.}~\bibnamefont {Babbush}}, \bibinfo {author} {\bibfnamefont {P.~J.}\ \bibnamefont {Love}},\ and\ \bibinfo {author} {\bibfnamefont {A.}~\bibnamefont {Aspuru-Guzik}},\ }\bibfield  {title} {\bibinfo {title} {Adiabatic quantum simulation of quantum chemistry},\ }\href@noop {} {\bibfield  {journal} {\bibinfo  {journal} {Scientific reports}\ }\textbf {\bibinfo {volume} {4}},\ \bibinfo {pages} {6603} (\bibinfo {year} {2014})}\BibitemShut {NoStop}%
\bibitem [{\citenamefont {Griffith}\ and\ \citenamefont {Orgel}(1957)}]{griffith1957ligand}%
  \BibitemOpen
  \bibfield  {author} {\bibinfo {author} {\bibfnamefont {J.}~\bibnamefont {Griffith}}\ and\ \bibinfo {author} {\bibfnamefont {L.}~\bibnamefont {Orgel}},\ }\bibfield  {title} {\bibinfo {title} {Ligand-field theory},\ }\href@noop {} {\bibfield  {journal} {\bibinfo  {journal} {Quarterly Reviews, Chemical Society}\ }\textbf {\bibinfo {volume} {11}},\ \bibinfo {pages} {381} (\bibinfo {year} {1957})}\BibitemShut {NoStop}%
\bibitem [{\citenamefont {Bredas}(2014)}]{bredas2014mind}%
  \BibitemOpen
  \bibfield  {author} {\bibinfo {author} {\bibfnamefont {J.-L.}\ \bibnamefont {Bredas}},\ }\bibfield  {title} {\bibinfo {title} {Mind the gap!},\ }\href@noop {} {\bibfield  {journal} {\bibinfo  {journal} {Materials Horizons}\ }\textbf {\bibinfo {volume} {1}},\ \bibinfo {pages} {17} (\bibinfo {year} {2014})}\BibitemShut {NoStop}%
\bibitem [{\citenamefont {Tiwary}\ and\ \citenamefont {Berne}(2016)}]{tiwary2016spectral}%
  \BibitemOpen
  \bibfield  {author} {\bibinfo {author} {\bibfnamefont {P.}~\bibnamefont {Tiwary}}\ and\ \bibinfo {author} {\bibfnamefont {B.}~\bibnamefont {Berne}},\ }\bibfield  {title} {\bibinfo {title} {Spectral gap optimization of order parameters for sampling complex molecular systems},\ }\href {https://doi.org/10.1073/pnas.1600917113} {\bibfield  {journal} {\bibinfo  {journal} {Proceedings of the National Academy of Sciences}\ }\textbf {\bibinfo {volume} {113}},\ \bibinfo {pages} {2839} (\bibinfo {year} {2016})}\BibitemShut {NoStop}%
\bibitem [{\citenamefont {Lupo~Pasini}\ \emph {et~al.}(2023)\citenamefont {Lupo~Pasini}, \citenamefont {Mehta}, \citenamefont {Yoo},\ and\ \citenamefont {Irle}}]{lupo2023two}%
  \BibitemOpen
  \bibfield  {author} {\bibinfo {author} {\bibfnamefont {M.}~\bibnamefont {Lupo~Pasini}}, \bibinfo {author} {\bibfnamefont {K.}~\bibnamefont {Mehta}}, \bibinfo {author} {\bibfnamefont {P.}~\bibnamefont {Yoo}},\ and\ \bibinfo {author} {\bibfnamefont {S.}~\bibnamefont {Irle}},\ }\bibfield  {title} {\bibinfo {title} {Two excited-state datasets for quantum chemical uv-vis spectra of organic molecules},\ }\href {https://doi.org/https://doi.org/10.1038/s41597-023-02408-4} {\bibfield  {journal} {\bibinfo  {journal} {Scientific Data}\ }\textbf {\bibinfo {volume} {10}},\ \bibinfo {pages} {546} (\bibinfo {year} {2023})}\BibitemShut {NoStop}%
\bibitem [{\citenamefont {Deshpande}\ \emph {et~al.}(2022)\citenamefont {Deshpande}, \citenamefont {Gorshkov},\ and\ \citenamefont {Fefferman}}]{deshpande2022importance}%
  \BibitemOpen
  \bibfield  {author} {\bibinfo {author} {\bibfnamefont {A.}~\bibnamefont {Deshpande}}, \bibinfo {author} {\bibfnamefont {A.~V.}\ \bibnamefont {Gorshkov}},\ and\ \bibinfo {author} {\bibfnamefont {B.}~\bibnamefont {Fefferman}},\ }\bibfield  {title} {\bibinfo {title} {Importance of the spectral gap in estimating ground-state energies},\ }\href {https://doi.org/https://doi.org/10.1103/PRXQuantum.3.040327} {\bibfield  {journal} {\bibinfo  {journal} {PRX Quantum}\ }\textbf {\bibinfo {volume} {3}},\ \bibinfo {pages} {040327} (\bibinfo {year} {2022})}\BibitemShut {NoStop}%
\bibitem [{\citenamefont {Somma}\ and\ \citenamefont {Boixo}(2013)}]{somma2013spectral}%
  \BibitemOpen
  \bibfield  {author} {\bibinfo {author} {\bibfnamefont {R.~D.}\ \bibnamefont {Somma}}\ and\ \bibinfo {author} {\bibfnamefont {S.}~\bibnamefont {Boixo}},\ }\bibfield  {title} {\bibinfo {title} {Spectral gap amplification},\ }\href {https://doi.org/https://doi.org/10.1137/120871997} {\bibfield  {journal} {\bibinfo  {journal} {SIAM Journal on Computing}\ }\textbf {\bibinfo {volume} {42}},\ \bibinfo {pages} {593} (\bibinfo {year} {2013})}\BibitemShut {NoStop}%
\bibitem [{\citenamefont {Arad}\ \emph {et~al.}(2017)\citenamefont {Arad}, \citenamefont {Landau}, \citenamefont {Vazirani},\ and\ \citenamefont {Vidick}}]{arad2017rigorous}%
  \BibitemOpen
  \bibfield  {author} {\bibinfo {author} {\bibfnamefont {I.}~\bibnamefont {Arad}}, \bibinfo {author} {\bibfnamefont {Z.}~\bibnamefont {Landau}}, \bibinfo {author} {\bibfnamefont {U.}~\bibnamefont {Vazirani}},\ and\ \bibinfo {author} {\bibfnamefont {T.}~\bibnamefont {Vidick}},\ }\bibfield  {title} {\bibinfo {title} {Rigorous rg algorithms and area laws for low energy eigenstates in 1d},\ }\href {https://doi.org/https://doi.org/10.1007/s00220-017-2973-z} {\bibfield  {journal} {\bibinfo  {journal} {Communications in Mathematical Physics}\ }\textbf {\bibinfo {volume} {356}},\ \bibinfo {pages} {65} (\bibinfo {year} {2017})}\BibitemShut {NoStop}%
\bibitem [{\citenamefont {Peruzzo}\ \emph {et~al.}(2014)\citenamefont {Peruzzo}, \citenamefont {McClean}, \citenamefont {Shadbolt}, \citenamefont {Yung}, \citenamefont {Zhou}, \citenamefont {Love}, \citenamefont {Aspuru-Guzik},\ and\ \citenamefont {O’brien}}]{peruzzo2014variational}%
  \BibitemOpen
  \bibfield  {author} {\bibinfo {author} {\bibfnamefont {A.}~\bibnamefont {Peruzzo}}, \bibinfo {author} {\bibfnamefont {J.}~\bibnamefont {McClean}}, \bibinfo {author} {\bibfnamefont {P.}~\bibnamefont {Shadbolt}}, \bibinfo {author} {\bibfnamefont {M.-H.}\ \bibnamefont {Yung}}, \bibinfo {author} {\bibfnamefont {X.-Q.}\ \bibnamefont {Zhou}}, \bibinfo {author} {\bibfnamefont {P.~J.}\ \bibnamefont {Love}}, \bibinfo {author} {\bibfnamefont {A.}~\bibnamefont {Aspuru-Guzik}},\ and\ \bibinfo {author} {\bibfnamefont {J.~L.}\ \bibnamefont {O’brien}},\ }\bibfield  {title} {\bibinfo {title} {A variational eigenvalue solver on a photonic quantum processor},\ }\href@noop {} {\bibfield  {journal} {\bibinfo  {journal} {Nature communications}\ }\textbf {\bibinfo {volume} {5}},\ \bibinfo {pages} {4213} (\bibinfo {year} {2014})}\BibitemShut {NoStop}%
\bibitem [{\citenamefont {Kuzmin}\ and\ \citenamefont {Silvi}(2020)}]{kuzmin2020variational}%
  \BibitemOpen
  \bibfield  {author} {\bibinfo {author} {\bibfnamefont {V.~V.}\ \bibnamefont {Kuzmin}}\ and\ \bibinfo {author} {\bibfnamefont {P.}~\bibnamefont {Silvi}},\ }\bibfield  {title} {\bibinfo {title} {Variational quantum state preparation via quantum data buses},\ }\href@noop {} {\bibfield  {journal} {\bibinfo  {journal} {Quantum}\ }\textbf {\bibinfo {volume} {4}},\ \bibinfo {pages} {290} (\bibinfo {year} {2020})}\BibitemShut {NoStop}%
\bibitem [{\citenamefont {Sim}\ \emph {et~al.}(2019)\citenamefont {Sim}, \citenamefont {Johnson},\ and\ \citenamefont {Aspuru-Guzik}}]{sim2019expressibility}%
  \BibitemOpen
  \bibfield  {author} {\bibinfo {author} {\bibfnamefont {S.}~\bibnamefont {Sim}}, \bibinfo {author} {\bibfnamefont {P.~D.}\ \bibnamefont {Johnson}},\ and\ \bibinfo {author} {\bibfnamefont {A.}~\bibnamefont {Aspuru-Guzik}},\ }\bibfield  {title} {\bibinfo {title} {Expressibility and entangling capability of parameterized quantum circuits for hybrid quantum-classical algorithms},\ }\href@noop {} {\bibfield  {journal} {\bibinfo  {journal} {Advanced Quantum Technologies}\ }\textbf {\bibinfo {volume} {2}},\ \bibinfo {pages} {1900070} (\bibinfo {year} {2019})}\BibitemShut {NoStop}%
\bibitem [{\citenamefont {Benedetti}\ \emph {et~al.}(2019)\citenamefont {Benedetti}, \citenamefont {Lloyd}, \citenamefont {Sack},\ and\ \citenamefont {Fiorentini}}]{benedetti2019parameterized}%
  \BibitemOpen
  \bibfield  {author} {\bibinfo {author} {\bibfnamefont {M.}~\bibnamefont {Benedetti}}, \bibinfo {author} {\bibfnamefont {E.}~\bibnamefont {Lloyd}}, \bibinfo {author} {\bibfnamefont {S.}~\bibnamefont {Sack}},\ and\ \bibinfo {author} {\bibfnamefont {M.}~\bibnamefont {Fiorentini}},\ }\bibfield  {title} {\bibinfo {title} {Parameterized quantum circuits as machine learning models},\ }\href@noop {} {\bibfield  {journal} {\bibinfo  {journal} {Quantum science and technology}\ }\textbf {\bibinfo {volume} {4}},\ \bibinfo {pages} {043001} (\bibinfo {year} {2019})}\BibitemShut {NoStop}%
\bibitem [{\citenamefont {Grimsley}\ \emph {et~al.}(2019)\citenamefont {Grimsley}, \citenamefont {Economou}, \citenamefont {Barnes},\ and\ \citenamefont {Mayhall}}]{grimsley2019adaptive}%
  \BibitemOpen
  \bibfield  {author} {\bibinfo {author} {\bibfnamefont {H.~R.}\ \bibnamefont {Grimsley}}, \bibinfo {author} {\bibfnamefont {S.~E.}\ \bibnamefont {Economou}}, \bibinfo {author} {\bibfnamefont {E.}~\bibnamefont {Barnes}},\ and\ \bibinfo {author} {\bibfnamefont {N.~J.}\ \bibnamefont {Mayhall}},\ }\bibfield  {title} {\bibinfo {title} {An adaptive variational algorithm for exact molecular simulations on a quantum computer},\ }\href@noop {} {\bibfield  {journal} {\bibinfo  {journal} {Nature communications}\ }\textbf {\bibinfo {volume} {10}},\ \bibinfo {pages} {3007} (\bibinfo {year} {2019})}\BibitemShut {NoStop}%
\bibitem [{\citenamefont {Ragone}\ \emph {et~al.}(2024)\citenamefont {Ragone}, \citenamefont {Bakalov}, \citenamefont {Sauvage}, \citenamefont {Kemper}, \citenamefont {Ortiz~Marrero}, \citenamefont {Larocca},\ and\ \citenamefont {Cerezo}}]{ragone2024lie}%
  \BibitemOpen
  \bibfield  {author} {\bibinfo {author} {\bibfnamefont {M.}~\bibnamefont {Ragone}}, \bibinfo {author} {\bibfnamefont {B.~N.}\ \bibnamefont {Bakalov}}, \bibinfo {author} {\bibfnamefont {F.~A.}\ \bibnamefont {Sauvage}}, \bibinfo {author} {\bibfnamefont {A.~F.}\ \bibnamefont {Kemper}}, \bibinfo {author} {\bibfnamefont {C.}~\bibnamefont {Ortiz~Marrero}}, \bibinfo {author} {\bibfnamefont {M.}~\bibnamefont {Larocca}},\ and\ \bibinfo {author} {\bibfnamefont {M.~V.~S.}\ \bibnamefont {Cerezo}},\ }\bibfield  {title} {\bibinfo {title} {A lie algebraic theory of barren plateaus for deep parameterized quantum circuits},\ }\href@noop {} {\bibfield  {journal} {\bibinfo  {journal} {Nature Communications}\ }\textbf {\bibinfo {volume} {15}} (\bibinfo {year} {2024})}\BibitemShut {NoStop}%
\bibitem [{\citenamefont {McClean}\ \emph {et~al.}(2018)\citenamefont {McClean}, \citenamefont {Boixo}, \citenamefont {Smelyanskiy}, \citenamefont {Babbush},\ and\ \citenamefont {Neven}}]{mcclean2018barren}%
  \BibitemOpen
  \bibfield  {author} {\bibinfo {author} {\bibfnamefont {J.~R.}\ \bibnamefont {McClean}}, \bibinfo {author} {\bibfnamefont {S.}~\bibnamefont {Boixo}}, \bibinfo {author} {\bibfnamefont {V.~N.}\ \bibnamefont {Smelyanskiy}}, \bibinfo {author} {\bibfnamefont {R.}~\bibnamefont {Babbush}},\ and\ \bibinfo {author} {\bibfnamefont {H.}~\bibnamefont {Neven}},\ }\bibfield  {title} {\bibinfo {title} {Barren plateaus in quantum neural network training landscapes},\ }\href@noop {} {\bibfield  {journal} {\bibinfo  {journal} {Nature communications}\ }\textbf {\bibinfo {volume} {9}},\ \bibinfo {pages} {4812} (\bibinfo {year} {2018})}\BibitemShut {NoStop}%
\bibitem [{\citenamefont {Holmes}\ \emph {et~al.}(2022)\citenamefont {Holmes}, \citenamefont {Sharma}, \citenamefont {Cerezo},\ and\ \citenamefont {Coles}}]{holmes2022connecting}%
  \BibitemOpen
  \bibfield  {author} {\bibinfo {author} {\bibfnamefont {Z.}~\bibnamefont {Holmes}}, \bibinfo {author} {\bibfnamefont {K.}~\bibnamefont {Sharma}}, \bibinfo {author} {\bibfnamefont {M.}~\bibnamefont {Cerezo}},\ and\ \bibinfo {author} {\bibfnamefont {P.~J.}\ \bibnamefont {Coles}},\ }\bibfield  {title} {\bibinfo {title} {Connecting ansatz expressibility to gradient magnitudes and barren plateaus},\ }\href@noop {} {\bibfield  {journal} {\bibinfo  {journal} {PRX quantum}\ }\textbf {\bibinfo {volume} {3}},\ \bibinfo {pages} {010313} (\bibinfo {year} {2022})}\BibitemShut {NoStop}%
\bibitem [{\citenamefont {Larocca}\ \emph {et~al.}(2025)\citenamefont {Larocca}, \citenamefont {Thanasilp}, \citenamefont {Wang}, \citenamefont {Sharma}, \citenamefont {Biamonte}, \citenamefont {Coles}, \citenamefont {Cincio}, \citenamefont {McClean}, \citenamefont {Holmes},\ and\ \citenamefont {Cerezo}}]{larocca2025barren}%
  \BibitemOpen
  \bibfield  {author} {\bibinfo {author} {\bibfnamefont {M.}~\bibnamefont {Larocca}}, \bibinfo {author} {\bibfnamefont {S.}~\bibnamefont {Thanasilp}}, \bibinfo {author} {\bibfnamefont {S.}~\bibnamefont {Wang}}, \bibinfo {author} {\bibfnamefont {K.}~\bibnamefont {Sharma}}, \bibinfo {author} {\bibfnamefont {J.}~\bibnamefont {Biamonte}}, \bibinfo {author} {\bibfnamefont {P.~J.}\ \bibnamefont {Coles}}, \bibinfo {author} {\bibfnamefont {L.}~\bibnamefont {Cincio}}, \bibinfo {author} {\bibfnamefont {J.~R.}\ \bibnamefont {McClean}}, \bibinfo {author} {\bibfnamefont {Z.}~\bibnamefont {Holmes}},\ and\ \bibinfo {author} {\bibfnamefont {M.}~\bibnamefont {Cerezo}},\ }\bibfield  {title} {\bibinfo {title} {Barren plateaus in variational quantum computing},\ }\href@noop {} {\bibfield  {journal} {\bibinfo  {journal} {Nature Reviews Physics}\ ,\ \bibinfo {pages} {1}} (\bibinfo {year} {2025})}\BibitemShut {NoStop}%
\bibitem [{\citenamefont {Spall}(2005)}]{spall2005introduction}%
  \BibitemOpen
  \bibfield  {author} {\bibinfo {author} {\bibfnamefont {J.~C.}\ \bibnamefont {Spall}},\ }\href@noop {} {\emph {\bibinfo {title} {Introduction to stochastic search and optimization: estimation, simulation, and control}}}\ (\bibinfo  {publisher} {John Wiley \& Sons},\ \bibinfo {year} {2005})\BibitemShut {NoStop}%
\bibitem [{\citenamefont {Lavrijsen}\ \emph {et~al.}(2020)\citenamefont {Lavrijsen}, \citenamefont {Tudor}, \citenamefont {M{\"u}ller}, \citenamefont {Iancu},\ and\ \citenamefont {De~Jong}}]{lavrijsen2020classical}%
  \BibitemOpen
  \bibfield  {author} {\bibinfo {author} {\bibfnamefont {W.}~\bibnamefont {Lavrijsen}}, \bibinfo {author} {\bibfnamefont {A.}~\bibnamefont {Tudor}}, \bibinfo {author} {\bibfnamefont {J.}~\bibnamefont {M{\"u}ller}}, \bibinfo {author} {\bibfnamefont {C.}~\bibnamefont {Iancu}},\ and\ \bibinfo {author} {\bibfnamefont {W.}~\bibnamefont {De~Jong}},\ }\bibfield  {title} {\bibinfo {title} {Classical optimizers for noisy intermediate-scale quantum devices},\ }in\ \href@noop {} {\emph {\bibinfo {booktitle} {2020 IEEE international conference on quantum computing and engineering (QCE)}}}\ (\bibinfo {organization} {IEEE},\ \bibinfo {year} {2020})\ pp.\ \bibinfo {pages} {267--277}\BibitemShut {NoStop}%
\bibitem [{\citenamefont {Farhi}\ \emph {et~al.}(2001)\citenamefont {Farhi}, \citenamefont {Goldstone}, \citenamefont {Gutmann}, \citenamefont {Lapan}, \citenamefont {Lundgren},\ and\ \citenamefont {Preda}}]{farhi2001quantum}%
  \BibitemOpen
  \bibfield  {author} {\bibinfo {author} {\bibfnamefont {E.}~\bibnamefont {Farhi}}, \bibinfo {author} {\bibfnamefont {J.}~\bibnamefont {Goldstone}}, \bibinfo {author} {\bibfnamefont {S.}~\bibnamefont {Gutmann}}, \bibinfo {author} {\bibfnamefont {J.}~\bibnamefont {Lapan}}, \bibinfo {author} {\bibfnamefont {A.}~\bibnamefont {Lundgren}},\ and\ \bibinfo {author} {\bibfnamefont {D.}~\bibnamefont {Preda}},\ }\bibfield  {title} {\bibinfo {title} {A quantum adiabatic evolution algorithm applied to random instances of an np-complete problem},\ }\href@noop {} {\bibfield  {journal} {\bibinfo  {journal} {Science}\ }\textbf {\bibinfo {volume} {292}},\ \bibinfo {pages} {472} (\bibinfo {year} {2001})}\BibitemShut {NoStop}%
\bibitem [{\citenamefont {Aspuru-Guzik}\ \emph {et~al.}(2005)\citenamefont {Aspuru-Guzik}, \citenamefont {Dutoi}, \citenamefont {Love},\ and\ \citenamefont {Head-Gordon}}]{aspuru2005simulated}%
  \BibitemOpen
  \bibfield  {author} {\bibinfo {author} {\bibfnamefont {A.}~\bibnamefont {Aspuru-Guzik}}, \bibinfo {author} {\bibfnamefont {A.~D.}\ \bibnamefont {Dutoi}}, \bibinfo {author} {\bibfnamefont {P.~J.}\ \bibnamefont {Love}},\ and\ \bibinfo {author} {\bibfnamefont {M.}~\bibnamefont {Head-Gordon}},\ }\bibfield  {title} {\bibinfo {title} {Simulated quantum computation of molecular energies},\ }\href@noop {} {\bibfield  {journal} {\bibinfo  {journal} {Science}\ }\textbf {\bibinfo {volume} {309}},\ \bibinfo {pages} {1704} (\bibinfo {year} {2005})}\BibitemShut {NoStop}%
\bibitem [{\citenamefont {Farhi}\ \emph {et~al.}(2014)\citenamefont {Farhi}, \citenamefont {Goldstone},\ and\ \citenamefont {Gutmann}}]{farhi2014quantum}%
  \BibitemOpen
  \bibfield  {author} {\bibinfo {author} {\bibfnamefont {E.}~\bibnamefont {Farhi}}, \bibinfo {author} {\bibfnamefont {J.}~\bibnamefont {Goldstone}},\ and\ \bibinfo {author} {\bibfnamefont {S.}~\bibnamefont {Gutmann}},\ }\bibfield  {title} {\bibinfo {title} {A quantum approximate optimization algorithm},\ }\href@noop {} {\bibfield  {journal} {\bibinfo  {journal} {arXiv preprint arXiv:1411.4028}\ } (\bibinfo {year} {2014})}\BibitemShut {NoStop}%
\bibitem [{\citenamefont {Nielsen}\ and\ \citenamefont {Chuang}(2010)}]{nielsen2010quantum}%
  \BibitemOpen
  \bibfield  {author} {\bibinfo {author} {\bibfnamefont {M.~A.}\ \bibnamefont {Nielsen}}\ and\ \bibinfo {author} {\bibfnamefont {I.~L.}\ \bibnamefont {Chuang}},\ }\href@noop {} {\emph {\bibinfo {title} {Quantum computation and quantum information}}}\ (\bibinfo  {publisher} {Cambridge university press},\ \bibinfo {year} {2010})\BibitemShut {NoStop}%
\bibitem [{\citenamefont {Nenciu}(1993)}]{nenciu1993linear}%
  \BibitemOpen
  \bibfield  {author} {\bibinfo {author} {\bibfnamefont {G.}~\bibnamefont {Nenciu}},\ }\bibfield  {title} {\bibinfo {title} {Linear adiabatic theory. exponential estimates},\ }\href@noop {} {\bibfield  {journal} {\bibinfo  {journal} {Communications in mathematical physics}\ }\textbf {\bibinfo {volume} {152}},\ \bibinfo {pages} {479} (\bibinfo {year} {1993})}\BibitemShut {NoStop}%
\bibitem [{\citenamefont {Cugini}\ \emph {et~al.}(2025)\citenamefont {Cugini}, \citenamefont {Nigro}, \citenamefont {Bruno},\ and\ \citenamefont {Gerace}}]{cugini2025exponential}%
  \BibitemOpen
  \bibfield  {author} {\bibinfo {author} {\bibfnamefont {D.}~\bibnamefont {Cugini}}, \bibinfo {author} {\bibfnamefont {D.}~\bibnamefont {Nigro}}, \bibinfo {author} {\bibfnamefont {M.}~\bibnamefont {Bruno}},\ and\ \bibinfo {author} {\bibfnamefont {D.}~\bibnamefont {Gerace}},\ }\bibfield  {title} {\bibinfo {title} {Exponential optimization of adiabatic quantum-state preparation},\ }\href@noop {} {\bibfield  {journal} {\bibinfo  {journal} {Physical Review Research}\ }\textbf {\bibinfo {volume} {7}},\ \bibinfo {pages} {L012074} (\bibinfo {year} {2025})}\BibitemShut {NoStop}%
\bibitem [{\citenamefont {Landau}\ and\ \citenamefont {Lifshitz}(2013)}]{landau2013quantum}%
  \BibitemOpen
  \bibfield  {author} {\bibinfo {author} {\bibfnamefont {L.~D.}\ \bibnamefont {Landau}}\ and\ \bibinfo {author} {\bibfnamefont {E.~M.}\ \bibnamefont {Lifshitz}},\ }\href@noop {} {\emph {\bibinfo {title} {Quantum mechanics: non-relativistic theory}}},\ Vol.~\bibinfo {volume} {3}\ (\bibinfo  {publisher} {Elsevier},\ \bibinfo {year} {2013})\BibitemShut {NoStop}%
\bibitem [{\citenamefont {Chakraborty}\ \emph {et~al.}(2022)\citenamefont {Chakraborty}, \citenamefont {Honda}, \citenamefont {Izubuchi}, \citenamefont {Kikuchi},\ and\ \citenamefont {Tomiya}}]{chakraborty2022classically}%
  \BibitemOpen
  \bibfield  {author} {\bibinfo {author} {\bibfnamefont {B.}~\bibnamefont {Chakraborty}}, \bibinfo {author} {\bibfnamefont {M.}~\bibnamefont {Honda}}, \bibinfo {author} {\bibfnamefont {T.}~\bibnamefont {Izubuchi}}, \bibinfo {author} {\bibfnamefont {Y.}~\bibnamefont {Kikuchi}},\ and\ \bibinfo {author} {\bibfnamefont {A.}~\bibnamefont {Tomiya}},\ }\bibfield  {title} {\bibinfo {title} {Classically emulated digital quantum simulation of the schwinger model with a topological term via adiabatic state preparation},\ }\href@noop {} {\bibfield  {journal} {\bibinfo  {journal} {Physical Review D}\ }\textbf {\bibinfo {volume} {105}},\ \bibinfo {pages} {094503} (\bibinfo {year} {2022})}\BibitemShut {NoStop}%
\bibitem [{\citenamefont {Hamma}\ and\ \citenamefont {Lidar}(2008)}]{hamma2008adiabatic}%
  \BibitemOpen
  \bibfield  {author} {\bibinfo {author} {\bibfnamefont {A.}~\bibnamefont {Hamma}}\ and\ \bibinfo {author} {\bibfnamefont {D.~A.}\ \bibnamefont {Lidar}},\ }\bibfield  {title} {\bibinfo {title} {Adiabatic preparation of topological order},\ }\href@noop {} {\bibfield  {journal} {\bibinfo  {journal} {Physical review letters}\ }\textbf {\bibinfo {volume} {100}},\ \bibinfo {pages} {030502} (\bibinfo {year} {2008})}\BibitemShut {NoStop}%
\bibitem [{\citenamefont {Ciavarella}\ \emph {et~al.}(2023)\citenamefont {Ciavarella}, \citenamefont {Caspar}, \citenamefont {Illa},\ and\ \citenamefont {Savage}}]{ciavarella2023state}%
  \BibitemOpen
  \bibfield  {author} {\bibinfo {author} {\bibfnamefont {A.~N.}\ \bibnamefont {Ciavarella}}, \bibinfo {author} {\bibfnamefont {S.}~\bibnamefont {Caspar}}, \bibinfo {author} {\bibfnamefont {M.}~\bibnamefont {Illa}},\ and\ \bibinfo {author} {\bibfnamefont {M.~J.}\ \bibnamefont {Savage}},\ }\bibfield  {title} {\bibinfo {title} {State preparation in the heisenberg model through adiabatic spiraling},\ }\href@noop {} {\bibfield  {journal} {\bibinfo  {journal} {Quantum}\ }\textbf {\bibinfo {volume} {7}},\ \bibinfo {pages} {970} (\bibinfo {year} {2023})}\BibitemShut {NoStop}%
\bibitem [{\citenamefont {Murnaghan}(1962)}]{murnaghan1962unitary}%
  \BibitemOpen
  \bibfield  {author} {\bibinfo {author} {\bibfnamefont {F.~D.}\ \bibnamefont {Murnaghan}},\ }\bibfield  {title} {\bibinfo {title} {The unitary and rotation groups},\ }\href@noop {} {\bibfield  {journal} {\bibinfo  {journal} {(No Title)}\ } (\bibinfo {year} {1962})}\BibitemShut {NoStop}%
\bibitem [{\citenamefont {Parrish}\ \emph {et~al.}(2019)\citenamefont {Parrish}, \citenamefont {Hohenstein}, \citenamefont {McMahon},\ and\ \citenamefont {Mart{\'\i}nez}}]{parrish2019quantum}%
  \BibitemOpen
  \bibfield  {author} {\bibinfo {author} {\bibfnamefont {R.~M.}\ \bibnamefont {Parrish}}, \bibinfo {author} {\bibfnamefont {E.~G.}\ \bibnamefont {Hohenstein}}, \bibinfo {author} {\bibfnamefont {P.~L.}\ \bibnamefont {McMahon}},\ and\ \bibinfo {author} {\bibfnamefont {T.~J.}\ \bibnamefont {Mart{\'\i}nez}},\ }\bibfield  {title} {\bibinfo {title} {Quantum computation of electronic transitions using a variational quantum eigensolver},\ }\href@noop {} {\bibfield  {journal} {\bibinfo  {journal} {Physical review letters}\ }\textbf {\bibinfo {volume} {122}},\ \bibinfo {pages} {230401} (\bibinfo {year} {2019})}\BibitemShut {NoStop}%
\bibitem [{\citenamefont {Lanczos}(1950)}]{lanczos1950iteration}%
  \BibitemOpen
  \bibfield  {author} {\bibinfo {author} {\bibfnamefont {C.}~\bibnamefont {Lanczos}},\ }\bibfield  {title} {\bibinfo {title} {An iteration method for the solution of the eigenvalue problem of linear differential and integral operators},\ }\href@noop {} {\  (\bibinfo {year} {1950})}\BibitemShut {NoStop}%
\bibitem [{\citenamefont {Osterkorn}\ \emph {et~al.}(2023)\citenamefont {Osterkorn}, \citenamefont {Meyer},\ and\ \citenamefont {Manmana}}]{osterkorn2023gap}%
  \BibitemOpen
  \bibfield  {author} {\bibinfo {author} {\bibfnamefont {A.}~\bibnamefont {Osterkorn}}, \bibinfo {author} {\bibfnamefont {C.}~\bibnamefont {Meyer}},\ and\ \bibinfo {author} {\bibfnamefont {S.~R.}\ \bibnamefont {Manmana}},\ }\bibfield  {title} {\bibinfo {title} {In-gap band formation in a periodically driven charge density wave insulator},\ }\href {https://doi.org/https://doi.org/10.1038/s42005-023-01346-2} {\bibfield  {journal} {\bibinfo  {journal} {Communications Physics}\ }\textbf {\bibinfo {volume} {6}},\ \bibinfo {pages} {245} (\bibinfo {year} {2023})}\BibitemShut {NoStop}%
\bibitem [{\citenamefont {Cubitt}\ \emph {et~al.}(2015)\citenamefont {Cubitt}, \citenamefont {Perez-Garcia},\ and\ \citenamefont {Wolf}}]{cubitt2015undecidability}%
  \BibitemOpen
  \bibfield  {author} {\bibinfo {author} {\bibfnamefont {T.~S.}\ \bibnamefont {Cubitt}}, \bibinfo {author} {\bibfnamefont {D.}~\bibnamefont {Perez-Garcia}},\ and\ \bibinfo {author} {\bibfnamefont {M.~M.}\ \bibnamefont {Wolf}},\ }\bibfield  {title} {\bibinfo {title} {Undecidability of the spectral gap},\ }\href {https://doi.org/https://doi.org/10.1038/nature16059} {\bibfield  {journal} {\bibinfo  {journal} {Nature}\ }\textbf {\bibinfo {volume} {528}},\ \bibinfo {pages} {207} (\bibinfo {year} {2015})}\BibitemShut {NoStop}%
\bibitem [{\citenamefont {Gosset}\ and\ \citenamefont {Huang}(2016)}]{gosset2016correlation}%
  \BibitemOpen
  \bibfield  {author} {\bibinfo {author} {\bibfnamefont {D.}~\bibnamefont {Gosset}}\ and\ \bibinfo {author} {\bibfnamefont {Y.}~\bibnamefont {Huang}},\ }\bibfield  {title} {\bibinfo {title} {Correlation length versus gap in frustration-free systems},\ }\href {https://doi.org/https://doi.org/10.1103/PhysRevLett.116.097202} {\bibfield  {journal} {\bibinfo  {journal} {Physical review letters}\ }\textbf {\bibinfo {volume} {116}},\ \bibinfo {pages} {097202} (\bibinfo {year} {2016})}\BibitemShut {NoStop}%
\bibitem [{\citenamefont {Can}\ \emph {et~al.}(2019)\citenamefont {Can}, \citenamefont {Oganesyan}, \citenamefont {Orgad},\ and\ \citenamefont {Gopalakrishnan}}]{can2019spectral}%
  \BibitemOpen
  \bibfield  {author} {\bibinfo {author} {\bibfnamefont {T.}~\bibnamefont {Can}}, \bibinfo {author} {\bibfnamefont {V.}~\bibnamefont {Oganesyan}}, \bibinfo {author} {\bibfnamefont {D.}~\bibnamefont {Orgad}},\ and\ \bibinfo {author} {\bibfnamefont {S.}~\bibnamefont {Gopalakrishnan}},\ }\bibfield  {title} {\bibinfo {title} {Spectral gaps and midgap states in random quantum master equations},\ }\href {https://doi.org/https://doi.org/10.1103/PhysRevLett.123.234103} {\bibfield  {journal} {\bibinfo  {journal} {Physical review letters}\ }\textbf {\bibinfo {volume} {123}},\ \bibinfo {pages} {234103} (\bibinfo {year} {2019})}\BibitemShut {NoStop}%
\bibitem [{\citenamefont {Hastings}(2007)}]{hastings2007area}%
  \BibitemOpen
  \bibfield  {author} {\bibinfo {author} {\bibfnamefont {M.~B.}\ \bibnamefont {Hastings}},\ }\bibfield  {title} {\bibinfo {title} {An area law for one-dimensional quantum systems},\ }\href {https://doi.org/10.1088/1742-5468/2007/08/P08024} {\bibfield  {journal} {\bibinfo  {journal} {Journal of statistical mechanics: theory and experiment}\ }\textbf {\bibinfo {volume} {2007}},\ \bibinfo {pages} {P08024} (\bibinfo {year} {2007})}\BibitemShut {NoStop}%
\bibitem [{\citenamefont {Landau}\ \emph {et~al.}(2015)\citenamefont {Landau}, \citenamefont {Vazirani},\ and\ \citenamefont {Vidick}}]{landau2015polynomial}%
  \BibitemOpen
  \bibfield  {author} {\bibinfo {author} {\bibfnamefont {Z.}~\bibnamefont {Landau}}, \bibinfo {author} {\bibfnamefont {U.}~\bibnamefont {Vazirani}},\ and\ \bibinfo {author} {\bibfnamefont {T.}~\bibnamefont {Vidick}},\ }\bibfield  {title} {\bibinfo {title} {A polynomial time algorithm for the ground state of one-dimensional gapped local hamiltonians},\ }\href {https://doi.org/https://doi.org/10.1038/nphys3345} {\bibfield  {journal} {\bibinfo  {journal} {Nature Physics}\ }\textbf {\bibinfo {volume} {11}},\ \bibinfo {pages} {566} (\bibinfo {year} {2015})}\BibitemShut {NoStop}%
\bibitem [{\citenamefont {Sachdev}(2011)}]{sachdev2011quantum}%
  \BibitemOpen
  \bibfield  {author} {\bibinfo {author} {\bibfnamefont {S.}~\bibnamefont {Sachdev}},\ }\href {https://doi.org/10.1017/CBO9780511973765} {\emph {\bibinfo {title} {Quantum Phase Transitions}}}\ (\bibinfo  {publisher} {Cambridge University Press},\ \bibinfo {address} {Cambridge},\ \bibinfo {year} {2011})\ \bibinfo {note} {online publication date: May 2011}\BibitemShut {NoStop}%
\bibitem [{\citenamefont {Hastings}\ and\ \citenamefont {Koma}(2006)}]{hastings2006spectral}%
  \BibitemOpen
  \bibfield  {author} {\bibinfo {author} {\bibfnamefont {M.~B.}\ \bibnamefont {Hastings}}\ and\ \bibinfo {author} {\bibfnamefont {T.}~\bibnamefont {Koma}},\ }\bibfield  {title} {\bibinfo {title} {Spectral gap and exponential decay of correlations},\ }\href {https://doi.org/https://doi.org/10.1007/s00220-006-0030-4} {\bibfield  {journal} {\bibinfo  {journal} {Communications in mathematical physics}\ }\textbf {\bibinfo {volume} {265}},\ \bibinfo {pages} {781} (\bibinfo {year} {2006})}\BibitemShut {NoStop}%
\bibitem [{\citenamefont {Kohn}(1964)}]{kohn1964theory}%
  \BibitemOpen
  \bibfield  {author} {\bibinfo {author} {\bibfnamefont {W.}~\bibnamefont {Kohn}},\ }\bibfield  {title} {\bibinfo {title} {Theory of the insulating state},\ }\href@noop {} {\bibfield  {journal} {\bibinfo  {journal} {Physical review}\ }\textbf {\bibinfo {volume} {133}},\ \bibinfo {pages} {A171} (\bibinfo {year} {1964})}\BibitemShut {NoStop}%
\bibitem [{\citenamefont {Szyma{\'n}ski}\ and\ \citenamefont {{\.Z}yczkowski}(2022)}]{szymanski2022universal}%
  \BibitemOpen
  \bibfield  {author} {\bibinfo {author} {\bibfnamefont {K.}~\bibnamefont {Szyma{\'n}ski}}\ and\ \bibinfo {author} {\bibfnamefont {K.}~\bibnamefont {{\.Z}yczkowski}},\ }\bibfield  {title} {\bibinfo {title} {Universal witnesses of vanishing energy gap},\ }\href@noop {} {\bibfield  {journal} {\bibinfo  {journal} {Europhysics Letters}\ }\textbf {\bibinfo {volume} {136}},\ \bibinfo {pages} {30003} (\bibinfo {year} {2022})}\BibitemShut {NoStop}%
\bibitem [{\citenamefont {Chung}\ and\ \citenamefont {Shapiro}(2023)}]{chung2023topological}%
  \BibitemOpen
  \bibfield  {author} {\bibinfo {author} {\bibfnamefont {J.-H.}\ \bibnamefont {Chung}}\ and\ \bibinfo {author} {\bibfnamefont {J.}~\bibnamefont {Shapiro}},\ }\bibfield  {title} {\bibinfo {title} {Topological classification of insulators: I. non-interacting spectrally-gapped one-dimensional systems},\ }\href@noop {} {\bibfield  {journal} {\bibinfo  {journal} {arXiv preprint arXiv:2306.00268}\ } (\bibinfo {year} {2023})}\BibitemShut {NoStop}%
\bibitem [{\citenamefont {Rai}\ \emph {et~al.}(2024)\citenamefont {Rai}, \citenamefont {Chen}, \citenamefont {Emonts},\ and\ \citenamefont {Tura}}]{rai2024spectral}%
  \BibitemOpen
  \bibfield  {author} {\bibinfo {author} {\bibfnamefont {K.~S.}\ \bibnamefont {Rai}}, \bibinfo {author} {\bibfnamefont {J.-F.}\ \bibnamefont {Chen}}, \bibinfo {author} {\bibfnamefont {P.}~\bibnamefont {Emonts}},\ and\ \bibinfo {author} {\bibfnamefont {J.}~\bibnamefont {Tura}},\ }\bibfield  {title} {\bibinfo {title} {Spectral gap optimization for enhanced adiabatic state preparation},\ }\bibfield  {journal} {\bibinfo  {journal} {arXiv preprint arXiv:2409.15433}\ }\href {https://doi.org/https://doi.org/10.48550/arXiv.2409.15433} {https://doi.org/10.48550/arXiv.2409.15433} (\bibinfo {year} {2024})\BibitemShut {NoStop}%
\bibitem [{\citenamefont {{Qiskit contributors}}(2023)}]{Qiskit}%
  \BibitemOpen
  \bibfield  {author} {\bibinfo {author} {\bibnamefont {{Qiskit contributors}}},\ }\href {https://doi.org/10.5281/zenodo.2573505} {\bibinfo {title} {Qiskit: An open-source framework for quantum computing}} (\bibinfo {year} {2023})\BibitemShut {NoStop}%
\bibitem [{\citenamefont {Cerezo}\ \emph {et~al.}(2023)\citenamefont {Cerezo}, \citenamefont {Larocca}, \citenamefont {Garc{\'\i}a-Mart{\'\i}n}, \citenamefont {Diaz}, \citenamefont {Braccia}, \citenamefont {Fontana}, \citenamefont {Rudolph}, \citenamefont {Bermejo}, \citenamefont {Ijaz}, \citenamefont {Thanasilp} \emph {et~al.}}]{cerezo2023does}%
  \BibitemOpen
  \bibfield  {author} {\bibinfo {author} {\bibfnamefont {M.}~\bibnamefont {Cerezo}}, \bibinfo {author} {\bibfnamefont {M.}~\bibnamefont {Larocca}}, \bibinfo {author} {\bibfnamefont {D.}~\bibnamefont {Garc{\'\i}a-Mart{\'\i}n}}, \bibinfo {author} {\bibfnamefont {N.~L.}\ \bibnamefont {Diaz}}, \bibinfo {author} {\bibfnamefont {P.}~\bibnamefont {Braccia}}, \bibinfo {author} {\bibfnamefont {E.}~\bibnamefont {Fontana}}, \bibinfo {author} {\bibfnamefont {M.~S.}\ \bibnamefont {Rudolph}}, \bibinfo {author} {\bibfnamefont {P.}~\bibnamefont {Bermejo}}, \bibinfo {author} {\bibfnamefont {A.}~\bibnamefont {Ijaz}}, \bibinfo {author} {\bibfnamefont {S.}~\bibnamefont {Thanasilp}}, \emph {et~al.},\ }\bibfield  {title} {\bibinfo {title} {Does provable absence of barren plateaus imply classical simulability? or, why we need to rethink variational quantum computing},\ }\href@noop {} {\bibfield  {journal} {\bibinfo  {journal} {arXiv preprint arXiv:2312.09121}\ } (\bibinfo {year} {2023})}\BibitemShut {NoStop}%
\bibitem [{\citenamefont {Arrazola}\ \emph {et~al.}(2022)\citenamefont {Arrazola}, \citenamefont {Di~Matteo}, \citenamefont {Quesada}, \citenamefont {Jahangiri}, \citenamefont {Delgado},\ and\ \citenamefont {Killoran}}]{arrazola2022universal}%
  \BibitemOpen
  \bibfield  {author} {\bibinfo {author} {\bibfnamefont {J.~M.}\ \bibnamefont {Arrazola}}, \bibinfo {author} {\bibfnamefont {O.}~\bibnamefont {Di~Matteo}}, \bibinfo {author} {\bibfnamefont {N.}~\bibnamefont {Quesada}}, \bibinfo {author} {\bibfnamefont {S.}~\bibnamefont {Jahangiri}}, \bibinfo {author} {\bibfnamefont {A.}~\bibnamefont {Delgado}},\ and\ \bibinfo {author} {\bibfnamefont {N.}~\bibnamefont {Killoran}},\ }\bibfield  {title} {\bibinfo {title} {Universal quantum circuits for quantum chemistry},\ }\href@noop {} {\bibfield  {journal} {\bibinfo  {journal} {Quantum}\ }\textbf {\bibinfo {volume} {6}},\ \bibinfo {pages} {742} (\bibinfo {year} {2022})}\BibitemShut {NoStop}%
\end{thebibliography}%

\end{document}